\pdfoutput=1
\RequirePackage[T1]{fontenc}
\documentclass[12pt]{article}

\usepackage[height=8.85in,width=6.45in]{geometry}

\usepackage[utf8]{inputenc}
\usepackage{amsmath}
\usepackage{amssymb}
\usepackage{mathtools}
\numberwithin{equation}{section}
\usepackage{slashed}
\usepackage{braket}
\usepackage{enumitem}
\usepackage[svgnames,psnames]{xcolor}
\usepackage[colorlinks,citecolor=DarkGreen,linkcolor=FireBrick,urlcolor=FireBrick,linktocpage,unicode,psdextra]{hyperref}
\usepackage{cite}
\usepackage{graphicx}
\usepackage{tikz}
\usepackage{tikz-cd}
\usetikzlibrary{calc,topaths,decorations,decorations.pathmorphing,arrows,decorations.markings,cd}
\tikzset{
->-/.style args={#1rotate#2}{decoration={markings, mark=at position #1 with {\arrow[scale=1.5,rotate = #2 ]{stealth}}}, postaction={decorate}}
}

\tikzset{line/.style={line width=0.25mm},
curve/.style={line,smooth,tension=1},
->-/.style={decoration={
  markings,
  mark=at position #1 with {\arrow[>=stealth]{>}}},postaction={decorate}},
-<-/.style={decoration={
  markings,
  mark=at position #1 with {\arrow[>=stealth]{<}}},postaction={decorate}},
}

\usepackage[bottom]{footmisc}

\usepackage{times}
\usepackage{courier}
\usepackage{bm}
\usepackage{subfig}

\usepackage{mdframed}

\renewenvironment{figure}[1][]{
  \begin{originalfigure}[#1]
    \begin{mdframed}[linecolor=black!0,backgroundcolor=black!1]
}{
    \end{mdframed}
  \end{originalfigure}
}

\def\cD{\mathcal{D}}

\def\cV{\mathcal{V}}
\def\B{\mathbb{B}}
\def\D{\mathbb{D}}
\def\ZC{\mathcal{Z}(\mathcal{C})}

\def\bC{\mathbb{C}}
\def\cZ{\mathcal{Z}}
\def\cH{\mathcal{H}}
\def\cA{\mathcal{A}}
\def\cC{\mathcal{C}}
\def\cL{\mathcal{L}}
\def\cM{\mathcal{M}}
\def\cO{\mathcal{O}}
\def\bB{\mathbb{B}}
\def\bD{\mathbb{D}}

\def\tr{\mathop{\mathrm{tr}}\nolimits}
\def\Hom{\mathop{\mathrm{Hom}}\nolimits}
\def\Vec{\mathop{\mathrm{Vec}}\nolimits}
\def\tot{\mathrm{total}}

\newcommand{\tub}{\mathrm{Tube}}
\newcommand{\rep}{\mathrm{Rep}}
\newcommand{\bim}[3]{{}_{#1}{#2}_{#3}}

\IfFileExists{ying}{
  \usetikzlibrary{external}
  \tikzexternalize[prefix=./figures/]
}

\newcommand{\TV}{\mathrm{TV}}

\newcommand{\plane}[1]{
  \begin{scope}[shift={({#1-0.75},-1.75)}]
    \draw[color=blue!70!green, dashed] (-0.1,-0.3) -- (-0.1,2.1) -- (1.6,3.8) -- (1.6,1.4) -- cycle;
  \end{scope}
}

\newcommand{\idem}[1]{P^{(#1)}}

\newcommand{\x}[4]{
  \begin{gathered}
  \begin{tikzpicture}[baseline=(X.base)]
  \draw [thick, 
  decoration = {markings, mark=at position 0.8 with {\arrow[scale=1]{stealth}}}, postaction=decorate] (-1,0) -- (1,0) node[right] (X) {$#3$};
  \draw [thick, 
    decoration = {markings, mark=at position 0.9 with {\arrow[scale=1]{stealth}}}, postaction=decorate] (0,-0.5) node[below]{$#1$} -- (0,0.5) node[above]{$#2$};
  \fill [
    ] (0,0) circle (0.05) node [above right] {\footnotesize $#4$};
  \end{tikzpicture}
  \end{gathered}
}

\newcommand{\xx}[3]{
  \begin{gathered}
  \begin{tikzpicture}
    \draw [thick, 
    decoration = {markings, mark=at position 0.8 with {\arrow[scale=1]{stealth}}}, postaction=decorate] (-1,0.5) -- (1,0.5) node[right]{$#2$};
    \draw [preaction={draw=white,line width=6pt}, thick, 
    decoration = {markings, mark=at position 0.9 with {\arrow[scale=1]{stealth}}}, postaction=decorate] (0,0) node[below]{~} -- (0,1.5) node[above]{$#1$};
    \draw [preaction={draw=white,line width=6pt}, thick, 
    decoration = {markings, mark=at position 0.8 with {\arrow[scale=1]{stealth}}}, postaction=decorate] (-1,1) -- (1,1) node[right]{$#3$};
  \end{tikzpicture}
  \end{gathered}
}

\usetikzlibrary{calc,shapes.geometric}
\pgfmathparse{atan2(0,1)}
\ifdim\pgfmathresult pt=0pt
  \tikzset{declare function={atanXY(\x,\y)=atan2(\y,\x);atanYX(\y,\x)=atan2(\y,\x);}}
\else                 
  \tikzset{declare function={atanXY(\x,\y)=atan2(\x,\y);atanYX(\y,\x)=atan2(\x,\y);}}
\fi

\begin{document}

\begin{titlepage}

\begin{flushright}
YITP-SB-2022-29
\end{flushright}

\vskip 3cm

\begin{center}

{\Large \bfseries  Asymptotic density of states in \\[1em]
2d CFTs with non-invertible symmetries }

\vskip 1cm
Ying-Hsuan Lin$^1$,
Masaki Okada$^2$,
Sahand Seifnashri$^3$, 
and Yuji Tachikawa$^2$
\vskip 1cm

\begin{tabular}{ll}
1 & Jefferson Physical Laboratory, Harvard University, Cambridge, MA 02138, USA\\
2 & Kavli Institute for the Physics and Mathematics of the Universe (WPI), \\
& University of Tokyo,  Kashiwa, Chiba 277-8583, Japan \\
3 & C. N. Yang Institute for Theoretical Physics 
and Simons Center for Geometry and Physics, \\
&Stony Brook University, Stony Brook, NY 11794-3636, USA
\end{tabular}

\vskip 2cm

\end{center}

\noindent 
It is known that the asymptotic density of states of a 2d CFT 
in an irreducible representation $\rho$ of a finite symmetry group $G$
is proportional to $(\dim\rho)^2$.
We show how this statement can be generalized when the symmetry can be non-invertible and is described by a fusion category $\mathcal{C}$. Along the way, we explain what plays the role of a representation of a group in the case of a fusion category symmetry; the answer to this question is already available in the broader mathematical physics literature but not yet widely known in hep-th.
This understanding immediately implies a selection rule on the correlation functions, and also allows us to derive the asymptotic density.

\end{titlepage}

\setcounter{tocdepth}{2}
\tableofcontents

\section{Introduction and summary}
\label{sec:introduction}

\paragraph{Backgrounds:} A fundamental result of Cardy \cite{Cardy:1986ie} says that 
the asymptotic density of states of a unitary conformal field theory (CFT) in two dimensions
is universally given in terms of its central charge.
More recently, Pal and Sun \cite{Pal:2020wwd} demonstrated that the 
asymptotic density of states of a 2d CFT\footnote{
Here we note some generalizations that appeared in the literature.
A generalization to higher dimensions was proposed in \cite{Harlow:2021trr},
the validity of which in the case of weakly coupled theories was studied in \cite{Cao:2021euf}.
A proof of the higher-dimensional statement in the context of algebraic quantum field theory was given in \cite{Magan:2021myk}.
More recently, a generalization to the case of continuous symmetry groups was performed in \cite{Kang:2022orq}.} in a specific irreducible representation $\rho$
of a faithful finite symmetry group $G$ is also given universally,
with a proportionality factor of $(\dim\rho)^2$, 
under a mild assumption on the nature of the 2d CFT in question.

We now know that the concept of symmetries goes beyond those described by groups.
In two dimensions, a symmetry operation $g\in G$ can be represented by a topological domain wall\footnote{
We use the phrases `domain walls', `defect lines', and `line operators' interchangeably. 
} $W_g$ across which the operation is performed.
In this case, the action of $W_g$ can be invertible, in the sense that 
there is a topological domain wall $W_{g^{-1}}$ for the inverse element
such that the fusion of two domain walls $W_g$ and  $W_{g^{-1}}$
becomes trivial.
Topological domain walls are, however, not necessarily invertible in this sense,
and we can regard the algebraic structure formed by the totality of topological domain walls as a generalized version of symmetry.
It is by now well established that this type of generalized symmetries---when finite---is described by fusion categories \cite{Fuchs:2002cm,Frohlich:2009gb,Carqueville:2012dk,Brunner:2013ota,Buican:2017rxc,Bhardwaj:2017xup,Vanhove:2018wlb,Chang:2018iay,Aasen:2020jwb}.\footnote{Although rational CFTs are important examples of theories with finite non-invertible symmetries, non-rational CFTs can also have finite non-invertible (sub)symmetries. Therefore, we do not have to restrict ourselves to rational CFTs.}

Given this development, it is natural to decompose 
the asymptotic density of states of a 2d CFT with a symmetry described by a fusion category\footnote{Here we are interested in unitary theories, hence we only consider \emph{unitary} fusion categories throughout the paper. \label{footnote:unitary}} $\cC$
according to which `irreducible representation of $\cC$' each state belongs to
in an appropriately generalized sense.
This is exactly the task we are going to achieve in this paper.

\paragraph{Two-dimensional approach: $\mathrm{Tube}(\cC)$}
One important question is finding the appropriate notion of representations for non-invertible symmetries. 
There are two different-looking approaches, which happen to be mathematically equivalent.
The first approach is purely two-dimensional.
In the group symmetry case, the basic operation is the action of $g\in G$ on a point operator, 
which can be implemented in terms of the domain wall as follows:
\begin{equation}
    \begin{tikzpicture}[scale=.5,baseline=(A.south)]
    \node [draw, fill=black, circle, inner sep=1pt] (A) at (0,0)  {};
    \draw (0,0) node [below] {$\cO(x)$};
    \draw [thick, decoration = {markings, mark=at position .5 with {\arrow[scale=1]{stealth[reversed]}}, mark=at position 1 with {\arrow[scale=1]{stealth[reversed]}}}, postaction=decorate] (0,0) circle (2);
    \draw (-2,0) node[left] {$g$};
    \end{tikzpicture} \quad = 
    \begin{tikzpicture}[scale=.5,baseline=(A.south)]
    \node [draw, fill=black, circle, inner sep=1pt] (A) at (0,0)  {};
    \draw (0,0) node [below] {$\cO'(x)$};
    \end{tikzpicture}
\end{equation}
Such actions of $g\in G$ satisfy $\rho(g)\rho(h)=\rho(gh)$.
This can be naturally generalized to the fusion category case, where we have actions 
\begin{equation}
    \begin{tikzpicture}[scale=.5,baseline=(A.south)]
    \node [draw, fill=black, circle, inner sep=1pt] (A) at (0,0)  {};
    \draw (0,0) node [below] {$\cO(x)$};
    \draw [thick, decoration = {markings, mark=at position .5 with {\arrow[scale=1]{stealth[reversed]}}, mark=at position 1 with {\arrow[scale=1]{stealth[reversed]}}}, postaction=decorate] (0,0) circle (2);
    \draw (-2,0) node[left] {$a$};
    \end{tikzpicture} \quad = 
    \begin{tikzpicture}[scale=.5,baseline=(A.south)]
    \node [draw, fill=black, circle, inner sep=1pt] (A) at (0,0)  {};
    \draw (0,0) node [below] {$\cO'(x)$};
    \end{tikzpicture}
    \label{fusion-algebra-action}
\end{equation}
by simple objects $a\in \cC$. Their actions then satisfy the fusion algebra \begin{equation}
\rho(a)\rho(b)=\sum_c N_{ab}^c \rho(c),
\end{equation} but the story does not end here. 

In the group symmetry case,  a standard fact is that there is a one-to-one correspondence between $g$-twisted operators and
$g'$-twisted operators when $g$ and $g'$ are conjugate. 
In the language of topological domain walls, this correspondence is implemented by 
\begin{equation} 
	\raisebox{-2.5em}{
    \begin{tikzpicture}[scale=.5]
    \draw [fill=black] (0,0) circle (.1);
    \draw [thick, decoration = {markings, mark=at position .3 with {\arrow[scale=1]{stealth}}, mark=at position .8 with {\arrow[scale=1]{stealth}}}, postaction=decorate] (0,0) -- (0,1) node[right] {$g$} -- (0,3) node[right] {$g'$} -- (0,4);
    \draw (0,0) node [below] {$\cO(x)$};
    \draw [thick, decoration = {markings, mark=at position .5 with {\arrow[scale=1]{stealth[reversed]}}, mark=at position 1 with {\arrow[scale=1]{stealth[reversed]}}}, postaction=decorate] (0,0) circle (2);
    \draw [fill=black] (0,2) circle (.1) node [above left] {};
    \draw (-2,0) node[left] {$h$};
    \end{tikzpicture}} \quad = ~
    \raisebox{-2em}{
    \begin{tikzpicture}[scale=.5]
    \draw [fill=black] (0,0) circle (.1);
    \draw [thick, decoration = {markings, mark=at position .55 with {\arrow[scale=1]{stealth}}}, postaction=decorate] (0,0) -- (0,2) node[right]{$g'$} -- (0,4);
    \draw (0,0) node [below] {$\cO'(x)$};
    \end{tikzpicture}} 
\end{equation}
where $g'=hgh^{-1}$.

In the fusion category case, we are then led to consider all possible actions of the form
\begin{equation} 
	\raisebox{-2.5em}{
    \begin{tikzpicture}[scale=.5]
    \draw [fill=black] (0,0) circle (.1);
    \draw [thick, decoration = {markings, mark=at position .3 with {\arrow[scale=1]{stealth}}, mark=at position .8 with {\arrow[scale=1]{stealth}}}, postaction=decorate] (0,0) -- (0,1) node[right] {$a$} -- (0,3) node[right] {$b$} -- (0,4);
    \draw (0,0) node [below] {$\cO(x)$};
    \draw [thick, decoration = {markings, mark=at position .5 with {\arrow[scale=1]{stealth[reversed]}}, mark=at position 1 with {\arrow[scale=1]{stealth[reversed]}}}, postaction=decorate] (0,0) circle (2);
    \draw [fill=black] (0,2) circle (.1) node [above left] {};
    \draw (-2,0) node[left] {$c$};
    \end{tikzpicture}} \quad = ~
    \raisebox{-2em}{
    \begin{tikzpicture}[scale=.5]
    \draw [fill=black] (0,0) circle (.1);
    \draw [thick, decoration = {markings, mark=at position .55 with {\arrow[scale=1]{stealth}}}, postaction=decorate] (0,0) -- (0,2) node[right]{$b$} -- (0,4);
    \draw (0,0) node [below] {$\cO'(x)$};
    \end{tikzpicture}},
    \label{lasso-intro}
\end{equation}
which include \eqref{fusion-algebra-action} as a special case where $a=b=e$.
One notable point is that they could transform a local operator (i.e.\ one with $a = e$) into a defect operator that is attached to a nontrivial topological line $b\in\cC$ and hence belongs to the $b$-twisted sector of the theory, when there is a $c\in \cC$ such that $c\otimes b$ contains $c$.
This cannot happen in the group theory case.
The operations \eqref{lasso-intro} form a finite-dimensional algebra known as the tube algebra, commonly denoted by $\mathrm{Tube}(\cC)$. We will work with representations of $\mathrm{Tube}(\cC)$ and show that it has all the nice properties to derive an asymptotic density formula. As a byproduct, we derive a selection rule on the correlation functions of local operators by organizing operators into representations of the tube algebra. This shows the advantage of working with the tube algebra over the fusion algebra of $\cC$.

\paragraph{Three-dimensional approach: $\ZC$}

The second approach is to use a three-dimensional perspective.
We start with the fact \cite{Kirillov:2010nh} that 
for any unitary\footnote{Strictly speaking, $\TV_\cC$ is constructed from a \emph{spherical} fusion category $\cC$. However, unitarity implies sphericity~\cite{Bhardwaj:2017xup}.} fusion category symmetry $\cC$,
there is a bulk 3d topological quantum field theory (TQFT) $\TV_\cC$ first introduced by Turaev, Viro, Barrett, and Westbury~\cite{Turaev:1992hq,Barrett:1993ab}
which has a `Dirichlet' topological boundary condition $\bD$ hosting topological lines whose behavior is described by $\cC$.
Furthermore, any 2d theory $Q$ with a symmetry given by $\cC$ can be 
lifted to a boundary condition $\bB_Q$ of the 3d theory $\TV_\cC$,
such that the 2d theory $Q$ can be recovered by putting the theory $\TV_\cC$ on an interval
whose two boundaries are $\bB_Q$ and $\bD$, respectively. 
When the symmetry $\cC$ is given by a group $G$,\footnote{In the sense that $\cC=\mathrm{Vec}_G^\omega$ for some 3-cocyle $\omega$.} 
we simply attach to $Q$ a bulk 3d $G$-gauge theory whose Dijkgraaf-Witten action matches the anomaly of $G$.\footnote{This 2d-3d coupling is related to the anomaly-inflow for theory $Q$ by gauging the symmetry $G$ in the bulk.}

Now, let us pick a symmetry operation $a\in \cC$ and
consider an operator $\cO$ in the $a$-twisted sector of the 2d theory $Q$.
In this description, $\cO$ lives on the boundary $\bB_Q$ and is connected to a bulk anyon $\mu$,
which then terminates on the other boundary $\bD$ and turns into the line operator $a$:
\begin{equation}
    \begin{tikzpicture}[scale=.5,baseline=(Y)]
    \draw [fill=black] (0,0) circle (.1);
    \draw [thick, decoration = {markings, mark=at position .55 with {\arrow[scale=1]{stealth}}}, postaction=decorate] (0,0) -- (0,2) node[right]{$a$} -- (0,4);
    \draw (0,0) node (Y) [below] {$\cO$};
    \end{tikzpicture}
\quad =\quad 
\begin{tikzpicture}[baseline=(X)]
\draw (1.35,0.8) node[below]{\small \color{red!75!DarkGreen} $\TV_\cC$};
\draw[color=DarkGreen] (0,0) -- (0,3) -- (1,4) -- (1,1) -- cycle;
\draw (0,0) node[below]{\color{DarkGreen} $\B_Q$};
\draw [color=red!75!DarkGreen, thick, decoration = {markings, mark=at position 0.6 with {\arrow[scale=1]{stealth}}}, postaction=decorate] (0.5,2) -- (1.5,2) node[below]{$\mu$} -- (2.5,2);
\draw [fill=DarkGreen] (0.5,2) circle (0.04) node (X)  [below] {\color{DarkGreen} $\cO$};
\draw [color=blue!70!green, thick, decoration = {markings, mark=at position 0.5 with {\arrow[scale=1]{stealth}}}, postaction=decorate] (2.5,2) -- (2.5,3) node[right]{$a$} -- (2.5,3.75);
\draw [fill=blue!70!green] (2.5,2) circle (0.04) node [below] {\color{blue!70!green} $x$};
\draw[color=blue!70!green, preaction={draw=white,line width=3pt}] (2,0) -- (2,3);
\draw[color=blue!70!green] (2,3) -- (3,4) -- (3,1) -- (2,0);
\draw (2,0) node[below]{\color{blue!70!green} $\D$};
\end{tikzpicture}
\end{equation}

In the group symmetry case, a local untwisted operator $\cO$ in an irreducible representation $\rho$ of $G$
is attached to the end of a bulk Wilson line in the representation $\rho$,
which is one of the anyon types of the bulk Dijkgraaf-Witten theory;
in other words, the anyon $\mu$ appearing in the description above is the representation $\rho$ itself.
Then, a useful way to label the `irreducible representation' of an operator $\cO$ in the general case is to use the type $\mu$ of the anyon.
The set of anyons of the bulk theory $\TV_\cC$ is known to be described by a modular fusion category\footnote{%
The terminology `modular tensor category' is also often used in physics literature instead.
In the mathematical literature, a finer distinction is often made in recent years,
where a `modular tensor category' can be non-semi-simple,
while a `modular fusion category' is assumed to be finite and semi-simple.
We follow this usage here. The authors thank the referee for the input.
} denoted by $\ZC$,
called the Drinfeld center (quantum double) of $\cC$. 
In summary, we were led to use simple objects $\mu\in\ZC$ to label the `irreducible sectors' of a 2d theory $Q$ with a fusion category symmetry $\cC$.

Recall that in the two-dimensional approach we earlier saw, we were instead led to study the irreducible representations of the tube algebra $\mathrm{Tube}(\cC)$.
Luckily for us, it was proved in \cite{evans1995ocneanu,Izumi:2000qa,Muger} 
the fundamental theorem that
the irreducible representations of $\mathrm{Tube}(\cC)$ 
and the simple objects of $\ZC$ are in a canonical one-to-one correspondence,
and therefore we can use whichever description is useful for us.

\paragraph{Main result:}

Now we can state our main result.
Assuming that the vacuum is unique and that the symmetry $\cC$ acts faithfully,
the asymptotic density of states of type $\mu$ in the $a$-twisted sector has a proportionality factor given by $\langle \mu,a\rangle \dim \mu$,
where $\langle \mu,a\rangle$ 
is an integer specifying  the number of boundary lines of type $a$
contained in a bulk line of type $\mu$ when it is brought to the boundary $\bD$,
and $\dim \mu$ is the quantum dimension of the anyon of type $\mu$.

When $\cC$ is given simply by a finite group $G$, 
the bulk theory described by $\ZC$ is simply the 3d $G$ gauge theory,
whose anyons are labeled by the pair $( [g], \rho)$ 
where $[g]$ is a conjugacy class of $G$ 
and $\rho$ is an irreducible representation of the commutant of $g$ in $G$.
Only the bulk line of type $\mu=([e],\rho)$ can end on operators in the untwisted (i.e.~$e$-twisted) sector,
for which $\langle \mu,1\rangle =\dim \mu = \dim \rho$,
reproducing the fundamental result of Pal and Sun \cite{Pal:2020wwd}.

\paragraph{Organization of the paper:}
As the most general result might be somewhat abstract to approach,
we organize our paper in the following cumulative manner. 
In Sec.~\ref{sec:group}, we start by reviewing the analysis of the group symmetry case in  \cite{Pal:2020wwd} 
in a language suitable for our purposes.
This will then be extended in Sec.~\ref{sec:MFC} to the case
when the symmetry is described by a modular fusion category $\cM$,
which happens for example in the case of diagonal rational conformal field theories (RCFTs).
In this case, the 2d theory of our interest can be thought of as arising from a finite slab of 
the 3d TQFT whose anyons are described by $\cM$ itself, 
with two boundary conditions on the slab hosting massless degrees of freedom.

Finally, in Sec.~\ref{sec:fusion},
we discuss the most general case when the symmetry is given by a fusion category.
There, we first describe how the notion of the representation of a group $G$ on the Hilbert space of states 
in terms of the action $U_g$ of a group element $g$ is generalized to
the concept of the tube algebra $\mathrm{Tube}(\cC)$ of a fusion category $\cC$,
and how its representations are labeled by objects of $\ZC$. 
As an immediate application of the tube algebra, we also derive a selection rule on correlation functions of local operators.
We then derive our central result in two ways,
first assuming the validity of our 3d interpretations,
and then using only the operations internal to the 2d description. 
We end our paper by explaining how the results of Sec.~\ref{sec:group} and Sec.~\ref{sec:MFC}
are recovered in the more general framework,
and by studying the case of a system with the Haagerup fusion category symmetry.

We have several Appendices. 
In Appendix~\ref{app:faithfulness},
 we formulate what we mean by the symmetry described by a fusion category $\cC$ 
on a 2d QFT to be faithful,
and show that the dimensions of operators in the $a$-twisted sector for any $a\neq1$ is strictly positive when the symmetry $\cC$ acts faithfully.
The assumption that the action of $\cC$ is faithful is necessary for our analysis.
Appendix~\ref{app:2dQFT.3dTQFT} contains a slightly more detailed description of how
a 2d theory $Q$ with a fusion category symmetry $\cC$ can be lifted to a boundary condition $\bB_Q$ of the bulk Turaev-Viro theory $\TV_\cC$.
Then in Appendix~\ref{app:half.braiding} we derive a generalized Verlinde formula describing the fusion of the bulk anyon $\mu\in \ZC$ and a boundary line operator $a\in \cC$ in terms of the half-linking data.
This will be needed in the two-dimensional derivation of our density formula in Sec.~\ref{sec:fusion}.
Finally, in Appendix~\ref{app:chain},
we consider the decomposition of the Hilbert states of an anyon chain based on a fusion category $\cC$ in the limit of an infinitely long chain.
We will see that the distribution of the irreducible sectors follows the same law, $\propto \langle \mu,a\rangle\dim \mu$,
as in the high-temperature limit of a CFT.

\section{Review of the case of finite group symmetry}
\label{sec:group}

Let us start by recalling the derivation of the asymptotic density of states 
in a specific irreducible representation $\rho$ of a symmetry group $G$.
We put our 2d CFT on a Euclidean $T^2$ of size $L\times T$.
Regarding the direction with the length $T$ as the Wick-rotated time direction,
we use the following graphical representation for the trace with an insertion of $g\in G$,
with the horizontal direction being spatial and the vertical direction being temporal:
\begin{equation}
\tr_{\cH} U_g \, e^{-(T/L)H} =
\vcenter{\hbox{\begin{tikzpicture}
\draw (-1,-.5) rectangle (1,.5);
\draw [thick, -stealth]  (-1,0) -- node(A)[below] {$_g$} (1,0);
\end{tikzpicture}}}
\end{equation} where $\cH$ is the Hilbert space of the theory,
$U_g$ is the action of $g$ on $\cH$, and
 $H$ is the Hamiltonian.

Let $\rho$ be an irreducible representation of $G$,
and $\cH^\rho$ be the space of states in the representation $\rho$.
As is well-known, the projector to this subspace is given by 
\begin{equation}
P_\rho = \frac{\dim\rho}{|G|} \sum_{g\in G}  \overline{\chi_\rho(g)} U_g,
\end{equation}
where $\chi_\rho(g)$ is the character of the representation $\rho$ evaluated on $g$. Therefore we have \begin{align}
\tr_{\cH^\rho } e^{-(T/L)H} &=
\frac{\dim\rho}{|G|}\sum_{g\in G} \overline{\chi_\rho(g)} \vcenter{\hbox{\begin{tikzpicture}
\draw (-1,-.5) rectangle (1,.5);
\draw [thick, -stealth]  (-1,0) -- node(A)[below] {$_g$} (1,0);
\end{tikzpicture}}} 
=
\frac{\dim\rho}{|G|}  \sum_{g\in G} \overline{\chi_\rho(g)} \vcenter{\hbox{\begin{tikzpicture}
\draw (-.5,-1) rectangle (.5,1);
\draw [thick, -stealth]  (0,-1) -- node(A)[right] {$_g$} (0,1);
\end{tikzpicture}}}\\
&=\frac{\dim\rho}{|G|}\sum_{g\in G} \overline{\chi_\rho(g)} \tr_{\cH_g}  e^{-(L/T)H}  \label{2.4}
\end{align} 
where $\cH_g$ is the Hilbert space of the theory on $S^1$ twisted by $g\in G$
and $H$ is the Hamiltonian acting on it.
Here and below, we use a single symbol $H$ to denote the Hamiltonian, irrespective of which sector it acts,
to simplify the notation.
To extract the asymptotic density of states,
we are interested in the regime where $T \ll L$.

We now assume that there is a unique lowest energy state in $\cH$ corresponding to the identity operator.
We also assume that the action of $G$ on our theory is faithful, 
in the sense we make precise in Appendix~\ref{app:faithfulness}.
Then, as shown in the same Appendix,
all operators in a $g$-twisted sector with $g\neq e$ have strictly positive dimensions.
Let us denote by $\Delta_g$ the dimension of the lowest dimension operator in the $g$-twisted sector. 
Then in the limit $L/T\to 0$, we have $\tr_{\cH_g} e^{-(L/T)H}\sim e^{-(L/T)(\Delta_g -c/12)}$,
and $\Delta_g \ge 0$ which is saturated only when $g=e$. 
Therefore, in \eqref{2.4}, the identity operator in the $g=e$ sector dominates the sum,
and we find \begin{equation}
\tr_{\cH^\rho } e^{-(T/L)H} \sim \frac{(\dim\rho)^2}{|G|} e^{2\pi(L/T) (c/12)}
\end{equation} where $c$ is the central charge of the theory.
We emphasize that we assumed that the theory is unitary, and the action of $G$ is faithful.
Converting the trace to the actual asymptotic density of states requires care, 
as was explained in detail in \cite{Pal:2020wwd},
but this detail does not concern us in this paper.

States in the twisted sector $\cH_g$ appeared during the derivation above.
Let us now consider the asymptotic density of states belonging to $\cH_g$.
We have an action of the commutant $C(g)$ of $g$ in $G$,
given by $C(g)=\{ h\in G \mid gh=hg\}$ on $\cH_g$.
When the $G$ symmetry is anomalous with the phase given by $\omega\in H^3(G,U(1))$,
this action of $C(g)$ comes with a projective phase $c_{g,\omega} \in H^2(C(g),U(1))$,
whose explicit form as a cocycle is given by \cite[(6.32)]{Dijkgraaf:1989pz} \begin{equation}
c_{g,\omega}(h_1,h_2)=\frac{\omega(g,h_1,h_2)\omega(h_1,h_2,g)}{\omega(h_1,g,h_2)}.
\label{eq:phase}
\end{equation}
Then $\cH_g$ can be decomposed into irreducible projective representations of $C(g)$ with this phase $c_{g,\omega}$.
Let us pick one such irreducible projective representation $\sigma$.
The corresponding projector is  given by \begin{equation} \label{G.projector}
P_\sigma = \frac{\dim\sigma}{|C(g)|} \sum_{h\in C(g)} \overline{\chi_\sigma(h)} U_h.
\end{equation}
Denoting the space of states belonging to $\sigma$ by $\cH^\sigma_g$, we have 
\begin{align}
\tr_{\cH^\sigma_g } e^{-(T/L)H} &=
 \frac{\dim\sigma}{|C(g)|} \sum_{h\in C(g)} \overline{\chi_\sigma(h)} 
 \vcenter{\hbox{\begin{tikzpicture}
\draw (-1,-.5) rectangle (1,.5);
\draw [thick, -stealth]  (-1,0) -- node(A)[above,pos=.2] {$_h$} (1,0);
\draw [thick, -stealth]  (0,-.5) -- node(B)[right,pos=.2] {$_g$} (0,.5);
\end{tikzpicture}}} 
=
\frac{\dim\sigma}{|C(g)|} \sum_{h\in C(g)} \overline{\chi_\sigma(h)}
\vcenter{\hbox{\begin{tikzpicture}
\draw (-.5,-1) rectangle (.5,1);
\draw [thick, -stealth]  (0,-1) -- node(A)[left,pos=.2] {$_h$} (0,1);
\draw [thick, -stealth]  (.5,0) -- node(B)[above,pos=.2] {$_g$} (-.5,0);
\end{tikzpicture}}}\\
&=\frac{\dim\sigma}{|C(g)|} \sum_{h\in C(g)} \overline{\chi_\sigma(h)}\tr_{\cH_h}  U_g \, e^{-(L/T)H} \label{2.8} 
\end{align}

Under the same assumptions as before, 
the identity operator in the untwisted sector dominates the sum in \eqref{2.8}
in our regime of interest.
As $U_g$ acts trivially on the identity operator, we see that \begin{equation}
\tr_{\cH^\sigma_g } e^{-(T/L)H} \sim \frac{(\dim\sigma)^2}{|C(g)|} e^{2\pi(L/T)(c/12)}.
\end{equation}
Summing over all irreducible projective representations $\sigma$ whose projective phases are given by $c_{g,\omega}$,  
we find that \begin{equation}
\tr_{\cH_g } e^{-(T/L)H} \sim e^{2\pi(L/T)(c/12)}.
\end{equation}
This means that the asymptotic density of the twisted sector states in  $\cH_g$
is independent of $g$,
and therefore is equal to the asymptotic density of the untwisted sector states in $\cH$.

\section{Extension to the case of modular fusion category symmetry}
\label{sec:MFC}

In this section, we derive the asymptotic density of states of a 2d CFT with a symmetry described by a modular fusion category (MFC) $\cM$.\footnote{Unitarity is always assumed for CFT, fusion category, and MFC throughout this paper.}
Again, we put our 2d CFT on a Euclidean $T^2$ of size $L\times T$ and adopt the graphical representation with the horizontal direction being spatial and the vertical direction being temporal.
For the sake of clarity, we take the following steps.
In Sec.~\ref{subsec:RCFT-untwisted}, taking the untwisted sector of an RCFT as an example, we explain our strategy to derive the asymptotic density.
In Sec.~\ref{subsec:MFC-derivation}, we generalize it to any 2d CFT with a symmetry described by an MFC.
In Sec.~\ref{subsec:RCFT-twisted}, we return to the case of RCFTs, and see how the general theory works, in particular for the twisted sectors of an RCFT.
Throughout this section, $a,b,c,\ldots\in\cM$ denote (isomorphic classes of) simple objects of $\cM$.
As for manipulations of the MFC data such as the quantum dimension $d_a=\dim a$, the fusion coefficient $N_{ab}^c$, and the modular $S$-matrix $S_{ab}$, refer to for example \cite{Moore:1989vd} or the Appendix~E of \cite{Kitaev:2005hzj}.

\subsection{The case of diagonal RCFTs: the untwisted sector}
\label{subsec:RCFT-untwisted}

Let us begin with a quick review of RCFTs and their topological lines.
As can be found in \cite{Verlinde:1988sn, Moore:1989vd}, an RCFT whose chiral algebra is $\cA$ can be described by an MFC $\cM_\cA$, whose simple objects $a,b,\ldots\in\cM_\cA$ label all the inequivalent irreducible representations $\{V_a\}_{a\in\cM_\cA}$ of $\cA$ \cite{Huang:2013jza}.

Furthermore, in the diagonal RCFT, we can consider topological lines labeled by $a\in \cM_\cA$, called Verlinde lines, which also form the same MFC $\cM_\cA$ \cite{Petkova:2000ip}.
Recall that in the diagonal RCFT, the (untwisted) Hilbert space\footnote{
We use the terms `sector' and `Hilbert space' interchangeably.}
associated to each time-slice, or spacial $S^1$, is given by
\begin{equation}
\cH_1=\bigoplus_{a}V_a\otimes\overline{V_a}\ .
\end{equation}
Then, the Verlinde line $\cL_b\ (b\in\cM_\cA)$ inserted along the spacial $S^1$ determines an action on $\cH_1$, which can be described as
\begin{equation}
\begin{array}{cccc}
\cL_b=
\raisebox{-5pt}{
\begin{tikzpicture}
\draw [thick, decoration = {markings, mark=at position 0.6 with {\arrow[scale=1]{stealth}}}, postaction=decorate] (-1,0) -- (1,0) node[right]{$b$};
\end{tikzpicture}
}
: & V_a\otimes\overline{V_a} & \to & V_a\otimes\overline{V_a}\ .\\
& \rotatebox{90}{$\in$} & & \rotatebox{90}{$\in$}\\
& |\phi_a\rangle & \mapsto & \dfrac{S_{ba}}{S_{1a}}|\phi_a\rangle
\end{array}
\end{equation}
In other words, we have
\begin{equation}
\tr_{\cH_1}\cL_b\ e^{-(T/L)H}=
\raisebox{-11pt}{
\begin{tikzpicture}
\draw [thick, decoration = {markings, mark=at position 0.6 with {\arrow[scale=1]{stealth}}}, postaction=decorate] (-1,0.5) -- (1,0.5) node[right]{$b$};
\draw (-1,0) rectangle (1,1);
\end{tikzpicture}
}
=\sum_{a}\frac{S_{ba}}{S_{1a}}\chi_a(q)\overline{\chi_a(q)}
\end{equation}
where $q=e^{-2\pi(T/L)}$ and $\chi_a(q)$ denotes the character of $V_a$\,.

Let $\cH_a$ denote the Hilbert space on the spacial $S^1$ with twisted by the boundary condition corresponding to the Verlinde line labeled by $a$.
Using the modular $S$-transformation ($q':=e^{-2\pi(L/T)}$), we have 
\begin{align}
\tr_{\cH_a}e^{-(T/L)H}&=
\raisebox{-12pt}{
\begin{tikzpicture}
\draw [thick, decoration = {markings, mark=at position 0.6 with {\arrow[scale=1]{stealth}}}, postaction=decorate] (0,0) -- (0,1) node[above]{$a$};
\draw (-1,0) rectangle (1,1);
\end{tikzpicture}
}
=
\raisebox{-26pt}{
\begin{tikzpicture}
\draw [thick, decoration = {markings, mark=at position 0.6 with {\arrow[scale=1]{stealth}}}, postaction=decorate] (1,1) -- (0,1) node[left]{$a$};
\draw (0,0) rectangle (1,2);
\end{tikzpicture}
}\\
&=\sum_{b}\frac{S_{\bar{a}b}}{S_{1b}}\chi_b(q')\overline{\chi_b(q')}
=\sum_{b,c,d}\frac{S_{\bar{a}b}}{S_{1b}}S_{bc}\overline{S_{bd}}\chi_c(q)\overline{\chi_d(q)}\\
&=\sum_{c,d}N_{ad}^c\chi_c(q)\overline{\chi_d(q)}\ .
\end{align}
From this, we can read off that
\begin{equation}
\cH_a=\bigoplus_{c,d}N_{ad}^cV_c\otimes\overline{V_d}\ .
\label{eq:RCFT-twisted-H}
\end{equation}

Now, we are ready to investigate the asymptotic behavior of the sector $V_a\otimes\overline{V_a}$ in the untwisted Hilbert space $\cH_1$.
First, we find that
\begin{equation}
P_1^{(a,a)}:=\sum_{b}S_{1a}\overline{S_{ba}}
\raisebox{-5pt}{
\begin{tikzpicture}
\draw [thick, decoration = {markings, mark=at position 0.6 with {\arrow[scale=1]{stealth}}}, postaction=decorate] (-1,0) -- (1,0) node[right]{$b$};
\end{tikzpicture}
}
\label{RCFT-untwisted-P}
\end{equation}
gives the projection $\cH_1\to V_a\otimes\overline{V_a}$\,, because
\begin{align}
\tr_{\cH_1}P_1^{(a,a)}e^{-(T/L)H}&=\sum_{b}S_{1a}\overline{S_{ba}}
\raisebox{-11pt}{
\begin{tikzpicture}
\draw [thick, decoration = {markings, mark=at position 0.6 with {\arrow[scale=1]{stealth}}}, postaction=decorate] (-1,0.5) -- (1,0.5) node[right]{$b$};
\draw (-1,0) rectangle (1,1);
\end{tikzpicture}
}\\
&=\sum_{b,c}S_{1a}\overline{S_{ba}}\frac{S_{bc}}{S_{1c}}\chi_c(q)\overline{\chi_c(q)}\\
&=\chi_a(q)\overline{\chi_a(q)}=\tr_{V_a\otimes\overline{V_a}}e^{-(T/L)H}.
\label{eq:RCFT-untwisted-P-proof}
\end{align}
Then, we can derive the asymptotic density of the sector $V_a\otimes\overline{V_a}$ in the untwisted Hilbert space $\cH_1$ under $T\ll L$ as follows:
\begin{align}
\tr_{V_a\otimes\overline{V_a}}e^{-(T/L)H}&=\sum_{b}S_{1a}\overline{S_{ba}}
\raisebox{-11pt}{
\begin{tikzpicture}
\draw [thick, decoration = {markings, mark=at position 0.6 with {\arrow[scale=1]{stealth}}}, postaction=decorate] (-1,0.5) -- (1,0.5) node[right]{$b$};
\draw (-1,0) rectangle (1,1);
\end{tikzpicture}
}\\
&=\sum_{b}S_{1a}\overline{S_{ba}}
\raisebox{-24pt}{
\begin{tikzpicture}
\draw [thick, decoration = {markings, mark=at position 0.6 with {\arrow[scale=1]{stealth}}}, postaction=decorate] (0.5,0) -- (0.5,2) node[above]{$b$};
\draw (0,0) rectangle (1,2);
\end{tikzpicture}
}
\sim S_{1a}\overline{S_{1a}}
\raisebox{-24pt}{
\begin{tikzpicture}
\draw [dashed, thick, decoration = {markings, mark=at position 0.6 with {\arrow[scale=1]{stealth}}}, postaction=decorate] (0.5,0) -- (0.5,2) node[above]{$1$};
\draw (0,0) rectangle (1,2);
\end{tikzpicture}
}
\sim S_{1a}\overline{S_{1a}}\chi_1(q')\overline{\chi_1(q')}\ .
\end{align}
In the steps `$\sim$' above, we used the assumption that there is a unique lowest energy state in $V_1\otimes\overline{V_1}\subset\cH_1$ corresponding to the identity operator.
In particular, note that (\ref{eq:RCFT-twisted-H}) shows that $V_1\otimes\overline{V_1}$ is not contained in $\cH_{b\neq1}$\,, and thus $b=1$ dominates in the first `$\sim$'.

This result can be reproduced from the asymptotic behavior of the characters.
Since
\begin{equation}
\chi_b(q')=\sum_{c}S_{bc}\chi_c(q)\ ,
\end{equation}
we have 
\begin{equation}
\chi_a(q)=\sum_{b}\overline{S_{ba}}\chi_b(q')\sim S_{1a}\chi_1(q')\ ,
\label{eq:character-asymptotic-behavior}
\end{equation}
which leads to
\begin{equation}
\tr_{V_a\otimes\overline{V_a}}e^{-(T/L)H}=\chi_a(q)\overline{\chi_a(q)}\sim S_{1a}\overline{S_{1a}}\chi_1(q')\overline{\chi_1(q')}\ .
\end{equation}
This is consistent with the earlier result.
Our purpose here was to introduce a method that applies to arbitrary 2d CFTs with symmetry described  by $\cM$,
not necessarily diagonal RCFTs.

\subsection{Derivation of the asymptotic density formula}
\label{subsec:MFC-derivation}

In this subsection, we will generalize the idea of Sec.~\ref{subsec:RCFT-untwisted} to any 2d CFT with a symmetry described by an MFC $\cM$.
Moreover, we will treat not only the untwisted Hilbert space as in Sec.~\ref{subsec:RCFT-untwisted}, but also twisted ones at the same time.

Let $\cH_a$ denote the Hilbert space associated to the spacial $S^1$ twisted by $a\in\cM$.
Our point of view of the action of topological line operators on a general CFT is that 
the diagrams drawn on the spacetime can be translated to operators acting on Hilbert spaces 
satisfying the diagrammatic relations dictated by the (modular) fusion categories.
For the fusion category case, the lines are not allowed to cross over each other,
but in the modular fusion category case considered here, 
one line can go over or go under another line.
Therefore, we can consider two kinds of actions of $b\in\cM$ on $\cH_a$, which can be graphically represented as
\begin{equation}
\raisebox{-13pt}{
\begin{tikzpicture}
\draw [thick, decoration = {markings, mark=at position 0.8 with {\arrow[scale=1]{stealth}}}, postaction=decorate] (-1,0) -- (1,0) node[right]{$b$};
\draw [preaction={draw=white,line width=6pt}, thick, decoration = {markings, mark=at position 0.9 with {\arrow[scale=1]{stealth}}}, postaction=decorate] (0,-0.5) -- (0,0.5) node[above]{$a$};
\end{tikzpicture}
}
\qquad \text{and} \qquad
\raisebox{-10pt}{
\begin{tikzpicture}
\draw [thick, decoration = {markings, mark=at position 0.9 with {\arrow[scale=1]{stealth}}}, postaction=decorate] (0,-0.5) -- (0,0.5) node[above]{$a$};
\draw [preaction={draw=white,line width=6pt}, thick, decoration = {markings, mark=at position 0.8 with {\arrow[scale=1]{stealth}}}, postaction=decorate] (-1,0) -- (1,0) node[right]{$b$};
\end{tikzpicture}
}.
\label{eq:two-kinds-of-actions}
\end{equation}
They have more explicit non-graphical expressions in the case of RCFTs, 
which we give in \eqref{XXXX} and \eqref{YYYY} in Sec.~\ref{subsec:RCFT-twisted}.

Note that these two kinds of actions commute because of the braiding:
\begin{equation}
\raisebox{-22pt}{
\begin{tikzpicture}
\draw [thick, decoration = {markings, mark=at position 0.8 with {\arrow[scale=1]{stealth}}}, postaction=decorate] (-1,0.5) -- (1,0.5) node[right]{$b$};
\draw [preaction={draw=white,line width=6pt}, thick, decoration = {markings, mark=at position 0.9 with {\arrow[scale=1]{stealth}}}, postaction=decorate] (0,0) -- (0,1.5) node[above]{$a$};
\draw [preaction={draw=white,line width=6pt}, thick, decoration = {markings, mark=at position 0.8 with {\arrow[scale=1]{stealth}}}, postaction=decorate] (-1,1) -- (1,1) node[right]{$b'$};
\end{tikzpicture}
}
=
\raisebox{-22pt}{
\begin{tikzpicture}
\draw [thick, decoration = {markings, mark=at position 0.8 with {\arrow[scale=1]{stealth}}}, postaction=decorate] (-1,1) -- (1,1) node[right]{$b$};
\draw [preaction={draw=white,line width=6pt}, thick, decoration = {markings, mark=at position 0.9 with {\arrow[scale=1]{stealth}}}, postaction=decorate] (0,0) -- (0,1.5) node[above]{$a$};
\draw [preaction={draw=white,line width=6pt}, thick, decoration = {markings, mark=at position 0.8 with {\arrow[scale=1]{stealth}}}, postaction=decorate] (-1,0.5) -- (1,0.5) node[right]{$b'$};
\end{tikzpicture}
}.
\end{equation}

Let us focus on the left one of (\ref{eq:two-kinds-of-actions}) first.
As a generalization of (\ref{RCFT-untwisted-P}), we define the operator
\begin{equation} \label{MFC.projector}
P_a^{(c,\ast)}:=\sum_{b}S_{1c}\overline{S_{bc}}
\raisebox{-14pt}{
\begin{tikzpicture}
\draw [thick, decoration = {markings, mark=at position 0.8 with {\arrow[scale=1]{stealth}}}, postaction=decorate] (-1,0) -- (1,0) node[right]{$b$};
\draw [preaction={draw=white,line width=6pt}, thick, decoration = {markings, mark=at position 0.9 with {\arrow[scale=1]{stealth}}}, postaction=decorate] (0,-0.5) -- (0,0.5) node[above]{$a$};
\end{tikzpicture}
}.
\end{equation}
We can show that $\{P_a^{(c,\ast)}\}_{c\in\cM}$ forms a complete set of orthogonal projections.
Indeed, that they are orthogonal projections follows from the computation
\begin{align}
P_a^{(c,\ast)}P_a^{(c',\ast)}&=\sum_{b,b'}S_{1c}\overline{S_{bc}}S_{1c'}\overline{S_{b'c'}}
\raisebox{-20pt}{
\begin{tikzpicture}
\draw [thick, decoration = {markings, mark=at position 0.8 with {\arrow[scale=1]{stealth}}}, postaction=decorate] (-1,0.5) -- (1,0.5) node[right]{$b'$};
\draw [thick, decoration = {markings, mark=at position 0.8 with {\arrow[scale=1]{stealth}}}, postaction=decorate] (-1,1) -- (1,1) node[right]{$b$};
\draw [preaction={draw=white,line width=6pt}, thick, decoration = {markings, mark=at position 0.93 with {\arrow[scale=1]{stealth}}}, postaction=decorate] (0,0) -- (0,1.5) node[above]{$a$};
\end{tikzpicture}
}\\
&=\sum_{b,b',b''}S_{1c}\overline{S_{bc}}S_{1c'}\overline{S_{b'c'}}N_{bb'}^{b''}
\raisebox{-13pt}{
\begin{tikzpicture}
\draw [thick, decoration = {markings, mark=at position 0.8 with {\arrow[scale=1]{stealth}}}, postaction=decorate] (-1,0) -- (1,0) node[right]{$b''$};
\draw [preaction={draw=white,line width=6pt}, thick, decoration = {markings, mark=at position 0.9 with {\arrow[scale=1]{stealth}}}, postaction=decorate] (0,-0.5) -- (0,0.5) node[above]{$a$};
\end{tikzpicture}
}\label{3.21}\\
&=\delta_{c,c'}P_a^{(c,\ast)}\ . \label{3.22}
\end{align}
Here, to go from \eqref{3.21} to \eqref{3.22}, we insert the Verlinde formula $N^{b''}_{bb'}=\sum_x S_{bx} S_{b'x} \overline{S_{b''x}}/S_{1x}$
and repeatedly use the fact $\sum_y \overline{S_{yx}}S_{yz}=\delta_{xz}$.
The completeness follows from
\begin{equation}
\sum_cP_a^{(c,\ast)}=\sum_{b,c}S_{1c}\overline{S_{bc}}
\raisebox{-14pt}{
\begin{tikzpicture}
\draw [thick, decoration = {markings, mark=at position 0.8 with {\arrow[scale=1]{stealth}}}, postaction=decorate] (-1,0) -- (1,0) node[right]{$b$};
\draw [preaction={draw=white,line width=6pt}, thick, decoration = {markings, mark=at position 0.9 with {\arrow[scale=1]{stealth}}}, postaction=decorate] (0,-0.5) -- (0,0.5) node[above]{$a$};
\end{tikzpicture}
}
=
\raisebox{-14pt}{
\begin{tikzpicture}
\draw [dashed, thick, decoration = {markings, mark=at position 0.8 with {\arrow[scale=1]{stealth}}}, postaction=decorate] (-1,0) -- (1,0) node[right]{$1$};
\draw [preaction={draw=white,line width=6pt}, thick, decoration = {markings, mark=at position 0.9 with {\arrow[scale=1]{stealth}}}, postaction=decorate] (0,-0.5) -- (0,0.5) node[above]{$a$};
\end{tikzpicture}
}
=\mathrm{id}_{\cH_a}.
\end{equation}
In the same way, the operators
\begin{equation}\label{MFC.projector2}
P_a^{(\ast,d)}:=\sum_{b}S_{1d}\overline{S_{bd}}
\raisebox{-12pt}{
\begin{tikzpicture}
\draw [thick, decoration = {markings, mark=at position 0.9 with {\arrow[scale=1]{stealth}}}, postaction=decorate] (0,-0.5) -- (0,0.5) node[above]{$a$};
\draw [preaction={draw=white,line width=6pt}, thick, decoration = {markings, mark=at position 0.8 with {\arrow[scale=1]{stealth}}}, postaction=decorate] (-1,0) -- (1,0) node[right]{$b$};
\end{tikzpicture}
}
\end{equation}
form another complete set $\{P_a^{(\ast,d)}\}_{d\in\cM}$ of orthogonal projections.

Now, we focus on the sector
\begin{equation}
\cH_a^{(c,d)}:=P_a^{(c,\ast)}P_a^{(\ast,d)}\cH_a\ .
\end{equation}
The asymptotic density of this sector $\cH_a^{(c,d)}$ under $T\ll L$ can be derived as follows.
First, we rewrite $\tr_{\cH_a^{(c,d)}}e^{-(T/L)H}$ as
\begin{align}
\tr_{\cH_a^{(c,d)}}e^{-(T/L)H}
&=\sum_{b,b'}S_{1c}\overline{S_{bc}}S_{1d}\overline{S_{b'd}}
\raisebox{-20pt}{
\begin{tikzpicture}[scale=.9]
\draw [thick, decoration = {markings, mark=at position 0.8 with {\arrow[scale=1]{stealth}}}, postaction=decorate] (-1,1) -- (1,1) node[right]{$b$};
\draw [preaction={draw=white,line width=6pt}, thick, decoration = {markings, mark=at position 0.9 with {\arrow[scale=1]{stealth}}}, postaction=decorate] (0,0) -- (0,1.5) node[above]{$a$};
\draw [preaction={draw=white,line width=6pt}, thick, decoration = {markings, mark=at position 0.8 with {\arrow[scale=1]{stealth}}}, postaction=decorate] (-1,0.5) -- (1,0.5) node[right]{$b'$};
\draw (-1,0) rectangle (1,1.5);
\end{tikzpicture}
}
=\sum_{b,b'}S_{1c}\overline{S_{bc}}S_{1d}\overline{S_{b'd}}
\raisebox{-30pt}{
\begin{tikzpicture}[scale=.9]
\draw [thick, decoration = {markings, mark=at position 0.8 with {\arrow[scale=1]{stealth}}}, postaction=decorate] (0.5,0) -- (0.5,2) node[above]{$b$};
\draw [preaction={draw=white,line width=6pt}, thick, decoration = {markings, mark=at position 0.9 with {\arrow[scale=1]{stealth}}}, postaction=decorate] (1.5,1) -- (0,1) node[left]{$a$};
\draw [preaction={draw=white,line width=6pt}, thick, decoration = {markings, mark=at position 0.8 with {\arrow[scale=1]{stealth}}}, postaction=decorate] (1,0) -- (1,2) node[above]{$b'$};
\draw (0,0) rectangle (1.5,2);
\end{tikzpicture}
}.
\end{align}
We now fuse $b$ and $b'$ and insert 
an orthonormal basis $\{\psi_j\}_j$ of $\mathrm{Hom}(b'',b\otimes b')$, to find
\begin{equation}
\tr_{\cH_a^{(c,d)}}e^{-(T/L)H}=
\sum_{b,b',b''}\sum_{j}S_{1c}\overline{S_{bc}}S_{1d}\overline{S_{b'd}}\sqrt{\frac{d_{b''}}{d_bd_{b'}}}
\raisebox{-25pt}{
\begin{tikzpicture}
\draw [thick] (0.75,0) -- (0.75,0.5);
\fill [black] (0.75,0.5) circle (0.07);
\draw (0.7,0.4) node [right] {$\psi_j^\dagger$};
\draw [thick] (0.75,1) circle [x radius=0.3, y radius=0.5];
\draw [thick, decoration = {markings, mark=at position 1 with {\arrow[scale=1]{stealth}}}, postaction=decorate] (0.97,1.36) -- (0.967,1.37);
\draw (0.993,1.29) node[right]{$b'$};
\draw [thick, decoration = {markings, mark=at position 1 with {\arrow[scale=1]{stealth}}}, postaction=decorate] (0.53,1.36) -- (0.533,1.37);
\draw (0.507,1.29) node[left]{$b$};
\draw [thick] (1.5,1) -- (1.14,1);
\draw [preaction={draw=white,line width=6pt}, thick, decoration = {markings, mark=at position 0.9 with {\arrow[scale=1]{stealth}}}, postaction=decorate] (0.95,1) -- (0,1) node[left]{$a$};
\fill [black] (0.75,1.5) circle (0.07);
\draw (0.7, 1.7) node [right] {$\psi_j$};
\draw [thick, decoration = {markings, mark=at position 0.9 with {\arrow[scale=1]{stealth}}}, postaction=decorate] (0.75,1.5) -- (0.75,2) node[above]{$b''$};
\draw (0,0) rectangle (1.5,2);
\end{tikzpicture}},
\end{equation}
where
\begin{equation}
d_b=\dim b=
\raisebox{-6pt}{
\begin{tikzpicture}
\draw [thick] (0,0) circle (0.3);
\draw [thick, decoration = {markings, mark=at position 1 with {\arrow[scale=1]{stealth}}}, postaction=decorate] (0.3,0) node[right]{$b$} -- (0.29,0.1);
\end{tikzpicture}
}
=\dfrac{S_{1b}}{S_{11}}
\end{equation}
is the quantum dimension of $b$.

We now use the assumptions that there is a unique lowest energy state in $\cH_1$ corresponding to the identity operator, and that the action of $\cM$ on our theory is faithful, in the sense in Appendix~\ref{app:faithfulness}.
From these assumptions, as shown in the same Appendix, all states in the twisted sectors $\cH_{b''\neq1}$ have strictly positive energy, and therefore $b''=1$ dominates, leading to \begin{multline}
\tr_{\cH_a^{(c,d)}}e^{-(T/L)H}\\
\sim\sum_{b}S_{1c}\overline{S_{bc}}S_{1d}\overline{S_{\bar{b}d}}\sqrt{\frac{1}{d_bd_{\bar{b}}}}
\raisebox{-25pt}{
\begin{tikzpicture}
\draw [thick, dashed] (0.75,0) -- (0.75,0.5);
\draw [thick] (0.75,1) circle [x radius=0.3, y radius=0.5];
\draw [thick, decoration = {markings, mark=at position 1 with {\arrow[scale=1]{stealth}}}, postaction=decorate] (0.53,1.36) -- (0.533,1.37);
\draw (0.507,1.29) node[left]{$b$};
\draw [thick] (1.5,1) -- (1.14,1);
\draw [preaction={draw=white,line width=6pt}, thick, decoration = {markings, mark=at position 0.9 with {\arrow[scale=1]{stealth}}}, postaction=decorate] (0.95,1) -- (0,1) node[left]{$a$};
\draw [thick, dashed, decoration = {markings, mark=at position 0.7 with {\arrow[scale=1]{stealth}}}, postaction=decorate] (0.75,1.5) -- (0.75,2) node[above]{1};
\draw (0,0) rectangle (1.5,2);
\end{tikzpicture}
}
=\sum_{b}S_{1c}\overline{S_{bc}}S_{1d}\overline{S_{\bar{b}d}}\frac{1}{d_b}\frac{S_{ba}}{S_{1a}}
\raisebox{-25pt}{
\begin{tikzpicture}
\draw [preaction={draw=white,line width=6pt}, thick, decoration = {markings, mark=at position 0.9 with {\arrow[scale=1]{stealth}}}, postaction=decorate] (1.5,1) -- (0,1) node[left]{$a$};
\draw (0,0) rectangle (1.5,2);
\end{tikzpicture}
}.
\end{multline}
Finally, the vacuum state in $\cH_1$ corresponding to the identity operator dominates according to our assumption.
Then, by the state-operator correspondence, the topological line $a$ is just wrapped around a disk where the identity operator, or nothing, is inserted, and ends up giving the factor $d_a$,
so we find 
\begin{multline}
\label{MFC.Asymptotic}
\tr_{\cH_a^{(c,d)}}e^{-(T/L)H}\\
\sim\sum_{b}S_{1c}\overline{S_{bc}}S_{1d}\overline{S_{\bar{b}d}}\frac{1}{d_b}\frac{S_{ba}}{S_{1a}}d_ae^{2\pi(L/T)(c/12)}
=S_{1c}S_{1d}N_{ad}^c\,e^{2\pi(L/T)(c/12)},
\end{multline}
where $c$ is the central charge of the theory.

\subsection{The case of diagonal RCFTs: application to twisted sectors}
\label{subsec:RCFT-twisted}

Let us get back to the case of diagonal RCFTs (Sec.~\ref{subsec:RCFT-untwisted}) and observe how our general theory for MFCs (Sec.~\ref{subsec:MFC-derivation}) works in this case.

For the untwisted sector $\cH_1$, it is easy to check that
\begin{equation}
\cH_1^{(a,a)}=P_1^{(a,\ast)}P_1^{(\ast,a)}\cH_1=P_1^{(a,a)}\cH_1=V_a\otimes\overline{V_a}
\end{equation}
and the asymptotic behavior of $\cH_1^{(a,a)}$ derived in Sec.~\ref{subsec:MFC-derivation} becomes
\begin{equation}
\tr_{\cH_1^{(a,a)}}e^{-(T/L)H}\sim S_{1a}S_{1a}N_{1a}^ae^{2\pi(L/T)(c/12)}=S_{1a}\overline{S_{1a}}e^{2\pi(L/T)(c/12)}.
\end{equation}
This is consistent with the result of Sec.~\ref{subsec:RCFT-untwisted}, because we have
\begin{equation}
\chi_1(q')\overline{\chi_1(q')}\sim e^{2\pi(L/T)(c/12)}
\label{eq:character-asymptotic-behavior2}
\end{equation}
from the assumption that there is a unique lowest energy state in $V_1\otimes\overline{V_1}\subset\cH_1$ corresponding to the identity operator.

\begin{figure}[t]
\centering
\begin{tikzpicture}
\draw (0,0.5) node[left]{$T^2$};
\draw (0,0) -- (0,3) -- (1,4) -- (1,1) -- cycle;
\fill [black] (0.5,2) circle (0.07) node [below] {$\cO_c$};
\draw [thick, decoration = {markings, mark=at position 0.6 with {\arrow[scale=1]{stealth}}}, postaction=decorate] (0.5,2) -- (1.25,2) node[below]{$c$} -- (2,2);
\draw [thick, decoration = {markings, mark=at position 0.5 with {\arrow[scale=1]{stealth}}}, postaction=decorate] (2,2) -- (2,3) node[right]{$a$} -- (2,4);
\draw [thick, decoration = {markings, mark=at position 0.55 with {\arrow[scale=1]{stealth}}}, postaction=decorate] (2,2) -- (2.7,2) node[below]{$d$} -- (3.5,2);
\fill [black] (2,2) circle (0.07) node [below] {$x_{cd}^a$};
\draw (3,0) -- (3,3) -- (4,4) -- (4,1) -- cycle;
\fill [black] (3.5,2) circle (0.07) node [below] {$\overline{\cO_d}$};
\draw [<->] (0,-0.3) -- (3,-0.3);
\draw (1.5,-0.3) node[below]{$I$};
\end{tikzpicture}
\caption{States in the sector $V_c\otimes\overline{V_d}\subset\cH_a$ can be realized in the 3d TQFT description as in this figure.
Associated with each end $T^2$ are holomorphic and antiholomorphic boundary conditions.
Take operators $\cO_c$ and $\overline{\cO_d}$ on the boundaries, corresponding to states in $V_c$ and $\overline{V_d}$, respectively.
They are connected by topological lines labeled by $c$ and $d$ meeting in the middle.
Choose $x_{cd}^a\in\mathrm{Hom}(a\otimes d,c)$ to fuse $c$ and $d$ into $a$.
Then $\cO_{cd}^{a;x}=(\cO_c,\cO_d,x_{cd}^a)$ is the defect operator corresponding to a state in the sector $V_c\otimes\overline{V_d}\subset\cH_a$ (See footnote \ref{folding}).
The degree of freedom in choosing $x_{cd}^a$ is $\dim \mathrm{Hom}(a\otimes d,c)=N_{ad}^c$, which leads to the multiplicity of the sector $V_c\otimes\overline{V_d}$ in $\cH_a$.}
\label{fig:3d-description-of-RCFT}
\end{figure}
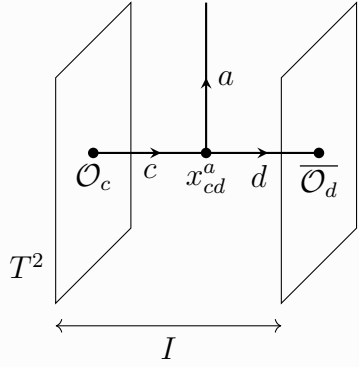

For a twisted sector $\cH_a$, recalling its expression (\ref{eq:RCFT-twisted-H}), it is expected that
\begin{equation}
\cH_a^{(c,d)}=P_a^{(c,\ast)}P_a^{(\ast,d)}\cH_a\overset{}{=}N_{ad}^cV_c\otimes\overline{V_d}.
\label{eq:RCFT-conjecture}
\end{equation}
We can explain this in light of the 3d TQFT description of the 2d RCFT.
Our diagonal RCFT with MFC symmetry $\cM_\cA$ on $T^2$ can be realized by compactifying the 3d TQFT with the same symmetry $\cM_\cA$ on $T^2\times I$, where $I$ is an interval in the compactified direction, with appropriate boundary conditions on each end $T^2$\cite{Fuchs:2002cm, Kapustin:2010if}\footnote{
This 3d TQFT description is a generalization of the well-known correspondence between the 3d Chern-Simons theory and the Wess-Zumino-Witten model \cite{Witten:1988hf, Moore:1989yh}.
A review can be found in Sec.~1 and Sec.~5.1 of \cite{Komargodski:2020mxz}.
The relation between this description of 2d RCFT with MFC symmetry $\cM_\cA$ by 3d TQFT with the same symmetry $\cM_\cA$ and the description of 2d QFT with a fusion category symmetry $\cC$ by 3d TQFT with the symmetry $\ZC$ is mentioned in Sec.~\ref{sec:tube.algebra}.
}.
In this description, any state in the sector $V_c\otimes\overline{V_d}\subset\cH_a$ can be realized as in Fig.~\ref{fig:3d-description-of-RCFT}.
Viewing this figure from the right side, we can see that the topological line $b$ on the left of (\ref{eq:two-kinds-of-actions}) acts on the topological line $c$ in the figure, and the line $b$ on the right of (\ref{eq:two-kinds-of-actions}) acts on the line $d$. This can be summarized as
\begin{equation}
\begin{array}{ccccccccc}
\raisebox{-13pt}{
\begin{tikzpicture}
\draw [thick, decoration = {markings, mark=at position 0.8 with {\arrow[scale=1]{stealth}}}, postaction=decorate] (-1,0) -- (1,0) node[right]{$b$};
\draw [preaction={draw=white,line width=6pt}, thick, decoration = {markings, mark=at position 0.9 with {\arrow[scale=1]{stealth}}}, postaction=decorate] (0,-0.5) -- (0,0.5) node[above]{$a$};
\end{tikzpicture}
}
: & V_c\otimes\overline{V_d} & \to & V_c\otimes\overline{V_d} & , \\
& \rotatebox{90}{$\in$} &  & \rotatebox{90}{$\in$} & \\
& |\phi\rangle & \mapsto & \dfrac{S_{bc}}{S_{1c}}|\phi\rangle & \\
\end{array}
\label{XXXX}
\end{equation}
and
\begin{equation}
\begin{array}{ccccccccc}
\raisebox{-10pt}{
\begin{tikzpicture}
\draw [thick, decoration = {markings, mark=at position 0.9 with {\arrow[scale=1]{stealth}}}, postaction=decorate] (0,-0.5) -- (0,0.5) node[above]{$a$};
\draw [preaction={draw=white,line width=6pt}, thick, decoration = {markings, mark=at position 0.8 with {\arrow[scale=1]{stealth}}}, postaction=decorate] (-1,0) -- (1,0) node[right]{$b$};
\end{tikzpicture}
}
: & V_c\otimes\overline{V_d} & \to & V_c\otimes\overline{V_d}\ .\\
& \rotatebox{90}{$\in$} & & \rotatebox{90}{$\in$}\\
& |\phi\rangle & \mapsto & \dfrac{S_{bd}}{S_{1d}}|\phi\rangle\\
\end{array}
\label{YYYY}
\end{equation}
Then, in the same way as (\ref{eq:RCFT-untwisted-P-proof}), we can verify that
\begin{align}
P_a^{(c,\ast)}&:\cH_a\to\bigoplus_{d}N_{ad}^cV_c\otimes\overline{V_d}\\
P_a^{(\ast,d)}&:\cH_a\to\bigoplus_{c}N_{ad}^cV_c\otimes\overline{V_d}\\
P_a^{(c,\ast)}P_a^{(\ast,d)}&:\cH_a\to\hspace{20pt} N_{ad}^cV_c\otimes\overline{V_d}
\end{align}
are the projections, and therefore (\ref{eq:RCFT-conjecture}) holds.

Now, the asymptotic behavior of $\cH_a^{(c,d)}$ derived in Sec.~\ref{subsec:MFC-derivation} was
\begin{equation}
\tr_{\cH_a^{(c,d)}}e^{-(T/L)H}\sim S_{1c}S_{1d}N_{ad}^ce^{2\pi(L/T)(c/12)}.
\end{equation}
Again, this can be reproduced from the asymptotic behavior of the characters.
From (\ref{eq:character-asymptotic-behavior}) and (\ref{eq:character-asymptotic-behavior2}), we have
\begin{align}
\tr_{\cH_a^{(c,d)}}e^{-(T/L)H}&=N_{ad}^c\tr_{V_c\otimes\overline{V_d}}e^{-(T/L)H}=N_{ad}^c\chi_c(q)\overline{\chi_d(q)}\\
&\sim N_{ad}^cS_{1c}\overline{S_{1d}}\chi_1(q')\overline{\chi_1(q')}\sim N_{ad}^cS_{1c}S_{1d}e^{2\pi(L/T)(c/12)}.
\end{align}
This is consistent with the above result.

\section{Extension to the case of fusion category symmetry}
\label{sec:fusion}

In this section, we generalize the results of the previous sections to the case of 2d CFTs with arbitrary fusion category symmetry $\cC$. In Sec.~\ref{sec:tube.algebra} we discuss some necessary backgrounds to derive the asymptotic formula. 
In Sec.~\ref{sec:formula.fusion.cat.symm} we use what we learned in Sec.~\ref{sec:tube.algebra} to derive
the asymptotic formula in two ways, one using the 3d perspective and another purely in 2d.
Then in Sec.~\ref{sec:examples} we revisit the group symmetry case treated in Sec.~\ref{sec:group}
and the MFC case treated in Sec.~\ref{sec:MFC} from the more general perspective of this section,
and discuss the case of the Haagerup fusion category explicitly.
Throughout the rest of the paper $a,b,c, \dots \in \cC$ stand for \emph{simple} objects, and $\sum_a$ should be understood as summing over the (isomorphism classes of) simple objects in $\cC$.

\subsection{Tube algebra and the Drinfeld center} \label{sec:tube.algebra}

In this subsection we study the action of non-invertible symmetries on point-like operators possibly attached to a topological line $a$, such as
\begin{equation}
    \raisebox{-2em}{\begin{tikzpicture}[scale=.5]
    \draw [fill=black] (0,0) circle (.1);
    \draw [thick, decoration = {markings, mark=at position .55 with {\arrow[scale=1]{stealth}}}, postaction=decorate] (0,0) -- (0,1) node[right]{$a$} -- (0,2);
    \draw (0,0) node [below] {$\cO(x)$};
    \end{tikzpicture}} \,.
\end{equation}
We call the point-like operator $\cO$ a `defect operator'\footnote{
It is also called a `disorder operator' or a `twist field'.} which is equivalent to a state in the $a$-twisted Hilbert space $\cH_a$ via the state-operator correspondence. Note that when $a=1$, $\cO$ is just an ordinary local operator. The action of the non-invertible symmetry $\cC$ on local operators forms an algebra that is the same as the fusion algebra of $\cC$. However, the action on defect operators forms a much richer algebra known as the \emph{tube algebra}, which is the subject of this subsection. The tube algebra is a finite-dimensional $\mathbb{C}^\ast$-algebra that was originally defined by Ocneanu~\cite{ocneanu1994chirality}, and its relation with the Drinfeld center was later established in \cite{evans1995ocneanu,Izumi:2000qa,Muger} (see also \cite{Williamson:2017uzx,Aasen:2020jwb} in the contexts of tensor networks and statistical lattice models).

\paragraph{Tube algebra $\mathrm{Tube}(\cC)$:}
The action of non-invertible symmetries on defect operators can be formulated in terms of the lasso actions~\cite{Chang:2018iay} that we describe below. 
Consider a defect operator $\cO \in \cH_a$ at the end of the topological line $a$.\footnote{For simplicity, we are using the state-operator correspondence to write $\cO \in \cH_a$. However, the lasso action is also defined for non-conformal QFTs.
} A lasso action of $c$ on $\cO$ is given by
\begin{equation} \label{tube.basis}
	\raisebox{-2.5em}{
    \begin{tikzpicture}[scale=.5]
    \draw [fill=black] (0,0) circle (.1);
    \draw [thick, decoration = {markings, mark=at position .3 with {\arrow[scale=1]{stealth}}, mark=at position .8 with {\arrow[scale=1]{stealth}}}, postaction=decorate] (0,0) -- (0,1) node[right] {$a$} -- (0,3) node[right] {$b$} -- (0,4);
    \draw (0,0) node [below] {$\cO(x)$};
    \draw [thick, decoration = {markings, mark=at position .5 with {\arrow[scale=1]{stealth[reversed]}}, mark=at position 1 with {\arrow[scale=1]{stealth[reversed]}}}, postaction=decorate] (0,0) circle (2);
    \draw [fill=black] (0,2) circle (.1) node [above left] {$x_{ab}^c$};
    \draw (-2,0) node[left] {$c$};
    \end{tikzpicture}} \quad = ~
    \raisebox{-2em}{
    \begin{tikzpicture}[scale=.5]
    \draw [fill=black] (0,0) circle (.1);
    \draw [thick, decoration = {markings, mark=at position .55 with {\arrow[scale=1]{stealth}}}, postaction=decorate] (0,0) -- (0,2) node[right]{$b$} -- (0,4);
    \draw (0,0) node [below] {$\cO'(x)$};
    \end{tikzpicture}} 
\end{equation}
with a choice of topological lines $b,c$ and a junction vector $x_{ab}^c \in \mathrm{Hom}(c \otimes a, b \otimes c)$.
This action maps $\cO \in \cH_a$ to $\cO' \in \cH_b$. 
We can also consider the action on the cylinder given by
\begin{equation}\label{tube.basis.cylinder}
  \x{a}{b}{c}{x_{ab}^c}
\end{equation}
which is equivalent to the lasso action \eqref{tube.basis} on operators by the state-operator correspondence.  Since this paper is about CFTs, we conflate \eqref{tube.basis} with \eqref{tube.basis.cylinder}.

Upon choosing a basis $x_{ab}^{c,\alpha}$ of $\Hom(c\otimes a, b\otimes c)$
where $\alpha=1,\dots,\mathrm{dim}\,\mathrm{Hom}(c \otimes a, b \otimes c)$,
the operations \eqref{tube.basis} form a basis for the tube algebra. An arbitrary element of the tube algebra is given by a linear operator
\begin{equation}
    U = \sum_{a,b,c,\alpha} U^{ab}_{c,\alpha} \, x_{ab}^{c,\alpha}~\in~ \mathrm{Tube}(\cC),
\end{equation}
for $U^{ab}_{c,\alpha} \in \bC$, acting on the \emph{total} Hilbert space
\begin{equation}
    \cH_\tot = \bigoplus_a \cH_a
\end{equation}
of all local and defect operators. In the above, we have slightly abused the notation and identified the basis junction vector $x_{ab}^{c,\alpha} \in \mathrm{Hom}(c \otimes a, b \otimes c)$ with the corresponding basis element of the tube algebra. 
Multiplication is given by the composition 
\begin{equation}
\begin{tikzpicture}[baseline=(X)]
\draw [color= {black}, thick, decoration = {markings, mark=at position 0.8 with {\arrow[scale=1]{stealth}}}, postaction=decorate] (-1,0.5) -- (1,0.5) node[right]{$c'$};
\draw [color= {black}, thick, decoration = {markings, mark=at position 0.9 with {\arrow[scale=1]{stealth}}, mark=at position 0.25 with {\arrow[scale=1]{stealth}}}, postaction=decorate] (0,0) node[below]{$a$} -- (0,1.5) node[above]{$a''$};
\draw [color= {black}, thick, decoration = {markings, mark=at position 0.8 with {\arrow[scale=1]{stealth}}}, postaction=decorate] (-1,1) -- (1,1) node[right]{$c$};
\draw [fill={black}] (0,1) node[above right]{\footnotesize \color{black} $y$} circle (0.04);
\draw [fill={black}] (0,.5) node (X) [above right]{\footnotesize \color{black} $x$} circle (0.04);
\end{tikzpicture}
=
\sum_{c'',z} \sqrt{\frac{d_{c''}}{d_c d_{c'}}} ~ \begin{tikzpicture}[baseline=(X)]
\draw [color= {black}, thick, decoration = {markings, mark=at position 0.8 with {\arrow[scale=1]{stealth}}}, postaction=decorate] (-1,0.75) -- (0,0.5) -- (1,0.75);
\draw [color= {black}] (0.4,0.45) node[right]{$c'$};
\draw [color= {black}, thick, decoration = {markings, mark=at position 0.9 with {\arrow[scale=1]{stealth}}, mark=at position 0.25 with {\arrow[scale=1]{stealth}}}, postaction=decorate] (0,0) node[below]{$a$} -- (0,1.5) node[above]{$a''$};
\draw [color= {black}, thick, decoration = {markings, mark=at position 0.8 with {\arrow[scale=1]{stealth}}}, postaction=decorate] (-1,0.75) -- (0,1) -- (1,0.75) ;
\draw [color= {black}] (0.4,1.15) node[right]{$c$};
\draw[color= {black}, thick] (-1.5,0.75)-- node[above right] {\footnotesize $\bar{z}$}  (-1,0.75);
\draw[color= {black}, thick] (1,0.75) -- (1.5,0.75);
\draw [color= {black}] (1.1,0.7) node[above]{\footnotesize $z$};
\draw [color= {black}] (1.5,0.8) node[right]{$c''$};
\draw [fill={black}] (-1,0.75) circle (0.04);
\draw [fill={black}] (1,0.75) circle (0.04);
\draw [fill={black}] (0,1) node[above right]{\footnotesize \color{black} $y$} circle (0.04);
\draw [fill={black}] (0,.5) node (X) [above right]{\footnotesize \color{black} $x$} circle (0.04);
\end{tikzpicture}
\end{equation}
where we sum $c''$ over simple objects appearing in $c\otimes c'$, and $z$ over an orthonormal basis of $\mathrm{Hom}(c\otimes c', c'')$.
The right-hand side can further be explicitly written in terms of the fusion $F$-matrices of $\cC$; see \cite{Williamson:2017uzx,Aasen:2020jwb} for a formula.
The lasso action describes the most general action of the fusion category symmetry $\cC$ on defect operators. Moreover, the composition of different lasso actions are governed by the tube algebra. The total Hilbert space $\cH_\tot$ is thereby organized into irreducible representations of the tube algebra.
We will identify these representations with objects of $\ZC$---the Drinfeld center of $\cC$---and explain their equivalence.

The tube algebra $\tub(\cC)$ contains a Dehn twist subalgebra generated by the modular $T$-transformation
\begin{equation}
  T = \sum_a
  \begin{gathered}
    \begin{tikzpicture}[baseline=(X.base)]
    \draw [thick, 
    decoration = {markings, mark=at position 0.8 with {\arrow[scale=1]{stealth}}}, postaction=decorate] (-1,0) -- (1,0) node[right] (X) {$a$};
    \draw [thick, 
      decoration = {markings, mark=at position 0.9 with {\arrow[scale=1]{stealth}}}, postaction=decorate] (0,-0.5) node[below]{$a$} -- (0,0.5) node[above]{$a$};
    \draw [thick, preaction={draw=white,line width=5pt}] (0,0.2) arc [start angle=0, end angle=-90, x radius=0.2, y radius=0.2];
    \draw [thick, preaction={draw=white,line width=5pt}] (0,-0.2) arc [start angle=180, end angle=90, x radius=0.2, y radius=0.2];
    \end{tikzpicture}
    \end{gathered}
\end{equation}
that commutes with $\tub(\cC)$, therefore each irreducible representation of $\tub(\cC)$ can be associated with a definite $T$-eigenvalue $e^{2\pi i s}$ where $s$ is the Lorentz spin of any defect operator transforming in the representation.  In the relation between $\tub(\cC)$ and the Drinfeld center $\ZC$ that will be discussed momentarily, the $T$-eigenvalues correspond to the topological spins of the bulk anyons. Therefore, we find a universal spin-selection rule determining the admissible spins in the different twisted sectors.

\paragraph{The bulk TQFT $\TV_\cC$ and the Drinfeld center $\ZC$:}

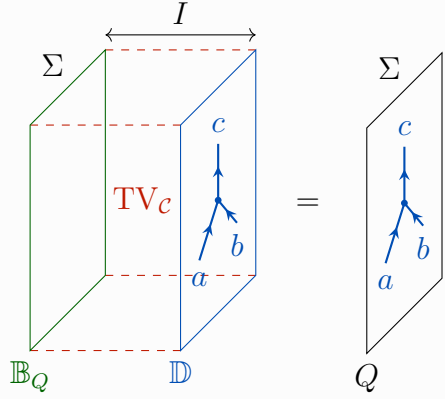
\begin{figure}[t]
    \centering
    \raisebox{-73pt}{\begin{tikzpicture}
\draw [<->] (1,4.2) -- (3,4.2);
\draw (2,4.2) node[above]{$I$};
\draw[color=red!75!DarkGreen, dashed] (0,0) -- (2,0);
\draw[color=red!75!DarkGreen, dashed] (0,3) -- (2,3);
\draw[color=red!75!DarkGreen, dashed] (1,4) -- (3,4);
\draw[color=red!75!DarkGreen, dashed] (1,1) -- (3,1);
\draw (1.5,2) node{\color{red!75!DarkGreen} $\TV_\cC$};
\draw[color=DarkGreen] (0,0) -- (0,3) -- (1,4) -- (1,1) -- cycle;
\draw(0.3,3.5) node[above] {$\Sigma$};
\draw (0,0) node[below]{\color{DarkGreen} $\B_Q$};
\draw [color=blue!70!green, thick, decoration = {markings, mark=at position 0.6 with {\arrow[scale=1]{stealth}}}, postaction=decorate] (2.5,2) -- (2.5,2.75) node[above]{$c$};
\draw [color=blue!70!green, thick, decoration = {markings, mark=at position 0.6 with {\arrow[scale=1]{stealth}}}, postaction=decorate] (2.25,1.2) node[below]{$a$} -- (2.5,2);
\draw [color=blue!70!green, thick, decoration = {markings, mark=at position 0.6 with {\arrow[scale=1]{stealth}}}, postaction=decorate] (2.75,1.7) node[below]{$b$} -- (2.5,2);
\fill [blue!70!green] (2.5,2) circle (0.045);
\draw[color=blue!70!green] (2,0) -- (2,3) -- (3,4) -- (3,1) -- cycle;
\draw (2,0) node[below]{\color{blue!70!green} $\D$};
\end{tikzpicture}}
    \quad $=$ ~
\raisebox{-73pt}{\begin{tikzpicture}
\draw(2,0) -- (2,3) -- (3,4) -- (3,1) -- cycle;
\draw [color=blue!70!green, thick, decoration = {markings, mark=at position 0.6 with {\arrow[scale=1]{stealth}}}, postaction=decorate] (2.5,2) -- (2.5,2.75) node[above]{$c$};
\draw [color=blue!70!green, thick, decoration = {markings, mark=at position 0.6 with {\arrow[scale=1]{stealth}}}, postaction=decorate] (2.25,1.2) node[below]{$a$} -- (2.5,2);
\draw [color=blue!70!green, thick, decoration = {markings, mark=at position 0.6 with {\arrow[scale=1]{stealth}}}, postaction=decorate] (2.75,1.7) node[below]{$b$} -- (2.5,2);
\fill [blue!70!green] (2.5,2) circle (0.045);
\draw(2.3,3.5) node[above] {$\Sigma$};
\draw (2,0) node[below]{$Q$};
\end{tikzpicture}}
    \caption{The 2d theory $Q$ on $\Sigma$ is obtained by putting the 3d TQFT $\TV_\cC$ on $\Sigma \times I$. We have the non-topological boundary condition $\B_Q$ on the left and the (Dirichlet) topological boundary condition $\D$ on the right. The topological lines $a,b,c,\dots \in \cC$ on the $\D$ boundary implement the $\cC$ symmetry of $Q$.}
    \label{fig:slab}
\end{figure}
Given a fusion category $\cC$, there is a 3d TQFT $\TV_\cC$ that admits a topological `Dirichlet' boundary condition $\bD$ hosting boundary lines described by $\cC$.
This TQFT was constructed by Turaev, Viro, Barrett, and Westbury in \cite{Turaev:1992hq,Barrett:1993ab}.\footnote{
This theory is realized as the low-energy limit of the Levin-Wen string-net lattice model \cite{Levin:2004mi,Lin:2020bak}.
It also has a state-sum construction which, by analogy with the group theory case, can be roughly interpreted as a discrete $\tilde{\cC}$-gauge theory, where $\tilde{\cC}$ describes topological surfaces in 3d with the $\cC$ fusion rule,
see \cite{Carqueville:2018sld} and in particular its Sec.~4.3.}
For the Dirichlet boundary condition, the $\tilde{\cC}$-surfaces can end on the boundary where after gauging they become topological lines forming $\cC$.
The anyons of $\TV_\cC$ form a modular fusion category (MFC) 
$\ZC$ \cite{Kirillov:2010nh,turaev2010two}, which is called the Drinfeld center (or quantum double) of $\cC$.
One notable feature of $\ZC$ is that \begin{equation}
\sum_{\mu\in\ZC} (\dim \mu)^2 = \left|\sum_{a\in\cC} (\dim a)^2 \right|^2 ,
\end{equation}
where $\mu$ runs over the isomorphism classes of simple objects of $\ZC$,
and $a$ over those of $\cC$.

As emphasized in \cite{Thorngren:2019iar,Gaiotto:2020iye}, any 2d QFT $Q$ with a fusion category symmetry $\cC$ can be lifted to a boundary condition $\B_Q$ for $\TV_\cC$. The boundary condition $\B_Q$ is such that we can recover the original 2d QFT $Q$ as a slab/interval reduction of $\TV_\cC$ with $\B_Q$ on one side and the Dirichlet boundary condition $\D$ on the other side, as depicted in Figure~\ref{fig:slab}. There and in the following, the topological boundary $\bD$ and the simple objects in $\cC$ are drawn in blue, the non-topological boundary $\bB_Q$ in green, and the bulk, including its anyons in $\ZC$ and their restrictions to $\bD$, in red.
We refer the reader to Appendix~\ref{app:2dQFT.3dTQFT} for an explicit construction of the boundary condition $\B_Q$. Recall that the boundary condition $\D$ is such that the boundary line defects form $\cC$.
These boundary lines become the $\cC$-symmetry lines of theory $Q$ after the interval reduction.

\begin{figure}[t]
    \centering
\raisebox{-73pt}{\begin{tikzpicture}
\draw [<->] (1,4.2) -- (3,4.2);
\draw (2,4.2) node[above]{$I$};
\draw (1.35,0.8) node[below]{\small \color{red!75!DarkGreen} $\TV_\cC$};
\draw[color=DarkGreen] (0,0) -- (0,3) -- (1,4) -- (1,1) -- cycle;
\draw (0,0) node[below]{\color{DarkGreen} $\B_Q$};
\draw [color=red!75!DarkGreen, thick, decoration = {markings, mark=at position 0.6 with {\arrow[scale=1]{stealth}}}, postaction=decorate] (0.5,2) -- (1.5,2) node[below]{$\mu$} -- (2.5,2);
\draw [fill=DarkGreen] (0.5,2) circle (0.04) node [below] {\color{DarkGreen} $\widetilde{\cO}$};
\draw [color=blue!70!green, thick, decoration = {markings, mark=at position 0.5 with {\arrow[scale=1]{stealth}}}, postaction=decorate] (2.5,2) -- (2.5,3) node[right]{$a$} -- (2.5,3.75);
\draw [fill=blue!70!green] (2.5,2) circle (0.04) node [below] {\color{blue!70!green} $x$};
\draw[color=blue!70!green, preaction={draw=white,line width=3pt}] (2,0) -- (2,3);
\draw[color=blue!70!green] (2,3) -- (3,4) -- (3,1) -- (2,0);
\draw (2,0) node[below]{\color{blue!70!green} $\D$};
\end{tikzpicture}}
    \quad $=$ ~
\raisebox{-73pt}{\begin{tikzpicture}
\draw [color=blue!70!green, thick, decoration = {markings, mark=at position 0.5 with {\arrow[scale=1]{stealth}}}, postaction=decorate] (2.5,2) -- (2.5,3) node[right]{$a$} -- (2.5,3.75);
\draw [fill=black] (2.5,2) circle (0.04) node [below] {$\cO$};
\draw(2,0) -- (2,3) -- (3,4) -- (3,1) -- cycle;
\draw (2,0) node[below]{$Q$};
\end{tikzpicture}}
    \caption{A defect operator $\cO\in\cH_a$ of the 2d theory $Q$ is composed of a bulk anyon $\mu\in\ZC$ and a pair of boundary (defect) operators. Via the state-operator correspondence, $\widetilde{\cO}$ belongs to the $\mu$-punctured disk Hilbert space of $\TV_\cC$ with boundary condition $\B_Q$, and $x$ belongs to $\mathrm{Hom}_\cC(F(\mu),a)$ where $F:\ZC \to \cC$ describes the fusion of bulk anyons with the topological boundary $\D$.
    }
    \label{fig:defect.operators}
\end{figure}
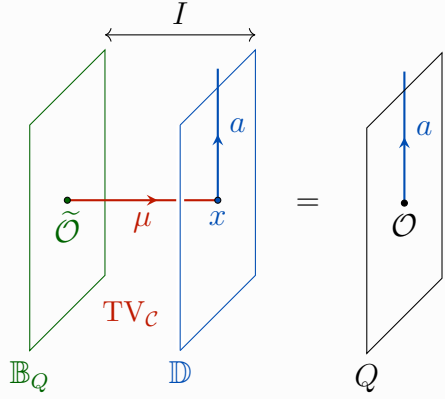

Using this slab construction, a defect operator $\cO \in \cH_a$ of theory $Q$ corresponds to a bulk anyon $\mu \in \ZC$ stretched between the two boundaries, and a pair of endpoint operators
\begin{equation}
    x\in W^\mu_a \qquad \text{and} \qquad \widetilde{\cO} \in \cV_\mu \,,
\end{equation}
as shown in Figure~\ref{fig:defect.operators}.\footnote{
When $Q$ is a diagonal RCFT, the two slab constructions---one with holomorphic and anti-holomorphic boundaries as in Figure~\ref{fig:3d-description-of-RCFT}, and the other with $B_Q$ and $\D$ in Figure~\ref{fig:defect.operators}---are related by folding, with the identifications
$x = x^a_{cd}, ~ \mu = c \otimes \bar d, ~ \widetilde O = \cO_c \overline{\cO}_d.$
\label{folding}
}
Here $W^\mu_a$ is the space of $\D$-boundary operators at the junction of $\mu$ and $a$, and $\cV_\mu$ is the space of $\B_Q$-boundary operators at the end of $\mu$.
Therefore, we find the decomposition
\begin{equation} \label{H.a.decomposition}
    \cH_a = \bigoplus_{\mu} W^\mu_a \otimes \cV_\mu\,.
\end{equation}
Using the state-operator correspondence, $\cV_\mu$ is identified with the $\mu$-punctured disk Hilbert space of $\TV_\cC$ with boundary condition $\B_Q$. Moreover, $W^\mu_a = \mathrm{Hom}_\cC(F(\mu),a)$, where $F:\ZC\to\cC$ describes the fusion of bulk anyons with the topological boundary $\D$.

\paragraph{2d description of the Drinfeld center $\ZC$:}

Bringing all the bulk lines to the boundary, we can formulate the Drinfeld center $\ZC$ in an intrinsically 2d description. First, we translate the crossing 
\begin{equation}
\vcenter{\hbox{\begin{tikzpicture}
\draw [thick, color= red!75!DarkGreen ,decoration = {markings, mark=at position 0.8 with {\arrow[scale=1]{stealth}}}, postaction=decorate] (-.5,-.5) -- (.5,.5) node[right]{$\mu$};
\draw [color={blue!70!green}, preaction={draw=white,line width=6pt}, thick, decoration = {markings, mark=at position 0.8 with {\arrow[scale=1]{stealth}}}, postaction=decorate] (.5,-.5) -- (-.5,.5) node[left]{$a$};
\end{tikzpicture}}}
\end{equation}
to 2d language.
Fusing $\mu$ to the boundary gives rise to an object $m\in \cC$ which is generically non-simple. 
Then, the figure above means that $m$ comes equipped with \emph{half-braidings}
\begin{equation}
R_{m,a}:=\vcenter{\hbox{\begin{tikzpicture}
\draw [thick, color= red!75!DarkGreen ,decoration = {markings, mark=at position 0.8 with {\arrow[scale=1]{stealth}}}, postaction=decorate] (-.5,-.5) -- (.5,.5) node[right]{$m$};
\draw [color={blue!70!green}, preaction={draw=white,line width=6pt}, thick, decoration = {markings, mark=at position 0.8 with {\arrow[scale=1]{stealth}}}, postaction=decorate] (.5,-.5) -- (-.5,.5) node[left]{$a$};
\end{tikzpicture}}} \in \Hom_\cC(m\otimes a,a\otimes m).
\end{equation}
These half-braidings should satisfy various consistency conditions, such as 
\begin{equation}
\vcenter{\hbox{\begin{tikzpicture}[yscale=1.2]
\draw [thick, color= red!75!DarkGreen ,decoration = {markings, mark=at position 0.9 with {\arrow[scale=1]{stealth}}}, postaction=decorate] (-.5,-.5) node[below] {$\vphantom{b}m$}  -- (.5,.5) node[above]{$m$};
\draw [color={blue!70!green}, preaction={draw=white,line width=6pt}, thick, decoration = {markings, mark=at position 0.8 with {\arrow[scale=1]{stealth}}}, postaction=decorate] (.8,-.5) node[below]{$b$} -- (-.2,.5) node[above]{$b$};
\draw [color={blue!70!green}, preaction={draw=white,line width=6pt}, thick, decoration = {markings, mark=at position 0.8 with {\arrow[scale=1]{stealth}}}, postaction=decorate] (.2,-.5) node[below]{$\vphantom{b}a$} -- (-.8,.5) node[above]{$a$};
\end{tikzpicture}}}
=
\sum_{c,z} \sqrt{\frac{d_c}{d_a d_b}} ~
\vcenter{\hbox{\begin{tikzpicture}[yscale=1.2,xscale=1.5]
\draw [thick, color= red!75!DarkGreen ,decoration = {markings, mark=at position 0.9 with {\arrow[scale=1]{stealth}}}, postaction=decorate] (-.5,-.5) node[below] {$\vphantom{b}m$} -- (.5,.5) node[above]{$m$};
\draw [color={blue!70!green}, thick,  preaction={draw=white,line width=6pt}, decoration = {markings, mark=at position 0.75 with {\arrow[scale=1]{stealth}}}, postaction=decorate] (0.2,-.2) -- (0,0) node[left]{$c$} --(-0.2,0.2);
\draw [color={blue!70!green}, thick, decoration = {markings, mark=at position 0.8 with {\arrow[scale=1]{stealth}}}, postaction=decorate] (-0.1,-.5)  node[below ]{$\vphantom{b}a$} --(0.2,-.2);
\draw [color={blue!70!green}, thick, decoration = {markings, mark=at position 0.8 with {\arrow[scale=1]{stealth}}}, postaction=decorate] (-0.2,.2) --(-.5,.5) node[above]{$a$}; 
\fill [blue!70!green] (-.2,.2) node[above]{\footnotesize $\bar{z}$} circle (0.045);
\fill [blue!70!green] (.2,-.2) node[above]{\footnotesize $z$} circle (0.045);
\draw [color={blue!70!green}, thick, decoration = {markings, mark=at position 0.8 with {\arrow[scale=1]{stealth}}}, postaction=decorate] (.5,-.5) node[below]{$b$} --(.2,-.2); 
\draw [color={blue!70!green}, thick, decoration = {markings, mark=at position 0.8 with {\arrow[scale=1]{stealth}}}, postaction=decorate]  (-.2,.2) -- (0.1,.5) node[above]{$b$};
\end{tikzpicture}}}
\end{equation}
where $c$ runs over the simple objects appearing in the product $a\otimes b$, and $z$ over an orthonormal basis of $\Hom(a\otimes b,c)$.
In fact, the pairs $(m,R_{m,a})$ of an object $m\in \cC$ and a compatible half-braiding $R_{m,a}$ give the objects of the Drinfeld center $\ZC$ through its standard definition.
It is a nice fact that $\ZC$, as defined in this manner internal to the 2d picture, 
has all the right properties to describe the bulk anyons of $\TV_\cC$.
Due to this construction, the process of bringing a bulk anyon to the boundary is described by the forgetful functor $F:\ZC\to \cC$, which sends $\mu\in \ZC$ to $m=F(\mu)\in \cC$.

\paragraph{Representations $\rep(\mathrm{Tube}(\cC))$ and the Drinfeld center $\ZC$:}

Now we argue that the bulk anyon $\mu$ in Figure~\ref{fig:defect.operators} (and in {} \eqref{H.a.decomposition}) labels the tube algebra representation under which the defect operator $\cO$ transforms.
First, note that because the $\cC$-symmetry lines live on $\D$, the lassos are inserted on $\D$ to implement the action of the tube algebra, and hence the tube algebra only acts on the degrees of freedom on the $\D$ boundary. 
Hence, it is manifest that the tube algebra cannot change the bulk anyon $\mu$, therefore different anyons $\mu$ correspond to different representations. In particular, $W^\mu := \bigoplus_a W^\mu_a$ is the representation space associated to $\mu$.
In fact, these spaces $W^\mu$ for simple objects $\mu\in \ZC$ are known to exhaust the irreducible representations of $\mathrm{Tube}(\cC)$,
and therefore there is a natural equivalence between $\mathrm{Rep}(\mathrm{Tube}(\cC))$ and $\ZC$.\footnote{
Let us point out that there are actually four equivalent entities:
1. $\mathrm{Rep}(\mathrm{Tube}(\cC))$,
2. $\ZC$,
3. Anyons in the Turaev-Viro TQFT $\TV_\cC$,
4. Anyons in the Levin-Wen string-net model associated with $\cC$.
\let\ref\relax
The equivalence $\ref{2}\sim\ref{3}$ is shown in \cite{Kirillov:2010nh,turaev2010two}, $\ref{3}\sim\ref{4}$ is shown in \cite{kirillov2011string}, $\ref{1}\sim\ref{3}$ is shown in \cite{evans1995ocneanu}, and finally $\ref{1}\sim\ref{2}$ is proven in \cite{Izumi:2000qa,Muger}. See also \cite{Aasen:2017ubm,Fuchs:2021zrx} for recent discussions of $\ref{1}\sim\ref{4}$. The equivalence $\ref{1}\sim\ref{3}$ will be relevant for 3d derivation of the asymptotic formula, and $\ref{1}\sim\ref{2}$ will be relevant for the 2d derivation in the next subsection.
}
We find that \eqref{H.a.decomposition} gives the decomposition of the $\cC$-twisted sectors of $Q$ into the irreducible representations of the tube algebra.

The tube algebra is a finite-dimensional $\mathbb{C}^\ast$-algebra, hence it has a decomposition into a direct sum of matrix algebras
\begin{equation}\label{matrixalg}
    \mathrm{Tube}(\cC) = \bigoplus_\mu \mathrm{Hom}_\bC\left( W^\mu, W^\mu \right)\,.
\end{equation}
Given any representation, the elements $1_\mu \in \mathrm{Hom}_\bC(W_\mu,W_\mu)$ are represented as a set of minimal central projectors into different direct summands consisting of copies of a single irreducible representation. 
We denote these projection operators by $P_\mu$; see \eqref{G.projector} and \eqref{MFC.projector} for examples.

\subsection{Selection rules on correlation functions} 
\label{sec:selection}

Before deriving the asymptotic density formula, we want to discuss an immediate application of the tube algebra. We derive a selection rule on the correlation functions of local operators of a (not necessarily conformal) 2d theory $Q$ with symmetry $\cC$. Consider the $n$-point function
\begin{equation} \label{n.point.function}
    \langle \cO_1(x_1) \cO_2(x_2) \cdots \cO_n(x_n) \rangle\,,
\end{equation}
where the local operator $\cO_i$ transforms under the representation $\mu_i \in \ZC$. From the discussion above, we see that the composite operator $\cO_1(x_1) \cO_2(x_2) \cdots \cO_n(x_n)$ transforms under the possibly-reducible representation $\mu_1 \otimes \mu_2 \otimes \dots \otimes \mu_n$. This is because we can use 
Figure~\ref{fig:defect.operators} and fuse the bulk anyons to find the representation under which a composite operator transforms. Therefore, $\mathrm{Rep}(\mathrm{Tube}(\cC))$ and $\ZC$ have the same fusion rule, and we arrive at the following selection rule:
\begin{quote}
The correlation function \eqref{n.point.function} can be non-zero only if the identity appears in the decomposition of $\mu_1 \otimes \mu_2 \otimes \dots \otimes \mu_n$ into irreducible representations. 
\end{quote}
Several comments are in order:

\begin{enumerate}
    \item The selection rule above is only valid if $\cC$ is not spontaneously broken. We say that $\cC$ is spontaneously broken if there exists a local operator $\cO$ (i.e.\ the order parameter) with a non-zero one-point function that transforms non-trivially under $\cC$.
    \item The spontaneous breaking of $\cC$ leads to the existence of multiple vacua. In other words, if the theory $Q$ has a unique vacuum $\ket{\Omega}$ then $\cC$ cannot be spontaneously broken. To prove this we need to show that
\begin{equation} \label{selection.rule}
    \langle \cO \rangle = \langle\Omega |\cO|\Omega \rangle = 0
\end{equation}
if $\cO$ transforms under an irreducible representation $\mu \neq 1$. We show this below by noting that $\ket{\Omega}$ and $\ket{\cO} := \cO|\Omega \rangle$ transform under different irreducible representations $1$ and $\mu$, and hence have zero overlap.

Since we assumed $Q$ has a unique vacuum, $\ket{\Omega}$ should correspond to the Hartle-Hawking state---namely, the state prepared on the circle by a path integral on the disk. Thus, $\ket{\Omega}$ transforms under the identity representation, and $\ket{\cO}$ under the representation $\mu$. To show that these two states have zero overlap, let us discuss their eigenvalues under the operator $U_\nu:\cH \to \cH$ given by wrapping the line $F(\nu) \in \cC$ on the cylinder. From Figure~\ref{fig:defect.operators}, it is clear that a state in representation $\mu$ has eigenvalue $S_{\nu\mu}/S_{1\mu}$ under the action of $\nu$. Thus, we have
\begin{equation}
    \langle\Omega | U_{\nu} | \cO \rangle = \frac{S_{\nu\mu}}{S_{1\mu}} \langle\Omega |\cO \rangle = \frac{S_{\nu1}}{S_{11}} \langle\Omega |\cO \rangle\,.
\end{equation}
From the unitarity of the $S$-matrix, we conclude that $\langle \cO \rangle=0$ unless $\mu=1$.\footnote{This is because we can multiply both sides of the equation by $\overline{S}_{1\nu}$ and conclude $\delta_{\mu,1} \langle \cO \rangle = \langle \cO \rangle$.}
    \item From the argument above, we see that the tube algebra representation of a local operator $\cO$ is completely determined by its eigenvalue under $U_\nu$.
    \item Finally, we want to point out that had we only considered the action of the fusion algebra of $\cC$, we would not have been able to determine the representation of composite operators. In other words, there is no canonical notion of tensor product between representations of arbitrary algebras, such as the fusion algebra of $\cC$. However, the tube algebra has additional structures that allow us to define the tensor product between its representations.
\end{enumerate}

\subsection{Derivation of the asymptotic density formula} \label{sec:formula.fusion.cat.symm}

Now we are ready to derive the asymptotic density formula for general fusion category symmetries. We first derive the formula using the 3d slab construction, and then give a purely 2d derivation using the projection operators that were mentioned below \eqref{matrixalg}.

\paragraph{Derivation in the 3d interpretation:}

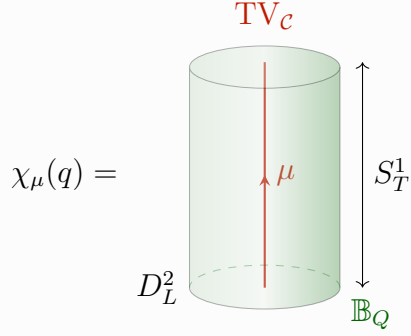
\begin{figure}[t]
    \centering
    $\chi_\mu(q)=$
\raisebox{-60pt}{\begin{tikzpicture}
\draw[color=red!75!DarkGreen, thick, decoration = {markings, mark=at position 0.5 with {\arrow[scale=1]{stealth}}}, postaction=decorate] (0,0) -- (0,1.5)node[right]{$\mu$} -- (0,3);
\draw (0,3.3) node[above]{\color{red!75!DarkGreen} $\TV_\cC$};
\draw [<->] (1.3,0) -- (1.3,3);
\draw (1.3,1.5) node[right]{$S^1_T$};
\draw (1,0) node[below right]{\color{DarkGreen} $\B_Q$};
\draw (-1,0) node[left]{$D^2_L$};
\node[style={draw, shape=cylinder, opacity=0.3, aspect=2, minimum height=+3.5cm,
minimum width=+2cm, left color=DarkGreen!30, right color=DarkGreen!90, middle color=DarkGreen!10,
shape border rotate=90, append after command={
  let \p{cyl@center} = ($(\tikzlastnode.before top)!0.5! (\tikzlastnode.after top)$),
      \p{cyl@x}      = ($(\tikzlastnode.before top)-(\p{cyl@center})$),
      \p{cyl@y}      = ($(\tikzlastnode.top)       -(\p{cyl@center})$)
  in (\p{cyl@center}) 
  edge[draw=none, opacity=0.2, fill=DarkGreen!10, to path={
    ellipse [x radius=veclen(\p{cyl@x})-1\pgflinewidth,
             y radius=veclen(\p{cyl@y})-1\pgflinewidth,
             rotate=atanXY(\p{cyl@x})]}] () }}] at (0,1.25) {};
\draw[color=DarkGreen, opacity=0.3, dashed] (1,0) arc [start angle=0, end angle=180, x radius=1, y radius=0.3];
\end{tikzpicture}}
    \caption{The partition function of $\TV_\cC$ on the solid torus $D_L^2 \times S_T^1$ with $\B_Q$ boundary condition and anyon $\mu$ in the middle. $D_L^2$ is the horizontal disk of circumference $L$, and $S^1_T$ is the vertical circle of size $T$. Since the boundary condition $\B_Q$ is conformal, this partition function depends only on $q=e^{-2\pi(T/L)}$.}
    \label{fig:solid.torus.partition.function}
\end{figure}

First, we identify the boundary condition $\B_Q$ as a state in the Hilbert space of the 3d TQFT $\TV_\cC$, also known as the \emph{partition vector} of $\B_Q$. We assume that $Q$ is a CFT, such that $\B_Q$ is a conformal boundary condition. Therefore, when the boundary is a Euclidean rectangular torus with modulus specified by $q=e^{-2\pi (T/L)}$, the partition vector only depends on $q$ and can be written as
\begin{equation} \label{partition.vector}
    \big | \B_Q(q) \big \rangle =
    \sum_\mu \chi_\mu(q) \ket{\mu} \in \cH(T^2)\,.
\end{equation}
Here $\cH(T^2)$ is the torus Hilbert space of $\TV_\cC$, and $\ket{\mu}$ is the orthonormal basis for this Hilbert space obtained by doing the path-integral on the solid torus with an anyon $\mu$ inside it.
Note that $\chi_\mu(q)=\langle \mu|\B_Q(q)\rangle$
is equal to the partition function of $\TV_\cC$ on a solid torus with $\B_Q$ boundary condition and the anyon $\mu$ inside it, as shown in Figure~\ref{fig:solid.torus.partition.function}.
Therefore,
\begin{equation}
    \chi_\mu(q) = \tr_{\cV_\mu} (e^{-(T/L)H_Q})\,,
\end{equation}
where $\cV_\mu$ is the Hilbert space of the theory $\TV_\cC$ on a disk
with an anyon $\mu$ inserted at the origin 
and the boundary condition $\B_Q$ placed at the circle boundary,
and $H_Q/L$ is the boundary Hamiltonian. We will compute the asymptotic density of states in terms of $\chi_\mu(q)$.

Define $\cH_a^\mu := W^\mu_a \otimes \cV_\mu$, which as mentioned after {} \eqref{H.a.decomposition} is the space of defect operators in the $a$-twisted sector that transform in representation $\mu$ of the tube algebra. We find
\begin{equation} \label{h.mu.a.trace}
    \tr_{\cH_a^\mu} (e^{-(T/L)H}) = \mathrm{dim}(W_a^\mu) \, \tr_{\cV_\mu} (e^{-(T/L)H_Q}) = \langle \mu, a \rangle \, \chi_\mu(q) \,,
\end{equation}
where we have denoted $\langle \mu, a \rangle := \mathrm{dim}(W_a^\mu)$, which are the fusion coefficients for the fusion of bulk anyons with the topological boundary $\D$.
Having derived the formula in \eqref{h.mu.a.trace}, we just need to find the asymptotic behavior of $\chi_\mu(q)$ in the $q \to 1$ limit.

Since $\B_Q$ is a conformal boundary condition, $\chi_\mu(q)$ transforms nicely under the modular transformations. In particular
\begin{equation}
    \chi_\mu(q) = \big\langle \mu \big | \B_Q(q) \big \rangle = \big\langle \mathbf{S} \cdot \mu \big | \mathbf{S} \cdot \B_Q(q) \big \rangle = \sum_\nu \overline{S}_{\mu\nu} \big\langle \nu \big | \B_Q(q') \big \rangle = \sum_\nu \overline{S}_{\mu\nu} \chi_\nu(q') \,,
\end{equation}
where $\mathbf{S}$ ($S_{\mu\nu}$) denotes the modular $S$-transformation (-matrix) of $\TV_\cC$, and $q' = e^{-2\pi(L/T)}$.
From the faithfulness assumption, there is a unique dimension-0 operator in the untwisted sector, as proven in Appendix~\ref{app:faithfulness}. Thus,
\begin{equation}
    \chi_\mu(q') = \tr_{\cV_\mu} (e^{-(L/T)H_Q}) \sim \delta_{\mu,1} \, e^{2\pi(L/T)(c/12)}
\end{equation}
in the limit $L \gg T$. Therefore, we find
\begin{equation} \label{chi.asymptotics}
    \chi_\mu(q) \sim S_{\mu1} \, e^{(L/T)(\pi c/6)} \,,
\end{equation}
which describes the asymptotic density of states in different representations of the tube algebra.

Finally, using \eqref{chi.asymptotics} and \eqref{h.mu.a.trace} we find that the asymptotic density of representation $\mu$ in the $a$-twisted sector behaves as
\begin{equation}
    \tr_{\cH_a^\mu} e^{-(T/L)H} \sim  \langle \mu,a \, \rangle S_{\mu1} \, e^{(L/T)(\pi c/6)}\,
\end{equation} in the regime $T \ll L$.
Note that $S_{\mu1} = S_{11} \dim \mu$, where $\dim \mu$ is the quantum dimension of $\mu$. As promised in Sec.~\ref{sec:introduction}, the density of states in the $a$-twisted sector transforming under the representation $\mu$ is proportional to $\langle \mu, a\rangle \dim \mu$, which generalizes the results of earlier sections.

\paragraph{Derivation internal to the 2d description:}

The derivation above, using the 3d perspective, is clear and concise, but it is rather hard to see
the relation to our analysis of the group symmetry case in Sec.~\ref{sec:group}
and of the MFC case in Sec.~\ref{sec:MFC}.
A more direct analog of what we did in these two previous sections can be performed as follows,
internal to the 2d description.

We first construct the projection operators $P_\mu$ that implement the decomposition \eqref{H.a.decomposition}. Via the forgetful tensor functor $F:\ZC \to \cC$ describing the fusion of bulk anyons with the topological boundary $\D$, there is an action of $\ZC$ on the theory $Q$. The action of $\mu \in \ZC$ on $\cH_a$ is graphically represented as
\begin{equation} \label{eq:action.of.Z}
\raisebox{-13pt}{
\begin{tikzpicture}
\draw [thick, color= red!75!DarkGreen ,decoration = {markings, mark=at position 0.8 with {\arrow[scale=1]{stealth}}}, postaction=decorate] (-1,0) -- (1,0) node[right]{$\mu$};
\draw [color={blue!70!green}, preaction={draw=white,line width=6pt}, thick, decoration = {markings, mark=at position 0.9 with {\arrow[scale=1]{stealth}}}, postaction=decorate] (0,-0.5) -- (0,0.5) node[above]{$a$};
\end{tikzpicture}
}.
\end{equation}
We claim that the projection operator into the sector $\cH^\mu_a = W^\mu_a \otimes \cV_\mu$ is given by
\begin{equation} \label{projectors}
    P^\mu_a := \sum_\nu S_{1\mu} \overline{S}_{\mu\nu} \raisebox{-13pt}{
\begin{tikzpicture}
\draw [thick, color= red!75!DarkGreen ,decoration = {markings, mark=at position 0.8 with {\arrow[scale=1]{stealth}}}, postaction=decorate] (-1,0) -- (1,0) node[right]{$\nu$};
\draw [color={blue!70!green}, preaction={draw=white,line width=6pt}, thick, decoration = {markings, mark=at position 0.9 with {\arrow[scale=1]{stealth}}}, postaction=decorate] (0,-0.5) -- (0,0.5) node[above]{$a$};
\end{tikzpicture}
}.
\end{equation}

As mentioned before, for a $\bC^\ast$-algebra such as the tube algebra, the irreducible representations are given by the \emph{minimal central idempotents} (MCI) acting as projectors into different irreducible representation types. Central idempotents are a set of central elements $1_\mu \in \mathrm{Tube}(\cC)$ satisfying $1_\mu 1_\nu = \delta_{\mu,\nu} 1_\mu$, and minimal central idempotents are central idempotents which cannot be written as a sum of central idempotents. Thus to prove the above claim, we need to show that $\sum_a P^\mu_a$ are the minimal central idempotents of the algebra.

To show that $\sum_a P^\mu_a$ are idempotents, note that
\begin{equation}
    P^\mu_a P^{\mu'}_{a'} = \delta_{a,a'} \sum_{\nu,\nu',\nu''} S_{1\mu} \overline{S}_{\mu\nu} S_{1\mu'} \overline{S}_{\mu'\nu'} N_{\nu\nu'}^{\nu''} \raisebox{-13pt}{
\begin{tikzpicture}
\draw [thick, color= red!75!DarkGreen ,decoration = {markings, mark=at position 0.8 with {\arrow[scale=1]{stealth}}}, postaction=decorate] (-1,0) -- (1,0) node[right]{$\nu''$};
\draw [color={blue!70!green}, preaction={draw=white,line width=6pt}, thick, decoration = {markings, mark=at position 0.9 with {\arrow[scale=1]{stealth}}}, postaction=decorate] (0,-0.5) -- (0,0.5) node[above]{$a$};
\end{tikzpicture}
} = \delta_{a,a'} \delta_{\mu,\mu'} P_a^\mu \,.
\end{equation}
To show that these projectors correspond to different representations, we need to show that they are central. This is manifest form the 3d slab construction since the fusion of bulk anyons with the topological boundary $\D$ commutes with the lasso action on the boundary; internally to the 2d description, it is implemented via the half-braiding of $\ZC$ with $\cC$:
\begin{equation}
\raisebox{-32pt}{
\begin{tikzpicture}
\draw [color=red!75!DarkGreen, thick, decoration = {markings, mark=at position 0.8 with {\arrow[scale=1]{stealth}}}, postaction=decorate] (-1,0.5) -- (1,0.5) node[right]{$\mu$};
\draw [color= {blue!70!green}, preaction={draw=white,line width=6pt}, thick, decoration = {markings, mark=at position 0.9 with {\arrow[scale=1]{stealth}}, mark=at position 0.25 with {\arrow[scale=1]{stealth}}}, postaction=decorate] (0,0) node[below]{$a$} -- (0,1.5) node[above]{$a'$};
\draw [color= {blue!70!green}, thick, decoration = {markings, mark=at position 0.8 with {\arrow[scale=1]{stealth}}}, postaction=decorate] (-1,1) -- (1,1) node[right]{$b$};
\draw [fill={blue!70!green}] (0,1) node[above right]{\footnotesize \color{blue!70!green} $x^b_{aa'}$} circle (0.04);
\end{tikzpicture}
}
= ~
\raisebox{-32pt}{
\begin{tikzpicture}
\draw [color=red!75!DarkGreen, thick, decoration = {markings, mark=at position 0.8 with {\arrow[scale=1]{stealth}}}, postaction=decorate] (-1,1) -- (1,1) node[right]{$\mu$};
\draw [color= {blue!70!green}, preaction={draw=white,line width=6pt}, thick, decoration = {markings, mark=at position 0.9 with {\arrow[scale=1]{stealth}}, mark=at position 0.2 with {\arrow[scale=1]{stealth}}}, postaction=decorate] (0,0) node[below]{$a$} -- (0,1.5) node[above]{$a'$};
\draw [color={blue!70!green}, thick, decoration = {markings, mark=at position 0.8 with {\arrow[scale=1]{stealth}}}, postaction=decorate] (-1,0.5) -- (1,0.5) node[right]{$b$};
\draw [fill={blue!70!green}] (0,0.5) node[below right]{\footnotesize \color{blue!70!green}
 $x^b_{aa'}$} circle (0.04);
\end{tikzpicture}
}.
\end{equation}
Since the number of representations of the tube algebra is the same as the number of MCIs constructed above, the central idempotents $\sum_a P^\mu_a$ have to be minimal.

Let us now begin the derivation of the asymptotic density formula. 
We will need to use various relations derived in Appendix~\ref{app:half.braiding}, which the reader might like to take a look at first.
We use the projector $P^\mu_a$ to find the asymptotic density of states in the sector $\cH^\mu_a$:
\begin{equation}
\tr_{\cH^\mu_a}e^{-(T/L)H} =\sum_{\nu}S_{1\mu}\overline{S}_{\mu\nu} ~
\raisebox{-30pt}{
\begin{tikzpicture}
\draw [color=red!75!DarkGreen, thick, decoration = {markings, mark=at position 0.8 with {\arrow[scale=1]{stealth}}}, postaction=decorate] (-1,0.75) -- (1,0.75) node[right]{$\nu$};
\draw [color={blue!70!green}, preaction={draw=white,line width=6pt}, thick, decoration = {markings, mark=at position 0.75 with {\arrow[scale=1]{stealth}}}, postaction=decorate] (0,0) -- (0,1.5) node[above]{$a$};
\draw (-1,0) rectangle (1,1.5);
\draw (1,0) node[left=2,above=5,anchor=north west]{\footnotesize $(q)$};
\end{tikzpicture}
}
=\sum_{\nu}S_{1\mu}\overline{S}_{\mu\nu} ~
\raisebox{-33pt}{
\begin{tikzpicture}
\draw [color=red!75!DarkGreen, thick, decoration = {markings, mark=at position 0.8 with {\arrow[scale=1]{stealth}}}, postaction=decorate] (0.75,0) -- (0.75,2) node[above]{$\nu$};
\draw [color={blue!70!green}, preaction={draw=white,line width=6pt}, thick, decoration = {markings, mark=at position 0.75 with {\arrow[scale=1]{stealth}}}, postaction=decorate] (1.5,1) -- (0,1) node[left]{$a$};
\draw (0,0) rectangle (1.5,2);
\draw (1.5,0) node[anchor=west]{\footnotesize $(q')$};
\end{tikzpicture}
}.
\end{equation}
Using the completeness relation \eqref{completeness} for the fusion of bulk anyons with the boundary, we find
\begin{equation} \label{expression}
\tr_{\cH^\mu_a}e^{-(T/L)H} = \sum_{\nu,b,x}S_{1\mu}\overline{S}_{\mu\nu} \, \sqrt{\frac{d_{b}}{d_\nu}}
\raisebox{-33pt}{
\begin{tikzpicture}
\draw [color={blue!70!green}, thick, decoration = {markings, mark=at position 0.65 with {\arrow[scale=1]{stealth}}}, postaction=decorate] (0.75,0) -- (0.75,0.5);
\draw [color=red!75!DarkGreen,thick] (0.75,1.5) arc[start angle=90, end angle=270, x radius=0.3, y radius=0.5];
\draw [fill={blue!70!green}] (0.75,0.5) circle (0.04);
\draw[color=blue!70!green] (0.75,0.5) node [right] {\small $x$};
\draw [color=red!75!DarkGreen,thick, decoration = {markings, mark=at position 1 with {\arrow[scale=1]{stealth}}}, postaction=decorate] (0.53,1.36) -- (0.533,1.37);
\draw[color=red!75!DarkGreen] (0.507,1.29) node[left]{$\nu$};
\draw [color={blue!70!green},preaction={draw=white,line width=6pt}, thick, decoration = {markings, mark=at position 0.55 with {\arrow[scale=1]{stealth}}}, postaction=decorate] (1.5,1) -- (0,1) node[left]{$a$};
\draw [fill={blue!70!green}] (0.75,1.5) circle (0.04);
\draw[color={blue!70!green}] (0.75, 1.5) node [right] {\small $x$};
\draw [color={blue!70!green}, thick, decoration = {markings, mark=at position 0.75 with {\arrow[scale=1]{stealth}}}, postaction=decorate] (0.75,1.5) -- (0.75,2) node[above]{$b$};
\draw (0,0) rectangle (1.5,2);
\draw (1.5,0) node[anchor=west]{\footnotesize $(q')$};
\end{tikzpicture}
} \sim
\sum_{\nu,x} \frac{ S_{1\mu}\overline{S}_{\mu\nu} }{\sqrt{d_\nu}}
\raisebox{-33pt}{
\begin{tikzpicture}
\draw [color={blue!70!green}, thick, dashed] (0.75,0) -- (0.75,0.5);
\draw [color=red!75!DarkGreen,thick] (0.75,1.5) arc[start angle=90, end angle=270, x radius=0.3, y radius=0.5];
\draw [fill={blue!70!green}] (0.75,0.5) circle (0.04);
\draw[color=blue!70!green] (0.75,0.5) node [right] {\small $x$};
\draw [color=red!75!DarkGreen,thick, decoration = {markings, mark=at position 1 with {\arrow[scale=1]{stealth}}}, postaction=decorate] (0.53,1.36) -- (0.533,1.37);
\draw[color=red!75!DarkGreen] (0.507,1.29) node[left]{$\nu$};
\draw [color={blue!70!green},preaction={draw=white,line width=6pt}, thick, decoration = {markings, mark=at position 0.55 with {\arrow[scale=1]{stealth}}}, postaction=decorate] (1.5,1) -- (0,1) node[left]{$a$};
\draw [fill={blue!70!green}] (0.75,1.5) circle (0.04);
\draw[color={blue!70!green}] (0.75, 1.5) node [right] {\small $x$};
\draw [color={blue!70!green}, thick, dashed] (0.75,1.5) -- (0.75,2);
\draw (0,0) rectangle (1.5,2);
\draw (1.5,0) node[anchor=west]{\footnotesize $(q')$};
\end{tikzpicture}
}.   
\end{equation}
where $x$ runs over an orthonormal basis of $W^\mu_a$--described around {} \eqref{orthogonality}. In the last step, we have used the result of Appendix~\ref{app:faithfulness} that for faithful symmetries there is a unique dimension-0 operator in the untwisted sector. Therefore, the states in the ($b \neq 1$)-twisted sectors have positive dimensions and do not contribute to the sum in the $q' \to 0$ limit.

Using the half-linking matrix $\Psi$ in {} \eqref{half.braiding}, we rewrite the right-hand side of \eqref{expression} as
\begin{equation} \label{expression.2}
\sum_{\nu,x} \frac{ S_{1\mu}\overline{S}_{\mu\nu} }{\sqrt{d_\nu}} \frac{\Psi_{a(\nu,xx)}}{\Psi_{a1}}
\raisebox{-33pt}{
\begin{tikzpicture}
\draw [color={blue!70!green},preaction={draw=white,line width=6pt}, thick, decoration = {markings, mark=at position 0.55 with {\arrow[scale=1]{stealth}}}, postaction=decorate] (1.5,1) -- (0,1) node[left]{$a$};
\draw (0,0) rectangle (1.5,2);
\draw (1.5,0) node[anchor=west]{\footnotesize $(q')$};
\end{tikzpicture}
} \sim
\sum_{\nu,x} \frac{ S_{1\mu}\overline{S}_{\mu\nu} }{\sqrt{S_{1\nu}}} \frac{\Psi_{a(\nu,xx)}}{d_a} \, d_a \, e^{(L/T)(\pi c/6)} \,.   
\end{equation}
Here we have used the fact that in the $q' \to 0$ limit, only the vacuum state contributes to the sum which has eigenvalue $d_a$ under the action of the topological line $a$. 
Using the generalized Verlinde formula \eqref{g.v.f}
\begin{equation}
    \sum_{\nu,x} \frac{ \Psi_{1(\nu,xx)} \overline{S}_{\mu\nu} \Psi_{a(\nu,xx)} }{S_{1\nu}} = \widetilde{N}_{\mu 1}^a = \langle  \mu , a \rangle \,,
\end{equation}
we simplify \eqref{expression.2} and find the asymptotic behavior 
\begin{equation}
     \tr_{\cH^\mu_a}e^{-(T/L)H} \sim  S_{1\mu} \langle \mu,a \rangle \, e^{(L/T)(\pi c/6)}\,
\end{equation}
when $T \ll L$.
This is consistent with the formula derived in the 3d construction.

Let us now sum over all possible irreducible representations $\mu$: \begin{equation}
     \tr_{\cH_a}e^{-(T/L)H} \sim  \sum S_{1\mu} \langle \mu,a \rangle \, e^{(L/T)(\pi c/6)}.
\end{equation}
Using the generalized Verlinde formula quoted above and $\Psi_{a1}=\sqrt{S_{11}} d_a$ 
as explained in Appendix~\ref{app:half.braiding}, we find \begin{equation}
\sim \dim a ~ e^{(L/T)(\pi c/6)}.
\end{equation}
This means that the asymptotic density of the whole of the $a$-twisted sector has a factor of the quantum dimension $\dim a$ of $a$, as already noted in \cite[Sec.~3.1]{Pal:2020wwd}.

\subsection{Concrete cases} \label{sec:examples}

\subsubsection{Finite group symmetry}
$\ZC$ for finite group $G$ and a general anomaly $\omega\in H^3(G,U(1))$, i.e.\ $\cC = \mathrm{Vec}_G^\omega$\,,
was determined in \cite{Roche:1990hs,Dijkgraaf:1990ne}.\footnote{
These references are somewhat hard to locate.
A nice summary can be found in \cite{deWildPropitius:1995cf}.
See its Sec.~1.5 for the non-anomalous case $\omega=0$ 
and its Sec.~3.1 for the more general case $\omega\neq 0$. See also \cite{coste2000finite}.}  The anyons in $\ZC$ are labeled by pairs $\mu=([g], \sigma)$, where $[g]$ runs over the conjugacy classes of $G$, and $\sigma$ over  the irreducible projective representations of the centralizer $C(g)$
with the projective phase $c_{g,\omega}\in H^2(C(g),U(1))$ given in \eqref{eq:phase}.
In this case, 
$\dim ([g],\sigma)=|G|/|C(g)| \dim \sigma$ and 
$\langle ([g],\sigma),g \rangle = \dim\sigma$,
and therefore our asymptotic density formula gives \begin{equation}
    \dim \mu \, \langle \mu, a\rangle = \frac{|G|}{|C(g)|} (\dim\sigma)^2,
\end{equation} 
in perfect agreement with what we saw in Sec.~\ref{sec:group}.

Let us consider the case of trivial anomaly $\omega$ and explicitly show that the irreducible representations of $\tub(\text{Vec}_G)$ are labeled by $([g], \sigma)$; in this case, $\sigma$ is a linear representation.  A natural basis for the elements of $\tub(\text{Vec}_G)$ is given by 
\begin{equation}
  e^g_h = \x{g}{hgh^{-1}}{h}{},
\end{equation}
with the composition rule
\begin{equation}
  e^g_h ~ e^{g'}_{h'} = \substack{\x{g}{hgh^{-1}~~~}{h}{} \\ \x{g'}{h'g'h'^{-1}}{h'}{}} = \delta_{g,h'g'h'^{-1}} ~ e^{g'}_{hh'}.
\end{equation}
Given $g \in G$, let $\sigma$ denote an irreducible representation of the centralizer $C(g)$.
Then
\begin{equation}
  \idem{\sigma}_g = \frac{\dim\sigma}{|C(g)|} \sum_{\substack{h \in C(g)}} \overline{\chi_\sigma(h)} ~ e^g_h = \frac{\dim\sigma}{|C(g)|} \sum_{\substack{h \in C(g)}} \overline{\chi_\sigma(h)} ~ \x{g}{g}{h}{}
\end{equation}
are idempotents because
\begin{equation}
  \begin{aligned}
  \idem{\sigma}_g \idem{\sigma'}_{g'} &= \frac{(\dim\sigma)^2}{|C(g)|^2} \sum_{\substack{h \in C(g)}} \sum_{\substack{h' \in C(g')}} \overline{\chi_\sigma(h)} ~ \overline{\chi_{\sigma'}(h')} ~ \delta_{g,g'} ~ e^{g'}_{hh'}
  \\
  &
  = \frac{(\dim\sigma)^2}{|C(g)|^2} ~ \delta_{\gamma,\gamma'} ~ \sum_{\substack{g \in \gamma \\ h, h' \in C(g)}} \overline{\chi_\sigma(h)} ~ \overline{\chi_{\sigma'}(h')} ~ e^g_{hh'}
  \\
  &
  = \frac{(\dim\sigma)^2}{|C(g)|^2} ~ \delta_{\gamma,\gamma'} ~ \sum_{\substack{h, h'' \in C(g)}} {\chi_\sigma(h^{-1}) ~ \chi_{\sigma'}(h''h) ~ e^g_{{h''}^{-1}}}
  \\
  &
  = \frac{\dim\sigma}{|C(g)|} ~ \delta_{\gamma,\gamma'} \delta_{\sigma,\sigma'} ~ \sum_{\substack{h'' \in C(g)}} \overline{\chi_\sigma({h''}^{-1})} ~ e^g_{{h''}^{-1}}
  = \delta_{\gamma,\gamma'} \delta_{\sigma,\sigma'} \idem{\sigma}_g,
  \end{aligned}
\end{equation}
where a character identity
\begin{equation}
  \sum_{h \in H} \chi_\sigma(h^{-1}) \, \chi_{\sigma'}(h'' h) = \frac{|H|}{\dim\sigma} \delta_{\sigma,\sigma'} \chi_\sigma(h''),
\end{equation}
with $H$ a finite group, $h, h'' \in H$, and $\sigma, \sigma' \in \rep(H)$, was used.\footnote{See for instance, \href{https://mathoverflow.net/questions/314913/a-character-identity}{https://mathoverflow.net/questions/314913/a-character-identity}.}
The $\idem{\sigma}_g$ is not central unless $[g]$ contains a single element, because otherwise, $\idem{\sigma}_g$ clearly does not commute with $e^{g'}_h$ for any $g' = hgh^{-1} \neq g$.  The remedy is to sum $g$ over a conjugacy class $\gamma$ of $G$,
\begin{equation}\label{groupmci}
  \idem{\gamma,\sigma} = \sum_{g \in \gamma} \idem{\sigma}_g = \frac{\dim\sigma}{|C(g)|} \sum_{\substack{g \in \gamma \\ h \in C(g)}} \overline{\chi_\sigma(h)} ~ e^g_h = \frac{\dim\sigma}{|C(g)|} \sum_{\substack{g \in \gamma \\ h \in C(g)}} \overline{\chi_\sigma(h)} ~ \x{g}{g}{h}{}.
\end{equation}
Indeed, $\idem{\gamma,\sigma}$ are central, as we compute
\begin{equation}
  \begin{aligned}
    \idem{\gamma,\sigma} ~ e^g_h &= \frac{\dim\sigma}{|C(g)|} \sum_{\substack{g' \in \gamma \\ h' \in C(g')}} \overline{\chi_\sigma(h')} ~ \delta_{g',hgh^{-1}} ~ e^{g}_{h'h} = \frac{\dim\sigma}{|C(g)|} \sum_{h' \in C(hgh^{-1})} \overline{\chi_\sigma(h')} ~ e^g_{h'h},
  \\
  e^g_h ~ \idem{\gamma,\sigma} &= \frac{\dim\sigma}{|C(g)|} \sum_{\substack{g'' \in \gamma \\ h'' \in C(g'')}} \overline{\chi_\sigma(h'')} ~ \delta_{g,g''} ~ e^{g''}_{hh''} = \frac{\dim\sigma}{|C(g)|} \sum_{h'' \in C(g)} \overline{\chi_\sigma(h'')} ~ e^g_{hh''},
  \end{aligned}
\end{equation}
and the expressions on the right can be identified upon the variable redefinition $h'' = h^{-1}h'h$.  We claim that $\idem{\gamma,\sigma}$ are the minimal central idempotents.  To see this, first note that the representation space of $\idem{\gamma,\sigma}$ is $W^{(\gamma,\sigma)} = \bigoplus_{g \in \gamma} W^{(g,\sigma)}$, which
has a basis labeled by $v^g_i$ where $g \in \gamma$ and $i=1, \dotsc, \dim \sigma$.  We can go between $v^g_i$ and $v^g_j$ for any pair of $i,j$ by an element $e^g_h$ with $h \in C(g)$, and also go between $W^{(g,\sigma)}$ and $W^{(hgh^{-1},\sigma)}$ by $e^g_h$.  Given the structure \eqref{matrixalg}, it is then clear that $\idem{\gamma,\sigma}$ is minimal.

\subsubsection{Modular fusion category symmetry}
If $\cC$ comes from a modular fusion category $\cM$, then $\mathcal{Z}(\cM)$ in this case is equivalent to $\cM \boxtimes \cM^\text{op}$ \cite[Prop.~7.13.8]{EGNO}.  
Symmetries described by MFCs were treated in section~\ref{sec:MFC}, where two sets of orthogonal projectors, $P_a^{(c,*)}$ and $P_a^{(*,d)}$, were constructed in \eqref{MFC.projector} and \eqref{MFC.projector2}.  
These projectors are not central, as is easily seen by considering their composition with a $\tub(\cC)$ element
\begin{equation}
  \x{a}{a'}{b}{}
\end{equation}
with $a' \neq a$ and $b$ non-invertible.\footnote{Here we assume that $\cC$ is non-invertible, since finite group symmetries have already been discussed.}
A simple way to fix this is to sum over $a$, but while $\sum_a P_a^{(c,*)}$ and $\sum_a P_a^{(*,d)}$ are central idempotents, they are not minimal.
It turns out that their compositions
\begin{equation}\label{MFCmci}
  P^{(c,d)} = \sum_a P_a^{(c,*)} P_a^{(*,d)} = \sum_{a,b,b'} S_{1c} \overline{S_{bc}} S_{1d} \overline{S_{b'd}} ~ \xx{a}{b}{b'}
\end{equation}
are minimal.
Note that under the folding operation described in footnote \ref{folding}, the operator \eqref{MFCmci} agrees with the general form of minimal central idempotents, $\sum_a P_a^\mu$ with $P_a^\mu$ defined in \eqref{projectors}, under the identifications
\begin{equation}
  \mu = c \otimes \bar d, \quad \nu = b \otimes \bar b'.
\end{equation}
Furthermore, the bulk-boundary map is given by $\langle \mu, a\rangle= N_{ca}^d$,
and therefore our general formula reduces to what we found in Sec.~\ref{sec:MFC}: \begin{equation}
\langle \mu,a\rangle  \dim \mu=  N_{ca}^d\dim c \dim d .
\end{equation}

The selection rule of Sec.~\ref{sec:selection} is trivially satisfied in a diagonal RCFT with MFC symmetry $\cM$.  Let us recall the general structure of defect Hilbert spaces \eqref{eq:RCFT-twisted-H} in an RCFT, where we wrote $\cO_{cd}^{a;x}$ to denote the $x$th direct summand of $V_c \otimes \overline{V}_d \in \cH_a$, with $x = 1, \dots, N_{ad}^c$.  If we label the simple objects in $\cZ(\cM)$ by $(c,d) \in \cM \otimes \cM^\text{op}$, then $\cO_{cd}^{a;x}$ belongs to the $\tub(\cM)$ representation corresponding to $(c,d)$.  The operator algebra has the structure
\begin{equation}
  \cO_{cd}^{a;x} \cO_{c'd'}^{a',x'} = 
  \bigoplus_{a'' \in a \otimes a', ~ c'' \in c \otimes c', ~ d'' \in d \otimes d'}
  \cO_{c''d''}^{a'';x''},
\end{equation}
and the selection rule just amounts to ignoring the $a'' \in a \otimes a'$ restriction.

\subsubsection{Haagerup fusion category symmetry}
At present, most of the known fusion category symmetries are based on finite groups,
modular fusion categories associated with affine Lie algebras,
or related to these two examples via gauging of an algebra object. 
In this sense, our analyses in Sec.~\ref{sec:group} and in Sec.~\ref{sec:MFC} usually suffice.
But there is a few cases where our more general analysis in this section is actually necessary.

A notable class of exceptions is the one originally found by Haagerup in \cite{HaagerupOriginal} in the context of subfactor theory, 
which was later generalized by Asaeda and Haagerup \cite{AsaedaHaagerup} 
and by Izumi \cite{IzumiClassification}.
Here we use a version $H_3$ obtained by Grossman and Snyder \cite{Grossman_2012} from a gauging of the original ones found by Haagerup.
A two-dimensional CFT having this symmetry was recently found numerically in \cite{Huang:2021nvb,Vanhove:2021zop}.
The category $H_3$ has six simple objects, $1$, $a$, $a^2$, $\rho$, $a\rho$, $a^2\rho$
with the fusion rule $a^3=1$, $a\rho=\rho a^2$, and $\rho^2=1+\rho+a\rho+a^2\rho$;
the quantum dimensions are given by $\dim a=1$ and $\dim \rho=(\sqrt{13}+3)/2$.
The $F$-symbols can be found in \cite{Titsworth,osborne2019fsymbols,Huang:2020lox}.
In the following, we adopt the `transparent' gauge of \cite{Huang:2020lox}.

The tube algebra and therefore the structure of the Drinfeld center of the Haagerup fusion category was determined by Izumi in \cite{Izumi}.
This example was further analyzed and generalized by Evans and Gannon in \cite{Evans:2010yr} and by Grossman and Izumi in \cite{GrossmanIzumi}.

The Drinfeld center of the Haagerup fusion category has 12 simple objects: $1$, $\pi_1$, $\pi_2$, $\sigma^{1,2,3}$, and $\mu^{1,\ldots,6}$,
such that \begin{equation}\label{F.Haagerup}
F(1)=1, \quad
F(\pi_1)=1\oplus X, \quad
F(\pi_2)=1\oplus1\oplus X, \quad
F(\sigma^i)=a\oplus a^2 \oplus X,\quad
F(\mu^j)=X
\end{equation} where $X=\rho\oplus a\rho \oplus a^2\rho$.\footnote{
  We adopt the labeling convention of \cite{Vanhove:2021zop} for $\mathcal{Z}(H_3)$ because earlier constructions of the Drinfeld center were based on $H_2$, which is related to $H_3$ by gauging the $\mathbb{Z}_3$ generated by $a$.  
  The forgetful functor for $H_2$ looks almost identical to that for $H_3$ except that $F(\pi_2)$ and $F(\sigma^3)$ are interchanged, since gauging $\mathbb{Z}_3$ exchanges charged untwisted sectors with uncharged twisted sectors.
}

With this, we can predict how the Hilbert space of the CFT numerically found in \cite{Huang:2021nvb,Vanhove:2021zop} should decompose under the Haagerup symmetry.  For intuition and clarity, we will often first discuss the representation theory of $\tub(H_3)_{u,u}$ for simples $u \in H_3$, and then lift them to $\rep(\tub(H_3))$.

In the untwisted Hilbert space $\cH_1$, there are 2 one-dimensional representations and 1 two-dimensional representation of the tube subalgebra $\tub(H_3)_{1,1}$, given by 
\begin{equation}
  \begin{aligned}
  1:
  & \quad a=1, ~~ \rho=\frac{\sqrt{13}+3}{2}, \qquad 
  \pi_1:
  \quad a=1, ~~ \rho=\frac{-\sqrt{13}+3}{2},
  \\
  \pi_2:
  & \quad 
    a=\begin{pmatrix}
      \omega
      \\
      & \omega^2
    \end{pmatrix},
    ~~
    \rho = \begin{pmatrix}
      & 1
      \\
      1
    \end{pmatrix},
  \end{aligned}
\end{equation}
where $\omega$ is a third root of unity.  These 3 representations lift to inside the 3 distinct representations of $\tub(H_3)$ with the same labels.

In the twisted Hilbert space $\cH_a$, there are 3 one-dimensional representations of the commutative tube subalgebra $\tub(H_3)_{a,a} = \text{Vec}(\mathbb{Z}_3)$,
given by
\begin{equation}
  \sigma^i:
  \quad \bim{a}{a}{a}=\x{}{a}{a}{}=e^{2\pi i s^i}=\omega^{-i},
\end{equation}
where $s^i$ denotes the Lorentz spin of states.
These 3 representations lift to inside the 3 distinct representations of $\tub(H_3)$ with the same labels.
For a fixed $i$, the Hilbert spaces $\cH_a^{\sigma^i}$ and $\cH_{a^2}^{\sigma^i}$ both lift to inside $\sigma^i$,
due to the existence of
\begin{equation}
  \bim{a}{\rho}{a^2} = \x{a}{a^2}{\rho}{} \in \tub(H_3).
\end{equation}
In the transparent gauge \cite{Huang:2020lox}, $\bim{a}{a}{a} \ \bim{a}{\rho}{a^2} = \bim{a}{\rho}{a^2} \ \bim{a^2}{a^2}{a^2}$, it is clear that the $\bim{a}{a}{a}$ and $\bim{a^2}{a^2}{a^2}$ eigenvalues are identical within the representation $\sigma^i$ of $\tub(H_3)$. 

To study the twisted Hilbert space $\cH_\rho$, 
we explicitly construct the tube subalgebra $\tub(H_3)_{\rho,\rho}$ using the basis
\begin{equation}
  e_{a_1,a_2} = 
\begin{gathered}
  \begin{tikzpicture}[scale=.8]
  \draw [line,-<-=.17,-<-=.5,-<-=.83] (2,0) -- (1.5,0) node [below] {$a_1$} -- (.5,0) node [below] {$a_2$}
  -- (-.5,0) node [below] {$a_1$} -- (-1,0);
  \draw [line,->-=.5] (0,-1) node [below] {$\rho$} -- (0,0);
  \draw [line,-<-=.5] (1,1) node [above] {$\rho$} -- (1,0);
  \end{tikzpicture}
\end{gathered}
\end{equation}
where
\begin{equation}
  (a_1, a_2) \in \{ (1,\rho), (\rho,1) \} \cup \{ (u, v) \mid u, v = \rho, a\rho, a^2\rho \},
\end{equation}
and applying the tube multiplication formula of \cite{Williamson:2017uzx} along with the explicit $F$-symbols \cite{Titsworth,osborne2019fsymbols,Huang:2020lox}.
The tube subalgebra $\tub(H_3)_{\rho,\rho}$ turns out to be \emph{commutative}, and therefore has 11 one-dimensional representations.  By a direct computation, the Dehn twist $e_{\rho,1}$ eigenvalues are found to be $e^{2\pi i s}$ with
\begin{equation}
  s = 0, ~0, ~0, ~\frac23, ~\frac13, ~\frac{2}{13}, ~\frac{6}{13}, ~\frac{5}{13}, ~\frac{8}{13}, ~\frac{7}{13}, ~\frac{11}{13},
\end{equation}
encoding the fractional Lorentz spins of states in $\cH_\rho$.
Among the 11 representations, 8 can be solely distinguished by $s$, and are lifted to representations of $\tub(H_3)$ as
\begin{equation}
  \begin{aligned}
  \sigma^1: & \quad s = \frac23, \qquad \sigma^2: \quad s = \frac13,
  \qquad
  \mu^j: \quad s \equiv \frac{2}{13}, \frac{6}{13}, \frac{5}{13}, \frac{8}{13}, \frac{7}{13}, \frac{11}{13}.
  \end{aligned}
\end{equation}
The remaining 3 representations with $s=0$ can be distinguished by e.g.\ their $e_{\rho,\rho}$-eigenvalues, and they lift to $1, \pi_1, \pi_2$.
The exact same analysis can be carried out for $\cH_{a\rho}$ and $\cH_{a^2\rho}$, but even without doing so, the results should obviously be the same since $\rho, a\rho, a^2\rho$ are permuted by the $\mathbb{Z}_3$ inner automorphism.
There are $\tub(H_3)$ elements, such as
\begin{equation}
  \x{\rho}{a\rho}{\rho}{}, \quad \x{\rho}{a^2\rho}{\rho}{},
\end{equation}
that weave $\cH_\rho, \cH_{a\rho}, \cH_{a^2\rho}$ together, which is why it is always the combination $X = \rho \oplus a\rho \oplus a^2\rho$ that appears in \eqref{F.Haagerup}.
Finally, other elements, such as 
\begin{equation}
  \x{\rho}{1}{\rho}{}, \quad \x{\rho}{a}{\rho}{}, \quad \x{\rho}{a^2}{\rho}{},
\end{equation}
make the connections to $\cH_1, \cH_a, \cH_{a^2}$, the details of which we omit.

As a simple application of the selection rule of Sec.~\ref{sec:selection}, consider the following fusion rule in $\cZ(H_3)$:
\begin{equation}
  \pi_1 \otimes \pi_2 = \pi_1 \oplus 2 \pi_2 \bigoplus_{i=1}^3 \sigma^i \bigoplus_{j=1}^6 \mu^j,
\end{equation}
which contains every simple object besides 1.  In the untwisted sector, this is trivialized by the $\mathbb{Z}_3$-charge conservation.  However, in the $\rho$-twisted sector this is not so obvious, but can be checked against the data of a topological field theory with $H_3$ symmetry \cite{huang2022topological,Huang:2021zvu}.

The numerical construction of this CFT in \cite{Huang:2021nvb} was based on the so-called `anyon chains' of \cite{Feiguin:2006ydp},
whose Hilbert space has a spin-chain-like structure constructed from the fusion category \cite{Buican:2017rxc,Vanhove:2018wlb,Aasen:2020jwb}.
In Appendix~\ref{app:chain} we show that the Hilbert spaces associated with the anyon chains also decompose with the same ratio of irreducibles in the long-chain limit.
This is in line with a very naive expectation since the asymptotic behavior of the Hilbert space of a CFT constructed from a chain would probe the Hilbert space of the original chain.
We should mention that the CFT limit seems subtle, since as $L \to \infty$ only states of sufficiently low energy $E < f(L)$ have meaning in the continuum.  Although $\lim_{L \to \infty} f(L) = \infty$, 
the $L \to \infty$ limit of the number of states $\#(E < f(L))$ as a fraction of all states may not be 1.
That said, we find the analysis in Appendix~\ref{app:chain} entertaining, which was why we included it.

\section*{Acknowledgments}
YL is supported by the Simons Collaboration Grant on the Non-Perturbative Bootstrap.
MO is supported by FoPM, WINGS Program, the University of Tokyo.
SS is supported in part by the Simons Foundation grant 488657 (Simons Collaboration on the Non-Perturbative Bootstrap).
YT is supported in part by WPI Initiative, MEXT, Japan at Kavli IPMU, the University of Tokyo, and by JSPS KAKENHI Grant-in-Aid (Kiban-S) No.16H06335. 

\appendix
\section{Dimension-0 operators and faithfulness}
\label{app:faithfulness}

Here we define a notion of faithfulness for fusion category symmetries. We show that our definition implies  the usual notion of faithfulness for the case of ordinary group symmetries.

For this purpose, we need to distinguish the objects of the fusion category $\cC$
and the topological line operators of the QFT.
Let $\cL$ be the category consisting of all topological line operators of the QFT under consideration.
For two line operators $L,L'\in \cL$, $\Hom_\cL(L,L')$ consists of all topological operators residing at the junction of the lines $L$ and $L'$.
We say that a 1+1d QFT has a fusion category symmetry $\cC$ if there is a faithful tensor functor $f:\cC\to \cL$, that is,
\begin{enumerate}
    \item For each element $a \in \cC$, there exists a topological line defect $f(a)\in \cL$ in the theory;
    \item For each $x \in \mathrm{Hom}_\cC(a,b)$, there exists a corresponding element $f(x)\in \Hom_\cL(f(a),f(b))$, and that $f$ is injective;
    \item And the fusion category structure of $\cC$ is preserved under the maps $a \mapsto f(a)$ and $x \mapsto f(x)$. 
\end{enumerate}
Now we are ready to define faithfulness:
\begin{quote}
A fusion category symmetry $\cC$ is said to act faithfully on a QFT if and only if 
the map $x \mapsto f(x)$ defined above is an isomorphism.
\end{quote}
Mathematically, it means that the tensor functor $f:\cC\to \cL$ is fully faithful.
Equivalently,  it means that $\cC$ is identified with a fusion subcategory~\cite[Definition 4.11.1]{EGNO} of $\cL$.

There are two consequences of the faithfulness that are important for us. 
First, we assume $\cC$ to be a fusion category, as opposed to a multifusion category,
meaning that $\Hom_\cC(1,1) \simeq \bC$.
As we assume $f(1)=1\in \cL$, we also have $\Hom_\cL(1,1)\simeq \bC$.\footnote{
$\cA:=\Hom_\cL(1,1)$ is in general a semisimple commutative algebra,
and therefore has the form $\cA=\bigoplus_i \bC e_i$ where
$e_i$ are orthogonal idempotents, i.e.~$(e_i)^2=e_i$, $e_i e_j=0$ ($i\neq j$).
Therefore, when $\cA\supsetneq \bC$, there are multiple idempotents,
and inserting one of them at a point decomposes the theory into separate sectors,
which are now commonly called \emph{universes}.
They can also be thought of as 1-form symmetries in two dimensions. 
For more details, see e.g.~the expository articles \cite{Sharpe:2021srf,Sharpe:2022ene}.
}
In the case of a CFT,
by using the state-operator correspondence, 
we see that there is a unique dimension-0 operator in the untwisted sector.

Next, for a simple object $a\in\cC$ not isomorphic to $1$, 
we have $\Hom_\cC(1,a)=\varnothing$ by definition.
When the action of $\cC$ is faithful,
we then have $\Hom_\cL(1,f(a))=\varnothing$.
In the case of a CFT,
this then means that there are no dimension-0 operators in the sector twisted by $f(a)$.

Let us now specialize to the case of group symmetries. 
In this case, $\cC = \mathrm{Vec}_G^\omega$ for some finite group $G$ and $\omega \in H^3(G,U(1))$, meaning that it describes a finite group symmetry $G$ with 
the 't Hooft anomaly specified by $\omega$.
In this case, faithfulness translates to the fact that for $g \neq h$, 
the topological lines $f(g)$ and $f(h)$ are distinct, which means that they need to act on the untwisted sector Hilbert space in distinct manners,
recovering the standard definition of the faithfulness for the action of a group $G$.

\section{Symmetric 2d QFTs as boundaries of 3d TQFTs} \label{app:2dQFT.3dTQFT}

In this rather technical appendix, we show that a 2d QFT $Q$ with fusion category symmetry $\cC$ can be lifted to a boundary condition $\B_Q$ of the 3d Turaev-Viro TQFT $\TV_\cC$.
The crucial property of the boundary condition $\B_Q$ is that we can recover the original 2d QFT $Q$ by putting $\TV_\cC$ on an interval with the left boundary specified by $\B_Q$
and the right boundary specified by the topological boundary condition $\bD_\cC$ hosting $\cC$.
We denote it schematically as $
Q=(\B_Q,\cC).
$

When $\cC=\Vec^\omega_G$, i.e.~a finite group $G$ with an anomaly specified by $\omega\in H^3(G,U(1))$,
the corresponding Turaev-Viro theory $\TV_\cC$ is the 3d Dijkgraaf-Witten gauge theory whose action is $\omega$.
Then our statement can be shown as follows: 
we first place the theory $Q$ at a boundary of a 3d bulk $G$-SPT, which is the ungauged Dijkgraaf-Witten theory with the action $\omega$;
we then gauge the $G$-symmetry in the bulk, which gives the boundary $\B_Q$.
Similarly, the ungauged Dijkgraaf-Witten theory has a boundary condition where the $G$ symmetry is completely broken. 
By gauging the bulk $G$ symmetry, we can define the Dirichlet boundary $\bD_\cC$. 
Let us now consider the gauged Dijkgraaf-Witten theory on an interval, with the two boundary conditions given by $\B_Q$ and $\bD_\cC$ respectively. 
Unwinding the definitions, this ends up a purely 2d operation where we diagonally gauge the $G$ symmetry of a combined system of the original theory $Q$ and a completely-broken $G$ theory, which simply gives back the theory $Q$.

For general $\cC$, we follow the same steps, but need to proceed more indirectly.
Let us start by recalling the basic properties of $\TV_\cC$.
It is known that a 3d TQFT (of Reshetikhin-Turaev type) has a topological boundary condition if and only if it is a Turaev-Viro TQFT $\TV_\cC$~\cite{Fuchs:2012dt,Freed:2020qfy,Kaidi:2021gbs}.
Its anyon is given by the Drinfeld center $\ZC$ of $\cC$, 
and  it has a topological boundary condition such that
the boundary line operators form $\cC$ 
 or $\cC^\mathrm{op}$,
depending on the choice of orientation.

By definition of $\ZC$ as the center of $\cC$,
our theory $Q$ not only has an action of $\cC$ but also of $\ZC$, via 
the forgetful functor $F:\ZC\to \cC$.
This can further be extended to the action of $\cC\boxtimes \ZC$,
since the action of $\ZC$ commutes with the action of $\cC$.
Therefore, $Q$ has a generally non-faithful action of $\cC\boxtimes \ZC$.
When we insert and draw lines from both $\cC$ and $\ZC$ to the theory $Q$ living on a 2d spacetime $M_\text{2d}$, 
the lines from $\ZC$ behave \emph{as if} living in the bulk of a short interval $\times M_\text{2d}$.
In other words,  $Q$ automatically behaves \emph{as if} the 3d theory $\TV_\cC$ is put on an interval $(\B_Q,\D_\cC)$.
This tautological construction does not, however, allow us to consider this hypothetical boundary condition $\B_Q$ on boundaries of more general 3-manifolds.

To construct $\B_Q$, 
we take the topological boundary condition whose lines form $\cC^\mathrm{op}$,
and stack it with our theory $Q$ with a $\cC$ action.
This now has the combined $\cC^\text{op} \boxtimes \cC$ symmetry,
whose `diagonal' $\cC$ symmetry we gauge.
We denote this process schematically as $\B_Q:= (Q \times \cC^\text{op})/\text{diag}$.
To show $Q=(\B_Q,\cC)$,
 we first clarify what we mean by `gauging the diagonal $\cC$ symmetry'.

As reviewed in \cite{Bhardwaj:2017xup}, gauging in the context of fusion category symmetries corresponds to choosing a (Frobenius) algebra $A$ in a fusion category $\cC$ and inserting its fine mesh into the spacetime. 
This changes the symmetry to be another fusion category $\cD=\cC/A$.
This gauging can be reversed, as $\cD$ has an algebra object $A'$ such that $\cC=\cD/A'$.

$\cC \boxtimes \cC^\mathrm{op}$ has a \emph{canonical algebra} $A_\text{can}$ given by viewing $\cC$ as a ($\cC \boxtimes \cC^\mathrm{op}$)-module category by its left and right actions  \cite[Definition~7.9.12]{EGNO},
with which we define the diagonal gauging, i.e.~$/\text{diag}:=/A_\text{can}$.
It is known that $(\cC^\text{op}\boxtimes \cC)/A_\text{can}=\ZC$ \cite[Prop.~7.13.8]{EGNO}.
The inverse gauging is done by an algebra object $A_\text{Lag}$ in $\ZC$ 
known as the Lagrangian algebra,
such that the category of $A_\text{Lag}$-modules in $\ZC$ is equivalent to $\cC$,
see \cite[Remark~4.9]{davydov2013witt}.

The property to be proved was $Q=(\B_Q,\cC)=(Q\times\cC^\text{op}/\text{diag}, \cC)$.
Both discrete gauging and interval reduction are finite topological manipulations, hence they commute with each other. 
Therefore, it suffices to show $Q=Q\times(\cC^\text{op},\cC)/\text{diag}$ instead,
where 
$(\cC^\text{op},\cC)$
is a TQFT obtained by putting $\TV_\cC$
on an interval whose two boundaries are specified by $\cC^\text{op}$ and $\cC$,
which turns out to be the regular 2d TQFT of~\cite{Huang:2021zvu}.
By inverse gauging,
the property to be proved is further equivalent to $Q/A_\text{Lag} = Q \times (\cC^\text{op},\cC)$,
where $Q$ on the left-hand side is now regarded as having a symmetry $\cC\boxtimes \ZC$,
which was possible because  $Q$ behaves \emph{as if} defined on an interval $\times M_\text{2d}$.
Now,  gauging an algebra $A$ of $\ZC$ corresponds to inserting the condensation surface defect $S_A$ in the middle of this interval~\cite{Kapustin:2010if,Gaiotto:2020iye,Komargodski:2020mxz}. 
In our case, $A=A_\text{Lag}$ is the Lagrangian algebra associated to the Dirichlet boundary condition $\cC$, 
and therefore $S_{A_\text{Lag}}$ is given by the fusion of the Dirichlet boundary condition with its inverse~\cite{Roumpedakis:2022aik}. 
Hence after inserting $S_{A_\text{Lag}}$ in the middle of the interval,
we end up splitting the 3d spacetime into two,
showing indeed $Q/A_\text{Lag} = Q \times (\cC^\text{op},\cC)$. 
This finishes the derivation of our claim.

This construction allows us to realize various gauging of $Q$ as the 3d theory $\TV_\cC$ on an interval with appropriate boundary conditions.
Indeed, as $Q=(\B_Q,\cC)$,
we clearly have $Q/A = (\B_Q,\cC/A)=(\B_Q,\cD)$,
where $\cD=\cC/A$ is now considered as a boundary condition of $\TV_\cC$
obtained by inserting a fine mesh of the algebra object $A$ 
to the original topological boundary condition $\cC$.
In fact, it is known that all topological boundary conditions of $\TV_\cC$ is of this form.
We can also write  $Q/A$ as $Q \times (\cC^\text{op},\cD)/\text{diag}$,
where $(\cC^\text{op},\cD)$ is a TQFT described in \cite{Huang:2021zvu}
constructed from $\cC$ and $A$.

\section{Half-linking and generalized Verlinde formula} \label{app:half.braiding}

In this Appendix, we define the half-linking matrix for the fusion of bulk anyons in the 3d TQFT $\TV_\cC$ with its topological boundary $\D$. Moreover, we show that this matrix satisfies a generalized Verlinde formula. In the following, we use blue for $\bD$ along with the simple objects of $\cC$ it hosts, and red for the bulk anyons in $\ZC$.

To begin, let us fix an orthonormal basis $\{x,y,z,\dots\}$ for each junction vector space $W^\mu_a = \mathrm{Hom}_\cC(F(\mu),a)$, where $F:\ZC \to \cC$ describes the fusion of bulk anyons with the boundary $\D$.  We choose the basis such that we have the following orthogonality relation\footnote{
The particular normalization above is chosen such that in the case where $\cC$ is an MFC---and thus $\ZC=\cC \boxtimes \overline{\cC}$---we recover the usual orthogonality and completeness relation of $\cC$ by folding. Note that the 3d TQFT with anyons $\ZC$ and ending on its topological boundary $\bD$ is related to the 3d TQFT with anyons $\cC$ by folding, where $\bD$ becomes a trivial interface in the latter. The normalization in \eqref{orthogonality} is related by folding to the standard normalization for the fusion of $\cC$ described, for instance, in \cite[Equation (195)]{Kitaev:2005hzj}. Another reason for this particular choice of normalization is to get the generalized Verlinde formula in \eqref{g.v.f} without any extra normalization factor.
} 
\begin{equation} \label{orthogonality}
\raisebox{-57pt}{
\begin{tikzpicture}
\draw[color=blue!70!green, dashed] (-0.1,-0.3) -- (-0.1,2.1) -- (1.6,3.8) -- (1.6,1.4) -- cycle;
\draw[color=blue!70!green, thick, decoration = {markings, mark=at position 0.65 with {\arrow[scale=1]{stealth}}}, postaction=decorate] (0,1) -- (0.1,1.1) node[above=2pt]{$a$} -- (0.5,1.5);
\draw[color=blue!70!green, thick, decoration = {markings, mark=at position 0.7 with {\arrow[scale=1]{stealth}}}, postaction=decorate] (1,2) -- (1.25,2.25) node[above=1.5pt]{$a'$} -- (1.5,2.5);
\draw[color=red!75!DarkGreen, thick, decoration = {markings, mark=at position 0.6 with {\arrow[scale=1]{stealth}}}, postaction=decorate] (0.5,1.5)[rotate=225] arc [start angle=0, end angle=180, x radius=0.3536, y radius=0.3536];
\draw[color=red!75!DarkGreen] (0.75,1.5) node[below right] {$\mu$}; 
\draw [fill={blue!70!green}] (0.5,1.5) node[above]{\small \color{blue!70!green}
 $x$} circle (0.04);
 \draw [fill={blue!70!green}] (1,2) node[above left = -2.5pt]{\small \color{blue!70!green}
 $y$} circle (0.04);
\end{tikzpicture}
}
~~= \delta_{a,a'}\delta_{x,y} \sqrt{\frac{d_\mu}{d_a}} \quad
\raisebox{-57pt}{
\begin{tikzpicture}
\draw[color=blue!70!green, dashed] (-0.1,-0.3) -- (-0.1,2.1) -- (1.6,3.8) -- (1.6,1.4) -- cycle;
\draw[color=blue!70!green, thick, decoration = {markings, mark=at position 0.6 with {\arrow[scale=1]{stealth}}}, postaction=decorate] (0,1) -- (0.70,1.70) node[above=4pt]{$a$} -- (1.5,2.5);
\end{tikzpicture}
}\,,
\end{equation}
and the completeness relation
\begin{equation} \label{completeness}
\raisebox{-57pt}{
\begin{tikzpicture}
\draw[color=blue!70!green, dashed] (-0.1,-0.3) -- (-0.1,2.1) -- (1.6,3.8) -- (1.6,1.4) -- cycle;
\draw[color=red!75!DarkGreen, preaction={draw=white,line width=4pt}, thick, decoration = {markings, mark=at position 0.6 with {\arrow[scale=1]{stealth}}}, postaction=decorate] (-0.2,0.3) -- (0.7,1.2) node[above=4pt]{$\mu$} -- (1.8,2.3);
\end{tikzpicture}
}
~~= \sum_{a;x} \sqrt{\frac{d_a}{d_\mu}} \quad
\raisebox{-57pt}{
\begin{tikzpicture}
\draw[color=blue!70!green, dashed] (-0.1,-0.3) -- (-0.1,2.1) -- (1.6,3.8) -- (1.6,1.4) -- cycle;
\draw[color=blue!70!green, thick, decoration = {markings, mark=at position 0.65 with {\arrow[scale=1]{stealth}}}, postaction=decorate] (0.5,1.5) -- (0.67,1.67) node[above=4pt]{$a$} -- (1,2);
\draw[color=red!75!DarkGreen, thick] (0.5,1.5)[rotate=45] arc [start angle=0, end angle=-90, x radius=0.3536, y radius=0.3536];
\draw[color=red!75!DarkGreen, thick] (1,2)[rotate=225] arc [start angle=0, end angle=90, x radius=0.3536, y radius=0.3536];
\draw[color=red!75!DarkGreen, preaction={draw=white,line width=4pt}, thick, decoration = {markings, mark=at position 0.8 with {\arrow[scale=1]{stealth}}}, postaction=decorate] (-0.2,0.3) -- (0.3,0.8) node[below=2.5pt]{$\mu$} -- (0.5,1);
\draw[color=red!75!DarkGreen, preaction={draw=white,line width=4pt}, thick, decoration = {markings, mark=at position 0.1 with {\arrow[scale=1]{stealth}}}, postaction=decorate] (1.5,2) node[below=3pt]{$\mu~~~$} -- (1.8,2.3);
\draw [fill={blue!70!green}] (0.5,1.5) node[left]{\small \color{blue!70!green}
 $x$} circle (0.04);
 \draw [fill={blue!70!green}] (1,2) node[above]{\small \color{blue!70!green}
 $x$} circle (0.04);
\end{tikzpicture}
}\,.
\end{equation}

We define the \emph{half-linking matrix} $\Psi$ as
\begin{equation} \label{psi}
    \Psi_{a(\mu,xy)} := \sqrt{S_{11}} ~~
\raisebox{-15pt}{
\begin{tikzpicture}

\draw[color=blue!70!green, thick, decoration = {markings, mark=at position 0.92 with {\arrow[scale=1]{stealth}}}, postaction=decorate] (0,0) circle [x radius =0.5, y radius=0.5];
\draw[color=blue!70!green] (0.45,-0.2) node[right] {$a$};
\draw[color=red!75!DarkGreen, preaction={draw=white,line width=4pt}, thick, decoration = {markings, mark=at position 0.6 with {\arrow[scale=1]{stealth[reversed]}}}, postaction=decorate] (0,0) arc [start angle=-90, end angle=90, x radius=0.5, y radius=0.5];
\draw[color=red!75!DarkGreen] (0.5,0.5) node[right] {$\mu$};
\draw [fill={blue!70!green}] (0,0) node[below]{\small \color{blue!70!green} $y$} circle (0.04);
\draw [fill={blue!70!green}] (0,1) node[left]{\small \color{blue!70!green} $x$} circle (0.04);
\end{tikzpicture}
} = \sqrt{S_{11}} ~~
\raisebox{-19pt}{
\begin{tikzpicture}
\draw[color=blue!70!green, thick, decoration = {markings, mark=at position 0.08 with {\arrow[scale=1]{stealth[reversed]}}}, postaction=decorate] (0,0) circle [x radius =0.5, y radius=0.5];
\draw[color=blue!70!green] (0.45,0.2) node[right] {$a$};
\draw[color=red!75!DarkGreen, preaction={draw=white,line width=4pt}, thick, decoration = {markings, mark=at position 0.6 with {\arrow[scale=1]{stealth}}}, postaction=decorate] (0,0) arc [start angle=90, end angle=-90, x radius=0.5, y radius=0.5];
\draw[color=red!75!DarkGreen] (0.5,-0.5) node[right] {$\mu$};
\draw [fill={blue!70!green}] (0,0) node[below]{\small \color{blue!70!green} $x$} circle (0.04);
\draw [fill={blue!70!green}] (0,-1) node[left]{\small \color{blue!70!green} $y$} circle (0.04);
\end{tikzpicture}
}\,.
\end{equation}
Note that $\Psi_{a(\mu,xy)}$ is just a complex number because we can shrink the configuration to a local boundary operator on $\D$. But since there is a unique \emph{local} operator---the identity---on the boundary, this number is identified with the coefficient of the identity operator that we get after shrinking. By taking the orientation reversal of \eqref{psi} and using unitarity, we find the complex conjugate of $\Psi$ to be
\begin{equation}
    \overline{\Psi}_{a(\mu,xy)} = \Psi_{\bar{a}(\mu,xy)} = \Psi_{a(\bar{\mu},yx)}\,.
\end{equation}
Note that $\Psi_{11}=\sqrt{S_{11}}$ and  $\Psi_{a1}=\Psi_{11}d_a$. Moreover, from the orthogonality relation we find $\Psi_{1(\mu,xy)}=\delta_{x,y} \sqrt{S_{1\mu}}$. We can use the half-linking matrix to compute the following relations
\begin{align} \label{half.braiding}
\raisebox{-19pt}{
\begin{tikzpicture}
\draw[color=blue!70!green, thick, decoration = {markings, mark=at position 0.2 with {\arrow[scale=1]{stealth[reversed]}}}, postaction=decorate] (-0.7,0.5) -- (1,0.5);
\draw[color=blue!70!green] (-0.7,0.5) node[left] {$a$};
\draw[color=red!75!DarkGreen, preaction={draw=white,line width=4pt}, thick, decoration = {markings, mark=at position 0.6 with {\arrow[scale=1]{stealth[reversed]}}}, postaction=decorate] (0,0) arc [start angle=-90, end angle=90, x radius=0.5, y radius=0.5];
\draw[color=red!75!DarkGreen] (0.5,0.5) node[above right] {$\mu$};
\draw [fill={blue!70!green}] (0,0) node[left]{\small \color{blue!70!green} $y$} circle (0.04);
\draw [fill={blue!70!green}] (0,1) node[left]{\small \color{blue!70!green} $x$} circle (0.04);
\end{tikzpicture}
} &= \frac{\Psi_{a(\mu,xy)}}{\Psi_{a1}} ~~
\raisebox{-10.5pt}{
\begin{tikzpicture}
\draw[color=blue!70!green, thick, decoration = {markings, mark=at position 0.5 with {\arrow[scale=1]{stealth[reversed]}}}, postaction=decorate] (-0.7,0.5) -- (1,0.5);
\draw[color=blue!70!green] (0,0.5) node[below] {$a$};
\end{tikzpicture}
}\,, \\
\raisebox{-15pt}{
\begin{tikzpicture}
\draw[color=blue!70!green, thick, decoration = {markings, mark=at position 0.92 with {\arrow[scale=1]{stealth}}}, postaction=decorate] (0,0) circle [x radius =0.5, y radius=0.5];
\draw[color=blue!70!green] (0.45,-0.2) node[right] {$a$};
\draw[color=red!75!DarkGreen, preaction={draw=white,line width=4pt}, thick, decoration = {markings, mark=at position 0.6 with {\arrow[scale=1]{stealth}}}, postaction=decorate] (0,0) -- (1,1);
\draw[color=red!75!DarkGreen] (0.5,0.5) node[above] {$\mu$};
\draw [fill={blue!70!green}] (0,0) node[below]{\small \color{blue!70!green} $x$} circle (0.04);
\end{tikzpicture}
} &= \sum_y \frac{\overline{\Psi}_{a(\mu,xy)}}{\sqrt{S_{1\mu}}} ~~
\raisebox{-15pt}{
\begin{tikzpicture}
\draw[color=red!75!DarkGreen, preaction={draw=white,line width=4pt}, thick, decoration = {markings, mark=at position 0.6 with {\arrow[scale=1]{stealth}}}, postaction=decorate] (0,0) -- (1,1);
\draw[color=red!75!DarkGreen] (0.5,0.5) node[above] {$\mu$};
\draw [fill={blue!70!green}] (0,0) node[below]{\small \color{blue!70!green} $y$} circle (0.04);
\end{tikzpicture}
}\,.
\end{align}

Since $\cC$ forms a module over $\ZC$ as we can fuse the bulk lines with the boundary lines, we have
\begin{equation}
\raisebox{-23pt}{
\begin{tikzpicture}
\draw[color=blue!70!green, thick, decoration = {markings, mark=at position 0.6 with {\arrow[scale=1]{stealth}}}, postaction=decorate] (0.3,-0.5) node[below]{$a$} -- (0.3,0.5);
\draw[color=red!75!DarkGreen, thick, decoration = {markings, mark=at position 0.6 with {\arrow[scale=1]{stealth}}}, postaction=decorate] (0,-0.5) node[below]{$\mu$} -- (0,0.5);
\end{tikzpicture}
} = \bigoplus_b \widetilde{N}_{\mu a}^b 
\raisebox{-23pt}{
\begin{tikzpicture}
\draw[color=blue!70!green, thick, decoration = {markings, mark=at position 0.6 with {\arrow[scale=1]{stealth}}}, postaction=decorate] (0.2,-0.5) node[below]{$b$} -- (0.2,0.5);
\end{tikzpicture}
}\,.
\end{equation}
The non-negative integers $\widetilde{N}_{\mu a}^b$ form a non-negative integer matrix representation (NIM-rep) of the fusion algebra of $\ZC$. 
The NIM-rep $\widetilde{N}_{\mu a}^b$ can be expressed in terms of the coefficients $\langle \mu,a \rangle$ capturing the fusion of $\mu$ to $\bD$ together with the fusion coefficients $n_{ca}^b$ of $\cC$,
\begin{equation}
    \widetilde{N}_{\mu a}^b = \sum_c \langle \mu,c \rangle \, n_{ca}^b\,.
\end{equation}
Now let us compute the following braiding configuration
\begin{equation}
  \begin{gathered}
\begin{tikzpicture}
\draw[color=blue!70!green, thick, decoration = {markings, mark=at position 0.92 with {\arrow[scale=1]{stealth}}}, postaction=decorate] (0,0) circle [x radius =0.5, y radius=0.5];
\draw[color=blue!70!green] (0.45,-0.2) node[right] {$a$};
\draw[color=red!75!DarkGreen, thick, decoration = {markings, mark=at position 0.5 with {\arrow[scale=1]{stealth}}}, postaction=decorate] (0,0) circle [x radius =0.7, y radius=0.7];
\draw[color=red!75!DarkGreen] (-0.7,0) node[left] {$\mu$};
\draw[color=red!75!DarkGreen, preaction={draw=white,line width=4pt}, thick, decoration = {markings, mark=at position 0.6 with {\arrow[scale=1]{stealth[reversed]}}}, postaction=decorate] (0,0) arc [start angle=-90, end angle=90, x radius=0.5, y radius=0.5];
\draw[color=red!75!DarkGreen] (0.5,0.5) node[above right] {$\nu$};
\draw [fill={blue!70!green}] (0,0) node[below]{\small \color{blue!70!green} $y$} circle (0.04);
\draw [fill={blue!70!green}] (0,1) node[left]{\small \color{blue!70!green} $x$} circle (0.04);
\end{tikzpicture}
  \end{gathered}
=  \sum_b \widetilde{N}_{\mu a}^b ~
  \begin{gathered}
\begin{tikzpicture}
\draw[color=blue!70!green, thick, decoration = {markings, mark=at position 0.92 with {\arrow[scale=1]{stealth}}}, postaction=decorate] (0,0) circle [x radius =0.5, y radius=0.5];
\draw[color=blue!70!green] (0.45,-0.2) node[right] {$b$};
\draw[color=red!75!DarkGreen, preaction={draw=white,line width=4pt}, thick, decoration = {markings, mark=at position 0.6 with {\arrow[scale=1]{stealth[reversed]}}}, postaction=decorate] (0,0) arc [start angle=-90, end angle=90, x radius=0.5, y radius=0.5];
\draw[color=red!75!DarkGreen] (0.5,0.5) node[above right] {$\nu$};
\draw [fill={blue!70!green}] (0,0) node[below]{\small \color{blue!70!green} $y$} circle (0.04);
\draw [fill={blue!70!green}] (0,1) node[left]{\small \color{blue!70!green} $x$} circle (0.04);
\end{tikzpicture}
  \end{gathered}
= \frac{S_{\mu\nu}}{S_{1\nu}} ~
\begin{gathered}
\begin{tikzpicture}
\draw[color=blue!70!green, thick, decoration = {markings, mark=at position 0.92 with {\arrow[scale=1]{stealth}}}, postaction=decorate] (0,0) circle [x radius =0.5, y radius=0.5];
\draw[color=blue!70!green] (0.45,-0.2) node[right] {$a$};
\draw[color=red!75!DarkGreen, preaction={draw=white,line width=4pt}, thick, decoration = {markings, mark=at position 0.6 with {\arrow[scale=1]{stealth[reversed]}}}, postaction=decorate] (0,0) arc [start angle=-90, end angle=90, x radius=0.5, y radius=0.5];
\draw[color=red!75!DarkGreen] (0.5,0.5) node[above right] {$\nu$};
\draw [fill={blue!70!green}] (0,0) node[below]{\small \color{blue!70!green} $y$} circle (0.04);
\draw [fill={blue!70!green}] (0,1) node[left]{\small \color{blue!70!green} $x$} circle (0.04);
\end{tikzpicture}
\end{gathered}.
\end{equation}
Above we have evaluated the left-hand side in two different ways of fusing $\mu$ with the boundary and shrinking $\mu$ on $\nu$. Therefore, we find
\begin{equation} \label{almost.verlinde.formula}
    \sum_b \widetilde{N}_{\mu a}^b \Psi_{b(\nu,xy)} = \frac{S_{\mu\nu}}{S_{1\nu}} \Psi_{a(\nu,xy)}\,.
\end{equation}
To derive the generalized Verlinde formula, we will show that the matrix $\Psi$ is `almost' unitary\footnote{
If it were unitary we would have $\delta_{x,z} \delta_{y,t}$ instead of $\delta_{x,t} \delta_{y,z}$ on the right-hand side of \eqref{almost.unitarity}, and have $\overline{\Psi}_{b(\mu,xy)}$ instead of $\overline{\Psi}_{b(\mu,yx)}$ in \eqref{almost.unitarity.2}.}
in the sense that we have
\begin{equation} \label{almost.unitarity}
    \sum_a \overline{\Psi}_{a(\mu,xy)} \Psi_{a(\nu,zt)} = \delta_{\mu,\nu} \delta_{x,t} \delta_{y,z}\,,
\end{equation}
and therefore
\begin{equation} \label{almost.unitarity.2}
    \sum_{\mu,x,y} \Psi_{a(\mu,xy)} \overline{\Psi}_{b(\mu,yx)} = \delta_{a,b}\,.
\end{equation}
Assuming \eqref{almost.unitarity} and \eqref{almost.unitarity.2}, 
we can easily convert \eqref{almost.verlinde.formula} into the generalized Verlinde formula
\begin{equation} \label{g.v.f}
    \widetilde{N}_{\mu a}^b = \sum_{\nu,x,y} \frac{\Psi_{a(\nu,xy)} S_{\mu\nu} \overline{\Psi}_{b(\nu,yx)}}{S_{1\nu}}\,.
\end{equation}
This generalized Verlinde formula is a special case of a more general Verlinde formula in the RCFT literature~\cite{Gaberdiel:2002qa,Fuchs:2002cm} which is true for NIM-reps associated with module categories over arbitrary MFCs. A formula similar to \eqref{g.v.f} has also appeared in \cite{Shen:2019wop}.

Our remaining task is then to establish the almost unitarity, i.e.\ \eqref{almost.unitarity} and \eqref{almost.unitarity.2}.
From \eqref{almost.verlinde.formula}, we see that $\vec{\Psi}_{\ast(\nu,xy)}$ is an eigenvector of the matrix
$[\widetilde{\bf N}_\mu]_{ab} = \widetilde{N}_{\mu a}^b$
with eigenvalue ${S_{\mu\nu}}/{S_{1\nu}}$. Using the fact that $(\widetilde{\bf N}_\mu)^T = \widetilde{\bf N}_{\bar{\mu}}$, we find
\begin{equation}
    (\vec{\Psi}_{\ast(\nu,xy)})^\dagger \cdot \widetilde{\bf N}_\mu \cdot \vec{\Psi}_{\ast(\rho,zt)} = \frac{S_{\mu\nu}}{S_{1\nu}} \, (\vec{\Psi}_{\ast(\nu,xy)})^\dagger \cdot \vec{\Psi}_{\ast(\rho,zt)} = \frac{S_{\mu\rho}}{S_{1\rho}} \, (\vec{\Psi}_{\ast(\nu,xy)})^\dagger \cdot \vec{\Psi}_{\ast(\rho,zt)}\,,
\end{equation}
which is true for all $\mu$. 
Since the matrix $\bf S$ is unitary we find that
\begin{equation}
    \sum_a \overline{\Psi}_{a(\nu,xy)} \Psi_{a(\rho,zt)} \propto \delta_{\nu,\rho}\,,
\end{equation}
and hence
\begin{equation} \label{kirby.relation}
\sum_a \Psi_{a1} ~
\raisebox{-15pt}{
\begin{tikzpicture}
\draw[color=blue!70!green, thick, decoration = {markings, mark=at position 0.92 with {\arrow[scale=1]{stealth}}}, postaction=decorate] (0,0) circle [x radius =0.5, y radius=0.5];
\draw[color=blue!70!green] (0.45,-0.2) node[right] {$a$};
\draw[color=red!75!DarkGreen, preaction={draw=white,line width=4pt}, thick, decoration = {markings, mark=at position 0.6 with {\arrow[scale=1]{stealth[reversed]}}}, postaction=decorate] (0,0) arc [start angle=-90, end angle=90, x radius=0.5, y radius=0.5];
\draw[color=red!75!DarkGreen] (0.5,0.5) node[above right] {$\mu$};
\draw [fill={blue!70!green}] (0,0) node[below]{\small \color{blue!70!green} $y$} circle (0.04);
\draw [fill={blue!70!green}] (0,1) node[left]{\small \color{blue!70!green} $x$} circle (0.04);
\end{tikzpicture}
} =  \frac{1}{\Psi_{11}} \sum_a \overline{\Psi}_{a1} \Psi_{a(\mu,xy)} \propto \delta_{\mu,1}\,.
\end{equation}
Now we are ready to prove the almost-unitarity of $\Psi$.

By the completeness relation for the bulk lines in $\ZC$, we have
\begin{equation}
    \sum_a \overline{\Psi}_{a(\mu,xy)} \Psi_{a(\nu,zt)} = \Psi_{11} \sum_a \Psi_{a1} ~
\begin{gathered}
\begin{tikzpicture}
\draw[color=blue!70!green, thick, decoration = {markings, mark=at position 0.92 with {\arrow[scale=1]{stealth}}}, postaction=decorate] (0,0) circle [x radius =0.7, y radius=0.7];
\draw[color=blue!70!green] (0.6,-0.3) node[right] {$a$};
\draw[color=red!75!DarkGreen, preaction={draw=white,line width=4pt}, thick, decoration = {markings, mark=at position 0.6 with {\arrow[scale=1]{stealth}}}, postaction=decorate] (0.3,0) arc [start angle=-90, end angle=90, x radius=0.3, y radius=0.6];
\draw[color=red!75!DarkGreen] (0.6,0.6) node[above right] {$\nu$};
\draw[color=red!75!DarkGreen, preaction={draw=white,line width=4pt}, thick, decoration = {markings, mark=at position 0.6 with {\arrow[scale=1]{stealth[reversed]}}}, postaction=decorate] (-0.3,0) arc [start angle=-90, end angle=90, x radius=0.3, y radius=0.6];
\draw[color=red!75!DarkGreen] (-0.1,0.6) node[above left] {$\mu$};
\draw [fill={blue!70!green}] (0.3,0) node[below]{\small \color{blue!70!green} $z$} circle (0.04);
\draw [fill={blue!70!green}] (0.3,1.2) node[above]{\small \color{blue!70!green} $t$} circle (0.04);
\draw [fill={blue!70!green}] (-0.3,0) node[below]{\small \color{blue!70!green} $y$} circle (0.04);
\draw [fill={blue!70!green}] (-0.3,1.2) node[above]{\small \color{blue!70!green} $x$} circle (0.04);
\end{tikzpicture}
\end{gathered}
= \Psi_{11} \sum_{a,\rho,\alpha} \Psi_{a1} \sqrt{\frac{d_\rho}{d_\mu d_\nu}}~
\begin{gathered}
\begin{tikzpicture}
\draw[color=blue!70!green, thick, decoration = {markings, mark=at position 0.92 with {\arrow[scale=1]{stealth}}}, postaction=decorate] (0,0) circle [x radius =0.7, y radius=0.7];
\draw[color=blue!70!green] (0.6,-0.3) node[right] {$a$};
\draw[color=red!75!DarkGreen, preaction={draw=white,line width=4pt}, thick, decoration = {markings, mark=at position 0.3 with {\arrow[scale=1]{stealth}},mark=at position 0.8 with {\arrow[scale=1]{stealth}}}, postaction=decorate] (0,0.25) arc [start angle=-30, end angle=30, x radius=0.7, y radius=0.7];
\draw[color=red!75!DarkGreen] (0,0.45) node[right] {\small $\rho$};
\draw[color=red!75!DarkGreen, thick, decoration = {markings, mark=at position 0.3 with {\arrow[scale=1]{stealth}},mark=at position 0.8 with {\arrow[scale=1]{stealth}}}, postaction=decorate] (-0.3,1.2) arc [start angle=180, end angle=360, x radius=0.3, y radius=0.25];
\draw (0.3,0) node[above right = -2pt]{\small \color{red!75!DarkGreen} $\nu$};
\draw (-0.3,0) node[above left = -2pt]{\small \color{red!75!DarkGreen} $\mu$};
\draw[color=red!75!DarkGreen, thick, decoration = {markings, mark=at position 0.3 with {\arrow[scale=1]{stealth}},mark=at position 0.8 with {\arrow[scale=1]{stealth}}}, postaction=decorate] (0.3,0) arc [start angle=0, end angle=180, x radius=0.3, y radius=0.25];
\draw (-0.3,1.2) node[left]{\small \color{red!75!DarkGreen} $\mu$};
\draw (0.3,1.2) node[right]{\small \color{red!75!DarkGreen} $\nu$};
\draw [fill={blue!70!green}] (0.3,0) node[below]{\small \color{blue!70!green} $z$} circle (0.04);
\draw [fill={blue!70!green}] (0.3,1.2) node[above]{\small \color{blue!70!green} $t$} circle (0.04);
\draw [fill={blue!70!green}] (-0.3,0) node[below]{\small \color{blue!70!green} $y$} circle (0.04);
\draw [fill={blue!70!green}] (-0.3,1.2) node[above]{\small \color{blue!70!green} $x$} circle (0.04);
\draw [fill=red!75!DarkGreen] (0,0.25) node[above left = -1.5pt]{\tiny \color{red!75!DarkGreen} $\alpha$} circle (0.04);
\draw [fill=red!75!DarkGreen] (0,0.95) node[below left = -2.5pt]{\tiny \color{red!75!DarkGreen} $\alpha$} circle (0.04);
\end{tikzpicture}
\end{gathered}.
\end{equation}
Since the right-hand side is only non-zero in the $\rho=1$ channel by \eqref{kirby.relation}, it can be simplified to 
\begin{equation}
\sum_a \overline{\Psi}_{a(\mu,xy)} \Psi_{a(\nu,zt)} = \delta_{\mu,\nu} \Psi_{11} \sum_a \frac{\Psi_{a1}}{d_\mu}~
\begin{gathered}
\begin{tikzpicture}
\draw[color=blue!70!green, thick, decoration = {markings, mark=at position 0.92 with {\arrow[scale=1]{stealth}}}, postaction=decorate] (0,0) circle [x radius =0.7, y radius=0.7];
\draw[color=blue!70!green] (0.6,-0.3) node[right] {$a$};
\draw[color=red!75!DarkGreen, thick, decoration = {markings,mark=at position 0.6 with {\arrow[scale=1]{stealth}}}, postaction=decorate] (-0.3,1.2) arc [start angle=180, end angle=360, x radius=0.3, y radius=0.25];
\draw (-0.3,0) node[above left = -2pt]{\color{red!75!DarkGreen} $\mu$};
\draw[color=red!75!DarkGreen, thick, decoration = {markings, mark=at position 0.6 with {\arrow[scale=1]{stealth}}}, postaction=decorate] (0.3,0) arc [start angle=0, end angle=180, x radius=0.3, y radius=0.25];
\draw (-0.3,1.2) node[below left]{\color{red!75!DarkGreen} $\mu$};
\draw [fill={blue!70!green}] (0.3,0) node[below]{\small \color{blue!70!green} $z$} circle (0.04);
\draw [fill={blue!70!green}] (0.3,1.2) node[above]{\small \color{blue!70!green} $t$} circle (0.04);
\draw [fill={blue!70!green}] (-0.3,0) node[below]{\small \color{blue!70!green} $y$} circle (0.04);
\draw [fill={blue!70!green}] (-0.3,1.2) node[above]{\small \color{blue!70!green} $x$} circle (0.04);
\end{tikzpicture}
\end{gathered}
= \delta_{\mu,\nu} \delta_{x,t} \delta_{y,z} \sum_a \Psi_{a1}^2\,.
\end{equation}
Now note that
\begin{equation}
    \sum_a \Psi_{a1}^2 = S_{11} \sum_a d_a^2 = \frac{\dim \cC}{\sqrt{\dim \ZC}} = 1\,,
\end{equation}
where the last step follows from the well-known statement that the global quantum dimension of $\ZC$ is the square of the global quantum dimension of $\cC$. Both \eqref{almost.unitarity} and \eqref{almost.unitarity.2} then follow.

\begin{figure}[t]
  \centering
  \[
  \begin{gathered}
    \begin{tikzpicture}
      \draw [color=red!75!DarkGreen, thick, decoration = {markings, mark=at position 0.55 with {\arrow[scale=1]{stealth}}}, postaction=decorate] (0,0) -- (1.5,0) node [below] {$\mu$} -- (3,0);
      \plane{0}
      \draw [fill={blue!70!green}] (0,0) node[above]{\small \color{blue!70!green} $x$} circle (0.04);
      \begin{scope}[shift={(3,0)}]
        \plane{0}
        \draw [fill={blue!70!green}] (0,0) node[above]{\small \color{blue!70!green} $y$} circle (0.04);
      \end{scope}
      \shade [draw, ball color=yellow, opacity = 0.1] (1.5,0) circle (2.25);
      \end{tikzpicture}
  \end{gathered}
  \quad=\quad
  \sum_a \frac{\Psi_{a(\mu,xy)}}{d_a^2} \quad
  \begin{gathered}
  \begin{tikzpicture}
    \plane{0}
    \begin{scope}[shift={(3,0)}]
      \plane{0}
    \end{scope}
    \node[style={draw, dashed, shape=cylinder, aspect=2, minimum height=3.6cm, minimum width=1cm, color=blue!70!green, shape border rotate=0}] at (1.5,0.1) {};
    \draw[color=blue!70!green, thick, thick, opacity=0.5, postaction=decorate] (1.65,-0.4) arc [start angle=-90, end angle=90, x radius=0.3, y radius=0.5];
    \draw[color=blue!70!green, thick, thick, decoration = {markings, mark=at position 0.5 with {\arrow[scale=1]{stealth}}}, postaction=decorate] (1.65,0.6) node [above=2pt] {$a$} arc [start angle=90, end angle=270, x radius=0.3, y radius=0.5];
    \shade [draw, ball color=yellow, opacity = 0.1] (1.5,0) circle (2.25);
  \end{tikzpicture}
  \end{gathered}
  \]
  \\
  \[
  \begin{gathered}
    \begin{tikzpicture}
      \draw [color=red!75!DarkGreen, thick, decoration = {markings, mark=at position 0.55 with {\arrow[scale=1]{stealth}}}, postaction=decorate] (0,0) -- (1.5,0) node [below] {$\mu$} -- (3,0);
      \plane{0}
      \draw[color=blue!70!green, thick, decoration = {markings, mark=at position 0.35 with {\arrow[scale=-1]{stealth}}}, postaction=decorate] (-0.8,-0.8) -- (-0.4,-0.4) node[below=2pt]{$b$} -- (0,0);
      \draw [fill={blue!70!green}] (0,0) node[above]{\small \color{blue!70!green} $x$} circle (0.04);
      \begin{scope}[shift={(3,0)}]
        \plane{0}
        \draw[color=blue!70!green, thick, decoration = {markings, mark=at position 0.35 with {\arrow[scale=-1]{stealth}}}, postaction=decorate] (-0.8,-0.8) -- (-0.4,-0.4) node[below=2pt]{$c$} -- (0,0);
        \draw [fill={blue!70!green}] (0,0) node[above]{\small \color{blue!70!green} $y$} circle (0.04);
      \end{scope}
      \end{tikzpicture}
  \end{gathered}
  \qquad\qquad \leftrightarrow \qquad\qquad
  \begin{gathered}
  \begin{tikzpicture}
    \plane{0}
    \draw[color=blue!70!green, thick, decoration = {markings, mark=at position 0.35 with {\arrow[scale=-1]{stealth}}}, postaction=decorate] (-0.8,-0.8) -- (-0.4,-0.4) node[below=2pt]{$b$} -- (0,0);
    \begin{scope}[shift={(3,0)}]
      \plane{0}
      \draw[color=blue!70!green, thick, decoration = {markings, mark=at position 0.35 with {\arrow[scale=-1]{stealth}}}, postaction=decorate] (-0.8,-0.8) -- (-0.4,-0.4) node[below=2pt]{$c$} -- (0,0);
    \end{scope}
    \node[style={draw, dashed, shape=cylinder, aspect=2, minimum height=3.6cm, minimum width=1cm, color=blue!70!green, shape border rotate=0}] at (1.5,0.1) {};
    \draw[color=blue!70!green, thick] (0,0) -- (3,0);
    \draw[color=blue!70!green, thick, thick, opacity=0.5, postaction=decorate] (1.65,-0.4) arc [start angle=-90, end angle=90, x radius=0.3, y radius=0.5];
    \draw[color=blue!70!green, thick, thick, decoration = {markings, mark=at position 0.35 with {\arrow[scale=1]{stealth}}}, postaction=decorate] (1.65,0.6) node [above=2pt] {$a$} arc [start angle=90, end angle=270, x radius=0.3, y radius=0.5];
    \draw [fill={blue!70!green}] (1.35,0) node[below left=-1pt]{\small \color{blue!70!green} $z$} circle (0.04);
  \end{tikzpicture}
  \end{gathered}
  \]
  \caption{ Boundary crossing relation in 2+1d TQFT.  The blue dashed lines sketch the surfaces of the gapped boundaries, such that the regions outside the slabs and inside the cylinders are empty.  Top: A special case that involves the half-linking matrices $\Psi_{a(\mu,xy)}$.  Bottom: The general case.}
  \label{fig:BoundaryCrossing}
\end{figure}
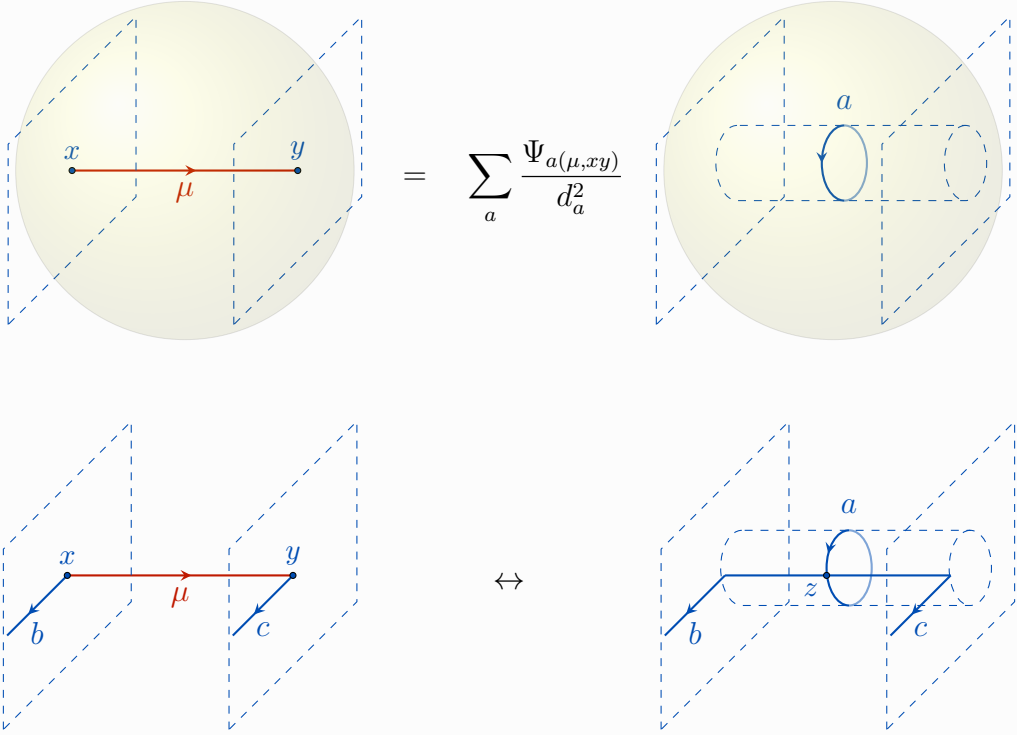

The invertibility of the half-linking matrix implies an isomorphism---the 2+1d version of the 1+1d \emph{boundary crossing relation} \cite{Huang:2021zvu}---as depicted on the top of Figure~\ref{fig:BoundaryCrossing}.  On the left-hand side, we perform a path integral in a shaded ball region $B_3$ to define a state $|\mu,xy\rangle_{S^2}$ on the bounding sphere $S^2$.  On the right-hand side, we do the same in $B'_3$ to define $|a\rangle_{S^2}$.  The inner product $\langle a | \mu,xy \rangle_{S^2}$ of the two states is computed by gluing the two balls along the sphere.  Since $B_3 \amalg_{S_2} B'_3 = S_3$, the glued configuration is a spherically-shaped gapped boundary with a half-linking between $a$ and $(\mu,xy)$ inside the $S^3$, and hence $\langle a | \mu;xy \rangle_{S^2} = \Psi_{a(\mu,xy)}$. The invertibility of $\Psi_{a(\mu,xy)}$ implies that a basis transformation $| \mu;xy \rangle_{S^2} = \sum_a \alpha_a |a\rangle_{S^2}$ exists, and if the partition functions of $S^2$- and $T^2$-shaped gapped boundaries are normalized to one, then the coefficients $\alpha_a$ can be fixed to be $\Psi_{a(\mu,xy)}/d_a^2$.  The bottom of Figure~\ref{fig:BoundaryCrossing} depicts a natural generalization of the boundary crossing relation that should involve generalized half-linking matrices.  This generalized relation provides an intuitive physical picture for the equivalence between $\ZC$ and $\tub(C)$.

\section{Decomposition of anyon chains into representations}
\label{app:chain}

Consider a periodic anyon chain with $L$ sites constructed out of a fusion category $\cC$ and a distinguished object $\rho$, with basis states given by the diagram
\begin{equation}
  \label{AnyonChainUntwisted}
\ket{a_1 a_2 \cdots a_L} =
\begin{gathered}
\begin{tikzpicture}[scale=.8]
\draw [line,-<-=.1,-<-=.3,-<-=.7,-<-=.9] (-1,0) node {$/\!/~$} -- (-.5,0) node [below] {$a_1$} -- (.5,0) node [below] {$a_2$}
-- (1.5,0) node [below] {$\dotsb$}
-- (2.5,0) node [below] {$a_L$} -- (3.5,0) node [below] {$a_1$} -- (4,0) node {$~/\!/$};
\draw [line,-<-=.5] (0,1) node [above] {$\rho$} -- (0,0);
\draw [line,-<-=.5] (1,1) node [above] {$\rho$} -- (1,0);
\draw [line,-<-=.5] (2,1) node [above] {$\rho$} -- (2,0);
\draw [line,-<-=.5] (3,1) node [above] {$\rho$} -- (3,0);
\end{tikzpicture}
\end{gathered}
\in \cH^{(L)}, \qquad\quad
\end{equation}
where $a_i \in \cC$, the fusion $a_i \otimes \rho$ contains $a_{i+1}$, and the slashes $/\!/$ represent periodic identification.  For simplicity, we assume that the fusion coefficients are either 0 or 1 to avoid specifying $\Hom(a_i \otimes \rho, a_{i+1})$ at junctions, but this assumption will not matter in the end.  The anyon chain has global symmetry $\cC$, where the action of $\sigma \in \cC$ is realized by fusing 
\begin{equation}
  \qquad\qquad\qquad O_\sigma = 
\begin{gathered}
\begin{tikzpicture}[scale=.8]
\draw [line,-<-=.5,] (-1,0) node {$/\!/~$} -- (1.5,0) node [below] {$\sigma$} -- (4,0) node {$~/\!/$};
\end{tikzpicture}
\end{gathered}
\in \Hom(\cH^{(L)}, \cH^{(L)})
\end{equation}
on \eqref{AnyonChainUntwisted} from below.
A $\tau$-twisted periodic anyon chain can be constructed by inserting $\tau \in \cC$ from below, with basis states given by 
\begin{equation}
  \label{AnyonChainTwisted}
\ket{\tau; a_1 a_2 \cdots a_L} =
\begin{gathered}
\begin{tikzpicture}[scale=.8]
\draw [line,-<-=.1,-<-=.3,-<-=.7,-<-=.9] (-1,0) node {$/\!/~$} -- (-.5,0) node [below] {$a_1$} -- (.5,0) node [below] {$a_2$}
-- (1.5,0) node [below] {$\dotsb$}
-- (2.5,0) node [below] {$a_{L+1}$} -- (3.5,0) node [below] {$a_1$} -- (4,0) node {$~/\!/$};
\draw [line,->-=.5] (0,-1) node [below] {$\tau$} -- (0,0);
\draw [line,-<-=.5] (1,1) node [above] {$\rho$} -- (1,0);
\draw [line,-<-=.5] (2,1) node [above] {$\rho$} -- (2,0);
\draw [line,-<-=.5] (3,1) node [above] {$\rho$} -- (3,0);
\end{tikzpicture}
\end{gathered}
\in \cH_\tau^{(L)}. \qquad\qquad
\end{equation}
In the presence of the $\tau$-twist, given an element $x \in \Hom(\sigma \otimes \tau, \tau' \otimes \sigma)$  of the tube algebra $\tub(\cC)$, we can fuse 
\begin{equation}
  \label{AnyonChainTube}
  \qquad\qquad\quad O_{\sigma; x}^{\tau, \tau'} = 
\begin{gathered}
\begin{tikzpicture}[scale=.8]
\draw [line,-<-=.5,] (-1,0) node {$/\!/~$} -- (1.5,0) node [below] {$\sigma$} -- (4,0) node {$~/\!/$};
\draw [line,->-=.25,->-=.75] (0,-1) node [below] {$\tau'$} -- (0,0) node [below right] {$x$} -- (0,1) node [above] {$\tau$};
\end{tikzpicture}
\end{gathered}
  \in \Hom(\cH_{\tau}^{(L)}, \cH_{\tau'}^{(L)})
\end{equation}
on \eqref{AnyonChainTwisted} from below.

We will decompose the Hilbert space $\cH_\tau^{(L)}$ of the $\tau$-twisted anyon chain
into representations of $\tub(\cC)$ in two steps.  First, we fuse all $\rho$'s together 
\begin{equation}
  \rho^{\otimes L} = \bigoplus_a \, [\mathbf{N}_\rho^L]_{1a} \, a,
\end{equation}
where $\mathbf{N}_\rho$ is a matrix with elements $[\mathbf{N}_\rho]_{ab} = N_{\rho a}^b$, and consider a new set of basis states
\begin{equation}
| a, n; \tau; a_1, a_2 \rangle =
\begin{gathered}
\begin{tikzpicture}[scale=.8]
\draw [line,-<-=.17,-<-=.5,-<-=.83] (-1,0) node {$/\!/~$} -- (-.5,0) node [below] {$a_1$} -- (.5,0) node [below] {$a_2$}
-- (1.5,0) node [below] {$a_1$} -- (2,0) node {$~/\!/$};
\draw [line,->-=.5] (0,-1) node [below] {$\tau$} -- (0,0);
\draw [line,-<-=.5] (1,1) node [above] {$a$} -- (1,0);
\end{tikzpicture}
\end{gathered}
\in \cH_\tau^{(L)},
\end{equation}
where $a \in \cC$ and $n = 1, \dotsc, [\mathbf{N}_\rho^L]_{1a}$.  We can further pass to another set of basis states
\begin{equation}
  | a, n; \tau; b, x \rangle = ~
  \x{\tau}{a}{b}{x_{\tau,a}^b},
\end{equation}
which we recognize to be elements of $\tub(\cC)$ together with the additional label $n$.
By now, the earlier assumption that the fusion coefficients are either 0 or 1 has become immaterial.

The second step is to decompose $\tub(\cC)_{*,a} = \bigoplus_{\tau,b} \Hom_\cC(\tau \otimes b, b \otimes a)$ as a left\footnote{
Namely, fusion from below.
} $\tub(\cC)$-module into irreducible representations labeled by $\mu \in \ZC$, which gives
\begin{equation}
  \tub(\cC)_{*,a} = \bigoplus_\mu \langle \mu, a \rangle \mu.
\end{equation}
The decomposition of states of the anyon chain with a $\tau$-twist is then
\begin{equation}\label{AnyonDecomp}
  \dim \cH_{\tau;\mu}^{(L)} = \langle \mu, \tau \rangle \sum_a \langle \mu, a \rangle \, [\mathbf{N}_\rho^L]_{1a}.
\end{equation}

To study the $L \to \infty$ limit, we apply Perron-Frobenius theory to the non-negative matrix $\mathbf{N}_\rho$.  There exists a non-negative eigenvector $v$ with non-negative eigenvalue $r$, whose absolute value is greater than or equal to those of all other eigenvalues; $r$ is called the spectral radius.  
We further assume that $\mathbf{N}_\rho$ is irreducible in the Perron-Frobenius sense: for any given pair $a, b \in \cC$, there exists some positive integer power $L$ such that $[\mathbf{N}_\rho^L]_{ab} \ge 1$.  
For an anyon chain, $\rho$ is typically chosen to be a generating object, which means that for any given $a \in \cC$, there exists a positive integer $L$ such that $[\mathbf{N}_\rho^L]_{1a} \ge 1$.
This is not quite the definition of irreducibility but motivates the assumption thereof.

Under the assumption of irreducibility, Perron-Frobenius theory tells us that, first, $r > 0$, and second, even though the eigenvalues with absolute value $r$ may not be unique, there exists a positive integer $h$ called the \emph{period}, such that every eigenvalue with absolute value $r$ takes the form $\omega_i r$, with $\omega_i$ an $h$-th root of unity.  Moreover, no two $\omega_i$ are identical, and the unique eigenvector $v$ with eigenvalue $r$ has strictly positive components.  With $\mathbf{N}_\rho$ coming from the fusion rule of $\cC$, we identify $v_a = \dim a$, and also note that $r \ge 1$.

In the simple situation where $h = 1$, the eigenvector $v$ dominates the large $L$ asymptotics, so
\begin{equation}
  \lim_{L\to\infty} 
  \frac{[\mathbf{N}_\rho^L]_{1a}}{r^L} 
  \propto \dim a.
\end{equation}
Using \eqref{AnyonDecomp}, we find the asymptotic decomposition
\begin{equation}\label{}
  \lim_{L\to\infty} 
  \frac{\dim \cH_{\tau;\mu}^{(L)}}{r^L} 
  \propto \langle \mu, \tau \rangle \sum_a \langle \mu, a \rangle \dim a = \langle \mu, a \rangle \dim \mu.
\end{equation}
If $h \neq 1$, we define an appropriate average and find the asymptotics
\begin{equation}
  \lim_{L\to\infty} \frac{1}{h} \sum_{\ell=L}^{L+h-1} \frac{[\mathbf{N}_\rho^\ell]_{1a}}{r^L} \propto \dim a,
\end{equation}
resulting in the same asymptotic decomposition
\begin{equation}
  \lim_{L\to\infty} \frac{1}{h} \sum_{\ell=L}^{L+h-1} \frac{\dim \cH_{\tau;\mu}^{(L)}}{r^L} \propto \langle \mu, a \rangle \dim \mu.
\end{equation}

\def\arxivfont{\rm}
\bibliographystyle{ytamsalpha}

\baselineskip=.98\baselineskip

\bibliography{ref}

\end{document}